\documentclass[10pt,a4paper]{book}
\usepackage[english]{babel}
\usepackage{graphicx}
\usepackage{dcolumn}
\usepackage{amsmath,mathrsfs,amssymb,amsthm}
\usepackage{hhline}
\usepackage{wasysym,enumerate,color}

\newcommand{\be}{\begin{equation}}
\newcommand{\ee}{\end{equation}}
\newcommand{\bea}{\begin{eqnarray}}
\newcommand{\eea}{\end{eqnarray}}

\newcommand{\ds}{\displaystyle}

\begin{document}

\null\vskip 2.09cm
\parindent 0cm
\begin{center}
\Huge{Manual for the flexible DM-NRG code \\
        Version 1.0.0}
%\author{\"O. Legeza, C. P. Moca,  A. I. T\'oth, I. Weymann and G. Zar\'and}
\vskip4cm
%\date{\today}
%\maketitle
{\Large Team members:
\vskip0.4cm
\"O. Legeza \\
C. P. Moca \\
A. I.  T\'oth \\
I. Weymann \\
G. Zar\'and  \\
}
\end{center}

%%%%%%%%%%%%%%%%%%%%%%%%%%%%%%%%%%%%%%%%%%%%%%%%%%%%%%%%%%%%%%%%%%%%%%%%%%%%%%%%
%                      Table of contents                                       %
%%%%%%%%%%%%%%%%%%%%%%%%%%%%%%%%%%%%%%%%%%%%%%%%%%%%%%%%%%%%%%%%%%%%%%%%%%%%%%%%

\tableofcontents

%%%%%%%%%%%%%%%%%%%%%%%%%%%%%%%%%%%%%%%%%%%%%%%%%%%%%%%%%%%%%%%%%%%%%%%%%%%%%%%%
%                      CHAPTER                                                 %
%%%%%%%%%%%%%%%%%%%%%%%%%%%%%%%%%%%%%%%%%%%%%%%%%%%%%%%%%%%%%%%%%%%%%%%%%%%%%%%%

\chapter{Introduction}
Quantum impurity  models describe interactions between some local
degrees of freedom  (e.g.\ a spin)  and a continuum of
non-interacting fermionic or bosonic states. The investigation of
quantum impurity models is a starting point towards the
understanding of more complex strongly correlated systems, but
quantum impurity models also provide the description of various
correlated mesoscopic structures, biological and chemical
processes, atomic physics  and describe phenomena such as
dissipation or dephasing. Prototypes of these models are the
Anderson impurity model, or the single- and multi-channel Kondo
models. The first two models are classic examples of Fermi liquid
models, while the multi-channel Kondo model is the most basic
example of a non-Fermi liquid system, and as such, it serves
possibly as the simplest realization of a quantum critical state.

The solution of these models for low energies was a major issue in
theoretical condensed matter research and led to the development
of various non-perturbative techniques. [The interested readers
are referred to the seminal book of Hewson \cite{hewson_book} and
to the extensive review of Cox and Zawadowski \cite{Cox}.]
However,  despite the extensive effort, many of the methods
developed are uncontrolled, while others can be applied only to a
subclass of models or to restricted regions of the parameter
space. Wilson's numerical renormalization method, originally
developed for the Kondo model, remained possibly the most reliable
method to study dynamical correlations of generic quantum impurity
models as well as their thermodynamic properties and finite size
spectra. It is still one of the most popular methods to study
quantum impurity models.
%G and its influence over the development of other methods is also significant.

Although Wilson's NRG has been used in its original form for a
longtime, a number of new developments took place recently: First,
a spectral sum-conserving  density matrix NRG approach (DM-NRG)
has been developed \cite{Anders,Delft}, which has very recently
been generalized  for non-Abelian symmetries
\cite{our_dmnrg_paper}. We remark that using symmetries as much as
possible is necessary in many cases to perform accurate enough
calculations. NRG has been also restructured as a matrix product
state approach~\cite{Weichselbaum07}. These new developments not
only made NRG much more reliable than Wilson's original
method~\cite{our_dmnrg_paper,Hofstetter}, but they made it
possible to extend NRG to study non-equilibrium phenomena
\cite{Anders_05,Anders_08}, and opened the way to use methods
familiar from the density matrix renormalization group (DMRG)
approach \cite{dmrg_review}.

In  this manual we do not intend to give a complete description of
the NRG machinery. Rather, we introduce some of the basic concepts that are
 needed to {\em use NRG} and to {\em use the code} we provide for quantum
impurity problems of interest. If you want to understand in
detail, how NRG and DM-NRG work, we advise you to consult Wilson's
original work~\cite{Wilson,Krishnamurthy}, and the references
listed above.

The code we describe in this manual is a free  density 
matrix numerical renormalization group (DM-NRG)
 code, which can be downloaded from the site
{\tt  http://www.phy.bme.hu/$\sim$dmnrg}. 
This code is a flexible NRG code, which uses user-defined 
non-Abelian symmetries dynamically, computes spectral functions,   
expectation values of local operators for user-defined impurity models.  
The code can use a uniform density of states as well as a 
user-defined density of states. The current version of 
the code assumes fermionic bath's. 
It uses any number of $U(1)$, $SU(2)$ charge $SU(2)$ or $Z_2$
symmetries, but the interested user can teach the code
other symmetries too. The code runs using a simple input file. 
We provide several example input files with the code as well as 
a few Mathematica files with which these input files can easily 
be constructed. An energy spectrum analyzer utility 
is also provided with the code to study finite size spectra too.

\chapter{ A short introduction to DM-NRG and to the use of symmetries}
\label{ch:NRG_intro}

In this chapter, we give a short overview of Wilson's numerical
renormalization group (NRG) and the density matrix NRG (DM-NRG).
As already mentioned in the introduction, here we introduce only
the basic concepts that are needed to {\em use NRG} and to {\em
use the code} we provide for quantum impurity problems, but we do
not attempt/intend to give a complete overview of the existing
literature. To learn more details about NRG and DM-NRG, we
recommend to read  Wilson's original work~\cite{Wilson,Krishnamurthy}.

\section{Wilson's NRG}

\subsection{The simplest example: The Kondo model}

Before introducing the general concepts, let us discuss the
simplest possible example, the Kondo model. The Kondo model
consists of a spin $S$ interacting locally with a non-interacting
conduction electron sea,
\be H_{Kondo} = \frac J 2 \vec S
\sum_{\sigma,\sigma'}\psi^\dagger_\sigma \vec
\sigma_{\sigma,\sigma'}  \psi_{\sigma'} + H_{\mathrm{cond}}\;.
\label{Kondo}
\ee
Here $J$ denotes the Kondo coupling, $\psi^\dagger_\sigma$ creates
a conduction electron at the impurity site, and $\vec
\sigma_{\sigma,\sigma'}$ is the vector of Pauli spin matrices. The
term, $H_{\mathrm{cond}}$, describes the conduction electron bath,
and its specific form is not very important for us.

\subsubsection{Wilson's mapping and iterative diagonalization}

\begin{figure}[t]
\centering
\includegraphics[width=8cm]{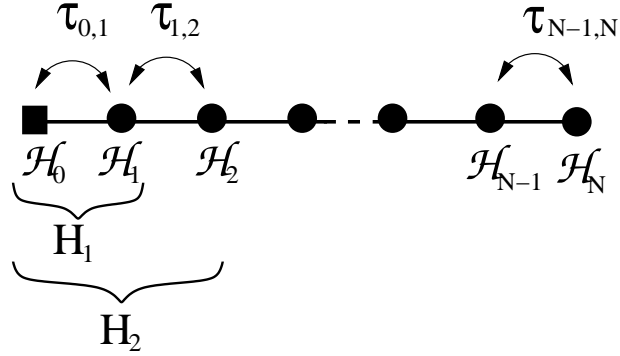}
\caption{ Wilson chain of length N. The impurity site is at the
origin and is represented by a square, while the sites are
represented as circles. For the impurity the Hamiltonian ${\cal
H}_0$ is identical with $H_0$. As we add sites the iterative
construction of the Hamiltonian at a given site $H_n$ is pictured.
} \label{fig:wilson_chain}
\end{figure}

From the point of view of local dynamics, $H_{\mathrm{cond}}$
contains a lot of redundant information: In fact, since our
fermions are non-interacting, the {\em local density of states}
$\varrho(\omega)$, i.e. the spectral function of the unperturbed
Green's function, $G_{\psi_\sigma,\psi^\dagger_\sigma}(\omega)$,
is generated by $H_{\mathrm{cond}}$ with $J=0$. Note that
$\varrho(\omega)$ determines completely the correlation functions
of $\psi^\dagger_\sigma$ and the spin dynamics  even for the
interacting system, $J\ne0$. The ingenious idea of Wilson was to
discretize logarithmically $\varrho(\omega)$ using a
discretization parameter $\Lambda>1$, and thus map approximately
the Hamiltonian \eqref{Kondo} to a semi-infinite chain (see Fig.
\ref{fig:dos} and Fig.~\ref{fig:wilson_chain}):

\begin{figure}[h]
\centering
\includegraphics[width=7cm]{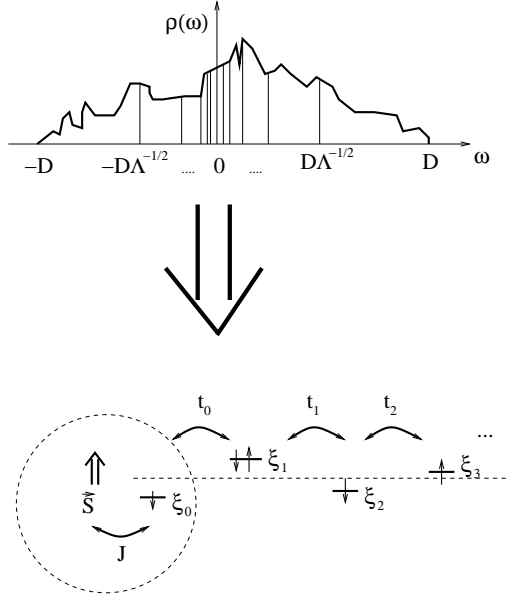}
\caption{Discretization of the conduction band density of states
on a logarithmic mesh and mapping onto the Wilson chain (see also
Fig.~\ref{fig:wilson_chain}).} \label{fig:dos}
\end{figure}

\be
 H^{Wilson}_{Kondo} = \frac J 2 \vec S \sum_{\sigma,\sigma'}f^\dagger_{0,\sigma} \vec
\sigma_{\sigma,\sigma'}  f_{0,\sigma'} + \sum_{n=0}^\infty
\sum_\sigma \xi_n \; f^\dagger_{n,\sigma}f_{n,\sigma} +
\sum_{n=0}^\infty \sum_\sigma t_n \;
(f^\dagger_{n,\sigma}f_{n+1,\sigma} + h.c.) \;.
\label{Wilson_Kondo} \ee

Here the spin interacts only with the fermion
$f^\dagger_{0,\sigma}$ at the end of the Wilson chain. The
operator $f^\dagger_{0,\sigma}$ creates a conduction electron at
the impurity site, and is essentially identical to the operator
$\psi^\dagger_\sigma$. The  on-site energies $\xi_n$ and the
hoppings $t_n$ depend solely on $\varrho(\omega)$ and can be
determined  recursively \cite{Wilson}. Our code takes care of this
part of the work: it determines numerically these constants if a
density of states $\varrho(\omega)$ is provided (see section
\ref{ch:output} and the description of the utility {\tt he}).

For a flat and symmetrical density of states,
$\varrho(\omega)=1/(2D)$, one has $\xi_n=0$ and the  $t_n$'s can
be determined analytically  ( $t_n\sim \Lambda^{-n/2}$)
\cite{Wilson}. Longer and longer chains give more and more
accurate description of the infinite chain. This observation led
Wilson and his co-workers to solve the Hamiltonian
\eqref{Wilson_Kondo} iteratively. Introducing the operator
 \be
H_n \equiv  \frac J 2 \vec S \sum_{\sigma,\sigma'}
f^\dagger_{0,\sigma} \vec \sigma_{\sigma,\sigma'}  f_{0,\sigma'} +
\sum_{m=0}^n \sum_\sigma \xi_m \; f^\dagger_{m,\sigma}f_{m,\sigma}
+ \sum_{m=0}^n \sum_\sigma t_m \;
(f^\dagger_{m,\sigma}f_{m+1,\sigma} + h.c.) \;. \ee
one has the obvious recursion relation,
\be
 H_{n+1} = H_n + {\tau}_{n,n+1} + {\cal H}_{n+1}\;,
 \label{eq:recursion}
\ee
with the notation, ${\tau}_{n,n+1}=t_n \sum_\sigma
(f^\dagger_{n,\sigma}f_{n+1,\sigma} + h.c.) $ and ${\cal
H}_{n+1}=\sum_\sigma \xi_{n+1} \;
f^\dagger_{n+1,\sigma}f_{n+1,\sigma}$ (see also
Fig.~\ref{fig:wilson_chain}). Then Wilson's procedure consists of
constructing from  the  low-energy eigenstates, $|u\rangle_n$, of
the operator  $H_n$ approximate eigenstates, $|\tilde
u\rangle_{n+1}$, of the operator $H_{n+1}$. To do this, one takes
a  definite number of the states $|u\rangle_n$ with the lowest
energies and generates new states from them by first adding an
empty site,  $|u\rangle_n \to |u\rangle_{n+1}$ and then  using the
operators $f^\dagger_{n+1,\sigma}$ to create new states as \be
|u\rangle_n \to \left (\begin{array}{l}
 |u\rangle_{n+1}  \\
 f^\dagger_{n+1,\uparrow}|u\rangle_{n+1}\\
 f^\dagger_{n+1,\downarrow}|u\rangle_{n+1} \\
 f^\dagger_{n+1,\uparrow}f^\dagger_{n+1,\downarrow}|u\rangle_{n+1}
\end{array}\right)\;\;.
 \;
\ee
Then one diagonalizes the Hamiltonian $H_{n+1}$ in this new basis
set. To construct the matrix elements of $H_{n+1}$ in this new
basis, one needs the following information
\begin{enumerate}
\item The eigenvalues $E_u^n$ of $H_{n}$,
\item  The matrix elements  of ${\cal H}_{n+1}$ between the {\em
    local states}, $|\mu\rangle$, constructed from the vacuum state $| 0\rangle$ as
\be
\{|\mu\rangle  \} \equiv
\left (\begin{array}{l}
 |0\rangle \\
 f^\dagger_{n+1,\uparrow}|0\rangle,\\
 f^\dagger_{n+1,\downarrow}|0\rangle\\
 f^\dagger_{n+1,\uparrow}f^\dagger_{n+1,\downarrow}|0\rangle
\end{array}\right)\;
\ee
\item  The matrix elements  of $f^\dagger_{n+1,\sigma}$ between the {\em
    local states},  $|\mu\rangle$,  and
\item The matrix elements  of $f^\dagger_{n,\sigma}$ between the {\em
    block states} $|u\rangle_n$.
\end{enumerate}
Diagonalizing the Hamiltonian  $H_{n+1}$ one then obtains the new eigenstates
$|\tilde u\rangle_{n+1}$, their eigenvalues,   $E_{\tilde u}^{n+1}$, and one can
trivially compute the matrix elements of $f^\dagger_{n+1,\sigma}$
between these states too, to proceed to  the next iteration.

If one is to compute  the spectral function of a local operator
$A$ acting at site $n=0$, then one has to keep track of the matrix
elements $_n\langle u|A|v\rangle_n $ of this operator, too.
Already the finite size spectrum, i.e.  the spectrum of $H_n$
contains a lot of precious information \cite{Wilson}. However,
once the matrix elements and the approximate spectrum of the
Hamiltonian  is at hand, one can go ahead and also compute thermal
expectation values or spectral functions from
it~\cite{Costi,Bulla}, and thus determine completely the
properties of a quantum impurity problem.

\subsubsection{Symmetries in the Kondo model}

The total spin operator
\begin{equation}
{\vec S}_{T} = {\vec S} +\frac{1}{2}\sum_{n=0}^{\infty} \sum_{\sigma, \sigma' \in\{\uparrow , \downarrow \}}
f_{n,\sigma}^{\dagger} {\vec \sigma}_{\sigma, \sigma'} f_{n,\sigma'}\label{eq:st}
\end{equation}
as well as the charge operator,
\be
 Q  =  \sum_{n=0}^{\infty} \left(
 \sum_{\sigma=\{\uparrow, \downarrow
 \}} f_{n, \sigma}^{\dagger}f_{n,\sigma}-1 \right)
\ee
commute with Eq.~\eqref{Wilson_Kondo}. This implies that  the
eigenstates of the Hamiltonian can be classified as {\em
multiplets}. Every multiplet $u$ will be characterized by a charge
quantum number, $Q(u)$ and a spin quantum number, $S(u)$. We shall
refer to these as {\em representation indices}. Internal states
within a multiplet are labeled by the $z$-component of the spin,
that we shall refer to as {\em labels}. Similarly, operators can
also be characterized by  quantum numbers. To give an example, in
this simple case, the components of
$\{f^\dagger_{n,\uparrow},f_{n,\downarrow}\}$ and
$\{f^\dagger_{n,\downarrow},-f_{n,\uparrow}\}$ transform under
spin rotations as the components $|\pm1/2\rangle $ of a spin $S_f
=1/2$ state, and have charges  $ Q_{f^\dagger}=1/2$ and $
Q_{f}=-1/2 $, respectively. The three operators formed from the
impurity spin, ${\cal S}_m\equiv\{-S^+/\sqrt{2},S^z,
S^-/\sqrt{2}\}$ transform, on the other hand, as the three
components of a spin $S_S=1$ state, $|m=0,\pm\rangle$, while they
have trivially charge $Q_S=0$. The operators discussed before
provide simple examples of irreducible tensor operator multiplets
$\{A\}$.   Notice the non-trivial prefactors in the previous
examples. These prefactors must always be worked out carefully,
since the  smallest sign mistake may alter the NRG results
significantly.

In the presence of symmetries, a powerful theorem, the
Wigner--Eckart theorem tells us that, in order to compute the
matrix elements of a member of an operator multiplet $\{A\}$
between two states within multiplets $u$ and $v$ a single matrix
element is needed, the so-called reduced matrix element
\cite{our_dmnrg_paper}:
\be \langle u||A||v\rangle \;. \ee

Wilson's procedure thus gets a bit modified. First of all, one
needs to classify not only the block states (multiplets) but also
the local states $\mu$ added at site $n+1$ by symmetries. Along
the iteration, one constructs from these new states using
group-theoretical methods (Clebsch-Gordan coefficients), which are
already eigenstates of $ \vec S_T^2$  and $ S^z$. Apart from this
twist, the discussion of the previous subsection is still valid,
and gets only slightly modified. We can thus make the following
statement: To construct the (reduced) matrix elements of $\langle
i || H_{n+1}||j\rangle$ in this basis, one needs the following
information
\begin{enumerate}
\item The eigenvalues $E_u^n$ of $H_{n}$,
\item  The matrix elements $\langle \mu ||{\cal H}_{n+1}||\nu\rangle$ between the {\em
    local multiplets}.
\item  The reduced matrix elements, $\langle \mu   ||f^\dagger_{n+1}||\nu\rangle$,
\item The reduced matrix elements  $_{n}\langle u ||f^\dagger_{n}||v\rangle_{n}$ between the {\em
    block multiplets}, $|u\rangle_n$.
\item Finally, to compute a spectral function of an operator multiplet
$A$, one also needs $_{n}\langle u ||A||v\rangle_{n}$.
\end{enumerate}
At this point, the reader does not need to know in detail, how the construction of ${H}_{n+1}$
takes place, since the code does that automatically,
but for more details, the interested reader is referred to  Ref.~\cite{our_dmnrg_paper}.

\subsubsection{ Kondo model with $U_{charge}(1)\times U_{spin}(1)$ symmetry}

Let us first illustrate the concepts introduced above  through the
simplest case, when every symmetry used is Abelian. In this case
every multiplet consists of a single state, which is characterized
by some quantum numbers. Let us thus consider the Kondo model with
two $U(1)$ symmetries, generated by the $z$-component of the total
spin, $S_z$  and the charge operators, $Q$. Now $S_z$  and  $Q$,
provide the quantum numbers according to which the multiplets of
the Hamiltonian are classified. The classification of the states
in this case is rather trivial and the computation of the reduced
matrix elements is also simple, since all the Clebsch-Gordan 
coefficients are 1  or
0. However, without spin $SU_{spin}(2)$ symmetry, the hoppings of
$f_{n,\uparrow}^\dagger $ and $f_{n,\downarrow}^\dagger $ are not
related by symmetry. There are therefore {\em two hopping
operators} which we need to keep track of, $f_{n,\uparrow}^\dagger
$ and $f_{n,\downarrow}^\dagger $.

%When an external magnetic field is applied to the system
%the total spin operator eq. (\ref{eq:st}) does no
%longer commute with the Zeemann term in the Hamiltonian so
%the magnetic field reduces the spin $SU_{spin}(2)$ symmetry down to $U_{spin}(1)$.
%Still, for simplicity, we consider the external magnetic field equal to zero.
%Here,  we show how to construct the initial states, as well as the
%reduced matrix elements for the Hamiltonian, when only a combination of
%$U_{charge}(1)\times U_{spin}(1)$  symmetries is used.   Now,
%the $z$-component of the total spin,

The initial block states and the matrix elements of the hopping
operators can be constructed immediately. A {\it Mathematica}
file, {\tt kondo\_U1\_charge\_U1\_spin.nb}, with all the numerical
details of this calculation as well as an input file is  provided
by the package. The corresponding block and the local states are
listed in Table \ref{table:block_states_one}.

\begin{table}[tb]
\centering
\begin{tabular}{lll}
\hline
\hline
 Block      &                               & \\
states &       $Q$                       &    $ S_z$\\
                                       &                &  \\
\hline
                                       &                & \\
$|1\rangle =  |\Downarrow \rangle $    & $-1$             &  $-\frac{1}{2}$\\
       &                               &  \\
$|2\rangle =  |\Uparrow\rangle $       &   $-1$           &  $\frac{1}{2}$\\
          &                              &  \\
$|3 \rangle =  f_{0,\downarrow}^\dagger|\Downarrow \rangle $   & 0             &  $-1$\\
          &                              &  \\
$|4 \rangle = \frac1{\sqrt{2}}\bigl(  f_{0,\downarrow}^\dagger |\Uparrow
\rangle -f_{0,\uparrow}^\dagger |\Downarrow \rangle \bigr) $   & 0             &  0\\
          &                              &  \\
$|5 \rangle =  \frac1{\sqrt{2}}\bigl(  f_{0,\downarrow}^\dagger |\Uparrow
\rangle +f_{0,\uparrow}^\dagger |\Downarrow \rangle \bigr) $   & 0             &  0\\
          &                              &  \\
$|6 \rangle =  f_{0,\uparrow}^\dagger |\Uparrow \rangle $   & 0             &  1\\
          &                              &  \\
$|7 \rangle =  f_{0,\uparrow}^\dagger f_{\downarrow}^\dagger |\Downarrow \rangle $   & 1             &  $-\frac{1}{2}$\\
          &                              &  \\
$|8 \rangle =  f_{0,\uparrow}^\dagger f_{\downarrow}^\dagger |\Uparrow \rangle $   & 1             &  $\frac{1}{2}$\\
          &                              &  \\
\hline \hline
\end{tabular}
\hskip2cm
\begin{tabular}{lll}
\hline
\hline
Added/local       &                               & \\
states &       $q$                      &     $s_z$\\
                                       &                &  \\
\hline
                                       &                & \\
$|1\rangle =  | 0 \rangle $    & $-1$             &  0 \\
       &                               &  \\
$|2\rangle =  f_{n+1,\downarrow}^\dagger |0\rangle $       &   0           &  $-\frac{1}{2}$\\
          &                              &  \\
$|3 \rangle =  f_{n+1,\uparrow}^\dagger| 0 \rangle $   & 0             &  $\frac{1}{2}$\\
          &                              &  \\
$|4 \rangle =  f_{n+1,\uparrow}^\dagger f_{n+1,\downarrow}^\dagger |0 \rangle $   & 1             &  0\\
          &                              &  \\

\hline \hline
\end{tabular}

\caption{Block states (left) and local/added states (right) for
the single channel Kondo model when $U_{charge}(1)\times
U_{spin}(1)$ symmetry is used. The first block states are formed
from the impurity spin and the conduction electron at site $n=0$
of the Wilson chain.} \label{table:block_states_one}
\end{table}

In the first iteration, the
reduced matrix elements of the Hamiltonian matrix between the block states
(Table \ref{table:block_states_one}) are given by
\be
_{0}\langle u ||{H}_{0}||v\rangle_{0} =
J
\left(\begin{array}{cccccccc}
0 &      0 &      0 &      0 &      0 &      0 &      0 &      0 \\
0 &      0 &      0 &      0 &      0 &      0 &      0 &      0 \\
0 &      0 &      1/4&     0 &      0 &      0 &      0 &      0 \\
0 &      0 &      0 &      -3/4 &  0 &    0 &      0 &      0 \\
0 &      0 &      0 &      0 &     1/4 &   0 &      0 &      0 \\
0 &      0 &      0 &      0 &      0 &      1/4 &   0 &      0 \\
0 &      0 &      0 &      0 &      0 &      0 &      0 &      0\\
0 &      0 &      0 &      0 &      0 &      0 &      0 &      0 \\

\end{array}\right)\;.
\ee
In general there might be 
off-diagonal matrix elements in this matrix. The code
takes care of this and diagonalizes $_{0}\langle u ||{H}_{0}||v\rangle_{0}$
before proceeding to the next iteration.

The matrix elements of the $f_{0,\uparrow}^\dagger$ are \be
_{0}\langle u ||f_{0,\uparrow}^{\dagger}||v\rangle_{0} =
\left(\begin{array}{cccccccc}

0 &   0 &   0 &   0 &   0 &   0 &   0 &   0 \\
0 &   0 &   0 &   0 &   0 &   0 &   0 &   0 \\
0 &   0 &   0 &   0 &   0 &   0 &   0 &   0 \\
-1/\sqrt{2} &   0 &   0 &   0 &   0 &   0 &   0 &   0 \\
1/\sqrt{2} &   0 &   0 &   0 &   0 &   0 &   0 &   0 \\
0 &   1 &   0 &   0 &   0 &   0 &   0 &   0 \\
0 &   0 &   1 &   0 &   0 &   0 &   0 &   0 \\
0 &   0 &   0 &   1/\sqrt{2} &   1/\sqrt{2}&   0 &   0 &   0 \\

\end{array}\right)\;,
\ee

while those of  $f_{0,\downarrow}^\dagger$ are given as \be
_{0}\langle u ||f_{0,\downarrow}^{\dagger}||v\rangle_{0} =
\left(\begin{array}{cccccccc}
0 &    0 &    0 &    0 &    0 &    0 &    0 &    0 \\

0 &    0 &    0 &    0 &    0 &    0 &    0 &    0 \\

1 &    0 &    0 &    0 &    0 &    0 &    0 &    0 \\

0 &    1/\sqrt{2} &    0 &    0 &    0 &    0 &    0 &    0 \\

0 &    1/\sqrt{2} &    0 &    0 &    0 &    0 &    0 &    0 \\

0 &    0 &    0 &    0 &    0 &    0 &    0 &    0 \\

0 &    0 &    0 &    1/\sqrt{2} &    -1/\sqrt{2} &   0 &    0 &    0 \\

0 &    0 &    0 &    0 &    0 &    -1 &   0 &    0 \\
\end{array}\right)\;.
\ee
These matrix elements are needed to construct the hopping terms in the first
iteration. In addition, one also needs the matrix elements of the hopping
operators $f_{n+1,\uparrow}^{\dagger}$ and $f_{n+1,\downarrow}^{\dagger}$
between the {\em local states}, listed  in Table
\ref{table:block_states_one}. These  are given by
\be
\langle \mu ||f_{n+1,\uparrow}^{\dagger}||\nu \rangle =
\left(\begin{array}{cccc}
0 &  0 &  0 & 0 \\
0 &  0 &  0 & 0 \\
1 &  0 &  0 & 0 \\
0 &  1 &  0 & 0 \\
\end{array}\right)\;,
\phantom{nnn}\langle \mu ||f_{n+1,\downarrow}^{\dagger}||\nu \rangle =
\left(\begin{array}{cccc}
0 &  0 &  0 & 0 \\
1 &  0 &  0 & 0 \\
0 &  0 &  0 & 0 \\
0 &  0 &  -1 & 0 \\
\end{array}\right)\;.
\ee

Finally, if we want to determine the auto correlation function
of $S_z$, e.g., then we also need to compute the matrix elements this
operator,
\be
_{0}\langle u ||S_z||v\rangle_{0} =
\left(\begin{array}{cccccccc}
-1/2 &    0 &    0 &    0 &    0 &    0 &    0 &    0 \\

0 &    1/2 &    0 &    0 &    0 &    0 &    0 &    0 \\

0 &    0 &    -1/2 &    0 &    0 &    0 &    0 &    0 \\

0 &    0 &    0 &    0 &    1/2 &    0 &    0 &    0 \\

0 &    0 &    0 &    1/2 &    0 &    0 &    0 &    0 \\

0 &    0 &    0 &    0 &    0 &    1/2 &    0 &    0 \\

0 &    0 &    0 &    0 &    0 &   0 &    -1/2 &    0 \\

0 &    0 &    0 &    0 &    0 &    0 &   0 &    1/2 \\
\end{array}\right)\;.
\ee
We remark at this point that, while $S_z$ is an irreducible tensor operator
and it has a spin quantum number 0, the operators $S_x$ and $S_y$
are not irreducible tensor operators with respect to the $U_{spin}(1)$
spin rotations, since they transform among  each other.
They are, however, linear combinations of the
irreducible tensor operators, $S_\pm$, which  have spin quantum numbers $\pm 1$
under rotations around the $z$-axis.

Once all this information  typed into the input file, {\tt input.dat},  the flexible
DM-NRG code is ready to compute the finite size spectrum, the spin's spectral function, and
the real part of the retarded spin-spin correlation function using both traditional
NRG as well as DM-NRG methods! An  example input file for this case
is provided in the directory {\tt input\_files}.

\subsubsection{Kondo model with $U_{charge}(1)\times SU_{spin}(2)$ symmetry}

\begin{table}[t]
\centering
\begin{tabular}{lll}
\hline
\hline
Block       &                               & \\
states &       $Q$                       &     $S$\\
       &                               &  \\
\hline
       &                               & \\
$|1\rangle =  |\Uparrow \rangle $   & $-1$             &  $\frac{1}{2}$\\
       &                               & \\
$|2\rangle =  \frac{1}{\sqrt{2}}( f_{0,\downarrow}^{\dagger}|\Uparrow\rangle -
f_{0,\uparrow}^{\dagger}|\Downarrow \rangle)$          &   0 & 0 \\
          &                              &  \\
$|3 \rangle =  f_{0,\uparrow}^\dagger|\Uparrow \rangle $   & 0             &  1\\
          &                              &  \\
$|4 \rangle =  f_{0,\uparrow}^\dagger f_{0,\downarrow}^\dagger |\Uparrow \rangle $   & 1             &  $\frac{1}{2}$\\
          &                              &  \\
\hline \hline
\end{tabular}
\hskip3cm
\begin{tabular}{lll}
\hline
\hline
Local       &                               & \\
states &       $q$                       &     $s$\\
       &                               &  \\
\hline
       &                               & \\
$|1\rangle =  |0\rangle $   & $-1$             &  0 \\
       &                               & \\
$|2\rangle =  f_{n+1,\uparrow}^{\dagger}|0 \rangle$          &   0 & $\frac{1}{2}$ \\
          &                              &  \\
$|3 \rangle =  f_{n+1,\uparrow}^\dagger f_{n+1\downarrow}^\dagger |0 \rangle $   & 1             &  0\\
          &                              &  \\

\hline \hline
\end{tabular}
\caption{ Block states (left) and local/added states (right) for
the single channel Kondo model when $U_{charge}(1)\times
SU_{spin}(2)$ symmetry is used. The first block states are formed
from the impurity spin and the conduction electron at site $n=0$
of the Wilson chain.} \label{table:block_states}
\end{table}

Let us now show on the example of the Kondo model, how non-Abelian symmetries
can be used. Let us thus enumerate all necessary information that is
needed to perform a calculation for the spin spectral function of the single
channel Kondo model with $U_{charge}(1)\times SU_{spin}(2)$ symmetry. A
{\it Mathematica} file  ({\tt kondo\_U1\_charge\_SU2\_spin.nb}) where these inputs have been computed is also provided by the
package.

In the first iteration we have four different block multiplets,
$u=1,\dots,4$ formed from the impurity spin and the conduction
electron at site $n=0$, while there are three added multiplets in
each iteration, $\mu=1,2,3$. These states and their quantum
numbers are listed in Table \ref{table:block_states} (only highest
weight states are given). The reduced matrix elements of the
Hamiltonian between the block states then read \be _{0}\langle u
||{H}_{0}||v\rangle_{0} = J \left(\begin{array}{cccc}
0 & 0& 0& 0  \\
0 & -3/4& 0& 0  \\
0 & 0& 1/4& 0  \\
0 & 0& 0& 0
\end{array}\right)\;.
\ee In this case, the matrix elements of $f^\dagger_{0,\uparrow}$
and $f^\dagger_{0,\downarrow}$ are related with  each other by
symmetry, and therefore they form a single hopping operator
multiplet $f^\dagger_0$ of spin $1/2$. The matrix elements of
$f^\dagger_0$ need to be determined using the  Wigner--Eckart
theorem, and are given by \be _{0}\langle u
||f^\dagger_{0}||v\rangle_{0} = \left(\begin{array}{cccc}
0 & 0& 0& 0  \\
1 & 0 & 0& 0  \\
1 & 0& 0 & 0  \\
0 & 1/\sqrt{2}& -\sqrt{3/2} & 0
\end{array}\right)\;.
\ee
To generate the hopping, we also need to know
the reduced matrix elements of $f^\dagger_{n+1}$ 
between the added (local)
states. These are computed as
\be
\langle \mu ||{f^\dagger_{n+1}}||\nu \rangle
\left(\begin{array}{ccc}
 0& 0& 0  \\
1 & 0  & 0  \\
0  & -\sqrt{2} & 0
\end{array}\right)\;.
\ee

Finally, to compute the spin-spin correlation function, we need to compute the
reduced matrix elements of the impurity spin.
As discussed before, the components of the spin operator are grouped in this case into a single
tensor operator, $\cal S$. The reduced matrix elements of this operator are computed as
\be
_{0}\langle u ||{\cal S}||v\rangle_{0} =
\left(\begin{array}{cccc}
\sqrt{3/4}  & 0& 0& 0  \\
0 & 0 & -\sqrt{3/4} & 0  \\
0 & 1/2 & 1/\sqrt{2} & 0  \\
0 & 0 & 0 & \sqrt{3/4}
\end{array}\right)\;.
\ee

To run the code with $U_{charge}(1)\times SU_{spin}(2)$ symmetry,
one only needs to type this information into the input file, {\tt
input.dat}, and then run the code. You can find the corresponding
input file, {\tt kondo\_model\_U1\_c\_SU2\_s.dat} in the directory
{\tt input\_files}.

\begin{table}[t]
\centering
\begin{tabular}{lll}
\hline
\hline
Block       &                               & \\
states &       $Q$                       &     $S$\\
     &                               &  \\
\hline
       &                               & \\
$|1\rangle =  \frac{1}{\sqrt{2}}(f_{0,\downarrow}^{\dagger}|\Uparrow\rangle-f_{0,\uparrow}^{\dagger}|\Downarrow \rangle ) $   &   0      &  0\\
       &                               & \\
$|2\rangle =  f_{0,\uparrow}^\dagger|\Uparrow \rangle$          &   0 & 1 \\
          &                              &  \\
$|3 \rangle =  f_{0,\uparrow}^\dagger f_{0,\downarrow}^{\dagger}|\Uparrow \rangle $   & $\frac{1}{2}$      &  $\frac{1}{2}$\\
          &                              &  \\
\hline \hline
\end{tabular}
\hskip3cm
\begin{tabular}{lll}
\hline
\hline
Local       &                               & \\
states &       $q$                       &     $s$\\
       &                               &  \\
\hline
       &                               & \\
$|1\rangle =  f_{n+1,\uparrow}^\dagger| 0 \rangle $   & 0             &  $\frac{1}{2}$ \\
       &                               & \\
$|2\rangle =  f_{n+1,\uparrow}^{\dagger} f_{n+1,\downarrow}^\dagger|0 \rangle$          &   $\frac{1}{2}$ & 0 \\
          &                              &  \\
\hline \hline
\end{tabular}
\caption{ Block states (left) and local/added states (right) for
the single channel Kondo model when $SU_{charge}(2)\times
SU_{spin}(2)$ symmetry is used. The first block states are formed
from the impurity spin and the conduction electron at site $n=0$
of the Wilson chain.} \label{table:block_states_su2_su2}
\end{table}

\subsubsection{Kondo model with $SU_{charge}(2)\times SU_{spin}(2)$ symmetry}

The Hamiltonian \eqref{Wilson_Kondo} is invariant under the action
of $SU_{charge}(2)$ rotations  in charge  space, $ {\cal U}_c =
e^{i {\vec \omega}_Q  {\vec Q} } $, generated by the operators
$Q^{x}=(Q^{+}+Q^{-})/2$, $Q^{y}=(Q^{+}-Q^{-})/2i$ and $Q^z$ with
\begin{eqnarray}\label{eq:generators_charge}
Q^{+} & = & \sum_{n=0}^{\infty} (-1)^n f_{n,\uparrow}^{\dagger}f_{n,\downarrow}^{\dagger}\;, \nonumber\\
Q^{z} & = & \frac{1}{2}\sum_{n=0}^{\infty} \left( \sum_{\mu=\{\uparrow, \downarrow \}} f_{n, \mu}^{\dagger}f_{n,\mu}-1\right)\;,\\
Q^{-} & = & \left(Q^{+}\right)^{\dagger}\,,\nonumber
\end{eqnarray}
 and parametrized by real, three-component vectors ${\vec\omega}_Q$.
Since the spin symmetry generators commute with the charge
symmetry generators, the Hamiltonian \eqref{Wilson_Kondo}
possesses a symmetry, $SU_{charge}(2)\times SU_{spin}(2)$.

Charge SU(2) symmetries must always be used with care, and one
must always carefully check if a given set of operators forms an
irreducible tensor operator. One can check, e.g., that the
correctly defined hopping operators read, $\gamma^\dagger_n =
\{(\gamma_n)_{\sigma\tau}\} = \{ f_{n,\uparrow}^{\dagger},\;
f_{n,\downarrow}^{\dagger},\; (-1)^n\; f_{n,\downarrow},\;
(-1)^{n+1}\;f_{n,\uparrow} \}$, and that they indeed transform as
a set of spin $S=1/2$ and charge spin $Q=1/2$ irreducible tensor
operators. Notice the curious sign-dependence of the last two
operators and their relative signs. Similar factors of $(-1)^n$
appear in the states generated from the highest weight states in
Table~\ref{table:block_states_su2_su2} through the action of the
operator $Q^-$.

In the first iteration we have three different block multiplets,
$u=1,\dots,3$ formed from the impurity spin and the conduction
electron at site $n=0$. There are two added (local) multiplets in
each iteration, $\mu=1,2$. These states and their quantum numbers
are listed in Table \ref{table:block_states_su2_su2} (only highest
weight states are given). The reduced matrix elements of the
Hamiltonian between the block states then read \be _{0}\langle u
||{H}_{0}||v\rangle_{0} = J \left(\begin{array}{ccc}
-3/4 & 0& 0  \\
0 & 1/4& 0  \\
0 & 0& 0\\
\end{array}\right)\;.
\ee
The matrix elements of $\gamma^\dagger_0$ need to be determined
using the Wiegner-Eckart theorem, and are given by
\be _{0}\langle u ||{\gamma}_{0}^\dagger||v\rangle_{0} =
\left(\begin{array}{ccc}
0               &0              &  -\sqrt{2}\\
0               &0               & -\sqrt{2}\\
\frac{1}{\sqrt2}   &-\sqrt{\frac{3}{2}} &0\\
\end{array}\right)\;.
\ee
To generate the hopping, we also need to know
the reduced matrix elements of $\gamma^\dagger_{1}$ 
between the added (local)
states. These are computed as
\be
\langle \mu ||{\gamma^\dagger_{1}}||\nu \rangle =
\left(\begin{array}{cc}
0                       &\sqrt{2}\\
-\sqrt{2}        &0\\
\end{array}\right)\;.
\ee

Finally, to determine the spin spectral function, we need to compute the
reduced matrix elements
of the impurity spin operator:

\be
_{0}\langle u ||{\cal S}||v\rangle_{0} =
\left(\begin{array}{cccc}
0 & -\sqrt{3/4} & 0  \\
1/2  & 1/\sqrt{2}& 0  \\
0 & 0 & \sqrt{3/4}
\end{array}\right)\;.
\ee

To use the code with $SU_{charge}(2)\times SU_{spin}(2)$ symmetry, one only
needs to type this information into the input file, {\tt input.dat}, and then
run the code. You can find the corresponding input file,
{\tt kondo\_model\_SU2\_c\_SU2\_s.dat}  in the directory {\tt input\_files},
and a Mathematica notebook to compute these matrix elements is also provided.

\subsection{Extension to arbitrary symmetries}

The procedure outlined above can be generalized to essentially any number and
type of  non-Abelian symmetries and to any Hamiltonian of the form:
\begin{equation}
H = {\cal H}_0 + \sum_{n=0}^{\infty} \left( {\tau}_{n,n+1} + {\cal H}_{n+1}\label{eq:chain_hamiltonian} \right )\;.
\end{equation}
The first term in  Eq.\ (\ref{eq:chain_hamiltonian}), ${\cal H}_0$,
describes the quantum  impurity coupled to the
`environment'. This is thus the 'interaction part' of the Hamiltonian.
The dynamics of the environment  is described by hopping terms that connect nearest-neighbors
only, ${\tau}_{n,n+1}$, and the on-site terms ${\cal H}_n$. As already stated
before, these on-site terms can account for the breaking of electron-hole symmetry or for the
presence of superconducting correlations \cite{bulla_supra}.

The above Hamiltonian can also be diagonalized iteratively using the recursion
relation, \eqref{eq:recursion}. The procedure is essentially the same as the
one outlined in the previous subsection, with the modification that  we now may
have $\Gamma$ different symmetries,  which commute with the Hamiltonian and
 each other, and which we shall label by an index $\gamma=1,\dots,\Gamma$. Correspondingly, multiplets will be labeled by
$\Gamma$ different quantum numbers, $\{Q^{\gamma}\}=\{Q^1,\dots,Q^\Gamma\}$,
with $Q^\gamma$ the label of the irreducible
representation of symmetry $\gamma$. States within
a degenerate multiplet can then be labeled by internal quantum numbers,
which, by analogy to the spin operators, we shall denote by  $Q_z^{\gamma}$.
Finally, the representation indices  and the internal quantum numbers can
both be  grouped into vectors, $\underline{Q} = \{Q^1, Q^2,...,Q^{\Gamma}\}$ and
$\underline {Q}^z =\{ Q_z^{1}, Q_z^{2},...,Q_z^{\Gamma} \}$.

Similar to  multiplets, a set of irreducible tensor operators, $\{A\}$ is also characterized
by some quantum numbers $\underline a$, and $\underline a^z$. The hopping
terms, ${\tau}_{n,n+1}$, can also be decomposed using a set of such irreducible
operators, which we shall call `hopping operators',
$f^\dagger\to C$, and label them by $\lambda$. Thus the hopping operator
multiplet, $C_\lambda$ has quantum numbers
$\underline c_\lambda$ and members of this multiplet are labeled by
$\underline c_\lambda^z$. Ordering the members of a hopping operator multiplet
into vectors, $\underline C_\lambda \equiv \{C_{c_\lambda,c_\lambda^z }\}$,
the hopping can be simply written as
\be
{\tau}_{n,n+1} = \sum_\lambda t_{n,\lambda} \left(\underline C_{n,\lambda} \cdot
\underline C_{n+1,\lambda}^\dagger  + h.c. \right)\;.
\ee
Note that here different hopping operator multiplets may have different
hopping amplitudes, and also, remember that now the  $\underline
C_{n,\lambda}$'s denote {\em creation operators}. An example for having two
operators has already been for the Kondo model with $U_{charge}(1)\times U_{spin}(1)$
symmetries, but in most cases several hopping operators must be used for multichannel models
too.

In our general framework the iteration step  $n\to n+1 $ reads now as follows:

\begin{itemize}
\item
Previous block states $u$ can be characterized by their
symmetry indices and labels, $\underline{Q}_u$, $\underline {Q}_u^z$,
$\left |u,  \underline{Q}_{u},
\underline {Q}_{u}^z\right >_{n} $. Similarly, local states $\mu$ are also characterized by their
 symmetry indices $\underline {q}_{\mu}= \{ q_{\mu}^{\gamma} \}$,
and internal labels $\underline {q}_{\mu}^z= \{ q_{\mu,
z}^{\gamma} \}$: $ \left | \mu, \underline{q}_{\mu},
\underline{q}_{\mu}^z \right >\;. $
\item The new basis states can
be constructed using Clebsch-Gordan coefficients as
\begin{equation}
\left |i,  \underline{Q}_{ i}, \underline {Q}_{ i}^z \right >_{n} =
\sum_{\underline {Q}_u^z, \underline {q}_{\mu}^z }
\left < \underline{q}_{\mu} \underline{q}_{\mu}^z; \underline{Q}_u \underline{Q}_u^z \right |
\left . \underline{Q}_{ i} \underline {Q}_{ i}^z  \right >
\left | u, \underline{Q}_{u}, \underline{Q}_u^z; \mu, \underline{q}_{\mu},
  \underline{q}_{\mu}^z
 \right >_{n-1} \;,
\label{eq:new_states}
\end{equation}
where we have also introduced a general notation for the Clebsch--Gordan
coefficients as a product of coefficients, one for each symmetry in the problem
\begin{equation}
\left< \underline{q}_{\mu}\underline{q}^{z}_{\mu}; \underline{Q}_u \underline{Q}^z_u\right|
\left. \underline{Q}_i \underline{Q}^z_i \right>\\
=\prod_{\gamma=1}^{\Gamma}
\left< q^{\gamma}_{\mu}q^{\gamma}_{\mu,z};Q^{\gamma}_u Q^{\gamma}_{u,z}
\left|Q^{\gamma}_i Q^{\gamma}_{i,z} \right>\right.\;.
\end{equation}

\item Knowing the reduced matrix elements, $_n\langle u\parallel
  C_{n,\lambda}\parallel v\rangle _n$ and $\langle \mu \parallel
  C_{n+1,\lambda}\parallel \nu \rangle$, and the eigenvalues $E_u^n$ from
  iteration $n$ one can construct the Hamiltonian in the new basis.
\item Diagonalizing this Hamiltonian, one can determine the new eigenstates,
$\left |\tilde u,  \underline{Q}_{\tilde u}, \underline {Q}_{\tilde u}^z\right >_{n+1} $
their energies, $E_u^{n+1}$, and the irreducible matrix elements, $_{n+1}\langle\tilde u\parallel
  C_{n+1,\lambda}\parallel \tilde v\rangle _{n+1}$. If a spectral function is
  computed then the corresponding operator's reduced matrix element,
$_{n+1}\langle\tilde u\parallel A\parallel \tilde v\rangle_{n+1}$  must also be
  computed.
\item After the iteration some of the high-energy states are also discarded
  before one proceeds to the next iteration.
\end{itemize}

Our flexible DM-NRG code implements this rather general procedure
dynamically, and in such a way that various symmetries can be taught to the code.
If you want to teach a new symmetry to the code, feel free to contact us
for help.

\section{The DM-NRG method}
\label{sec:DM-NRG}

Wilson's procedure gives accurate results for the finite size spectrum,
and it can also be used to compute spectral functions \cite{Costi}. However,
there are several problems with the simple extension of Wilson's method. On one
hand, in the original approach, spectral functions have been computed
using only the approximate ground states in a given iteration. This may give
incorrect results if the fine structure of the ground state shows up at
subsequent iterations only. Hofstetter proposed to use a density matrix NRG
(DM-NRG) method to treat this problem \cite{Hofstetter}. Furthermore,
in the original method one simply  chops off parts of the Hilbert space, and
therefore spectral sum rules are not respected. This problem has been solved
only recently, through the introduction of a {\em complete basis set}
\cite{Anders_05}. In this approach, rather than constructing approximate eigenstates
for a chain of length $n$, one constructs approximate eigenstates
of $H_n$ that live on a chain of length $N>n$, by adding {\em environment
states}
\be
\left |u,  \underline{Q}_{ u}, \underline {Q}_{ u}^z \right >_{n}
\to \left |u,  \underline{Q}_{ u}, \underline {Q}_{ u}^z; e \right >_{n}
\ee
where $e$ labels all possible  states of the environment, i.e.,
chain sites from site $n+1$ to $N$.

\subsubsection{The reduced density matrix}

Physical quantities are then computed from the density matrix, which is
approximated as
\bea
\varrho &=& \sum_{n=0}^N \varrho_n\;,\\
\varrho_n&=&
\frac 1 Z \sum_e \sum_{u\in {\rm Discarded}}{\sum_{{\underline{Q}^z_u}}}
{e^{-\beta E_u^n}}
\left|u, \underline{Q}_u,\underline{Q}^z_{u};e\right>_{n}\Big._{n}
\left<u,\underline{Q}_u,\underline{Q}^z_{u};e\right|\;,
\label{eq:DM}
\eea
where the partition function is defined as
 \be
{\cal Z} = \sum_{u}
{\rm dim}^{N-n}_{\rm loc}\; {\rm dim}(u)\; e^{-\beta E_u^n} \;,
\label{eq:partition_function}
\ee
where $ {\rm dim}(u)$ is the dimension of the multiplet
$u$, and  ${\rm  dim}_{\rm loc} = \sum_\mu {\rm dim}(\mu)$ denotes the
dimension of the local basis at a given site of the chain. The second
summation in Eq.~\eqref{eq:partition_function} runs over states that have been
discarded in iteration $n\to n+1$.

The calculation of a spectral function then consists of a forward sweep,
where we determine all eigenstates in a given iteration, and a backward sweep,
where we determine  $\varrho$ and the truncated reduced density matrices, defined as
\be
R^{[n]}\equiv {\rm Tr}_{N-n}\{\sum_{m\ge n} \varrho_m \}\;.
\ee
The details of the corresponding
recursion relations were given in Ref.~\cite{our_dmnrg_paper}.

\subsubsection{Spectral function calculations in the flexible-DM-NRG framework}

The main purpose of the code  is to compute the retarded Green's function
of two local operators, $A$ and $B$.
This is defined for two irreducible tensor operators as
\begin{equation}
G^{R}_{A_{\underline a, \underline a_z},B^\dagger_{\underline b, \underline b_z}}(t) =
-i \left < \left [ A_{\underline a, \underline a_z} (t),
B^\dagger_{\underline b, \underline b_z}(0) \right ]   \right > \Theta \left( t \right).
\end{equation}
where the quantum numbers $\underline a, \underline a_z$ and $\underline b,
\underline b_z$ refer to the quantum numbers of operators $A^\dagger$
and $B^\dagger$. For fermionic operators one must replace the commutator above
by an anticommutator. By symmetry,  this Green's function is non-zero
only if the spectral operators $A$ and $B$ transform accordingly to the same
representation,
\begin{equation}
 G^{R}_{A_{\underline a, \underline a_z},B^\dagger_{\underline b, \underline b_z}}(t)
=G^{R}_{A,B^\dagger} (t)\; \delta_{\underline a, \underline b}\;
\delta _{\underline a_z, \underline b_z}\; \label{eq:green_function_extended}\;.
\end{equation}

Although it is a cumbersome expression, for completeness, let us  give here the
expression  obtained generalizing the procedure of Ref.\cite{Anders_05}.
\begin{multline}
G^{R}_{A,B^{\dagger}}(z)=
\ds{\sum^{N}_{n=0}}\;\sum_{i\in D, K}\;\ds{\sum_{(j,k)\notin(\rm K,\rm K)}}
\;\big . _n\left <i \right .\parallel R^{[n]}\parallel \left . j\right >\big . _n\\
 \times\left[
\frac{\big ._n \left < k \right . \parallel A^{\dagger }_{}\parallel \left . j \right >\big . _n^{*}
\; \big . _n \left < k \right . \parallel B^{\dagger }_{} \parallel \left . i\right>\big . _n}
{z+\frac{1}{2}(E^n_i +E^n_j)-E^n_k}\frac{\dim(k)}{\dim(a)}
-\xi\;\frac{\big . _n \left< j \right . \parallel B^{\dagger }_{} \parallel \left . k \right>\big . _n
\;\big . _n \left<i \right . \parallel A^{\dagger }_{}\parallel \left . k \right>\big . _n^*}
{ z-\frac{1}{2}(E^n_i +E^n_j)-E^n_k } \frac{\dim (i)}{\dim (a)} \right]\;.
\label{eq:green_function_explicite}
\end{multline}
Remarkably, this formula contains exclusively the reduced matrix elements
and the dimensions of the various multiplets and operator multiplets.
Here the second sum is over all the multiplets $i,j,k$ of the given iteration
subject to the restriction
that $j,k$ do not belong to kept states at the same time. No summation is
 needed for states within the multiplets. The sign factor $\xi$ is $\xi=1$ for
 bosonic operators, while it is $\xi=-1$ for fermionic operators,
and $\dim(a)={\prod_{\gamma=1}^{\Gamma}}\dim (a^{\gamma})$ is the dimension of the operator
multiplet $A^\dagger_{\underline a, \underline a_z}$. The code computes the
spectral function using this formula, and the full Green's function is
then computed by doing a Hilbert transform numerically.

\subsubsection{Smoothing procedure}

In general, the spectral functions  are given as weighted sums of $\delta$-functions of the form:
\begin{eqnarray}
 {\cal A}_{A, B^\dagger}(\omega) &= &-\frac{1}{\pi}\Im m \;G^R_{A, B^\dagger}(\omega+i\,0^+)\\
                                &= & \sum_i w_i\; \delta(\omega -\omega_i)
\end{eqnarray}
where the weighting coefficients
$w_i$ can be expressed from Eq. (\ref{eq:green_function_explicite}) after analytical
continuation to the real axis. To get a smooth spectral function, however, the
delta functions  need to be replaced by some broadening kernels $K(\omega, \omega_i)$
that fulfill the condition
\be
\int d\omega\; K(\omega, \omega_i) = 1 \; .
\ee
This condition guarantees that spectral sum rules remain valid after the broadening
procedure. In our code we use two types of kernels to generate the real axis
spectral function: (a) the {\em log-Gaussian} kernel \cite{bulla_kernel} and
 (b) the {\em interpolative kernel} that we constructed
 to avoid certain difficulties with the interpolation around  $\omega=0$.

For the log-Gaussian  method the following  kernel is used
\be
 K_{log}(\omega, \omega_i) \equiv \frac{1}{b \sqrt{\pi}}
e^{-b^2/4} e^{-(\ln \omega -\ln \omega_i)^2/b^2}\frac{1}{\omega_i}
\label{eq:kernel_log}.
\ee
This kernel works rather well at $T=0$ calculations, however, it extrapolates
to the $\omega\to\pm 0$ values in a singular way. Therefore, it has problems
at finite temperatures and also in cases where spectral losses lead to an
artificial jump in the spectral function at $\omega=0$.

The interpolative scheme avoids this difficulty by using a kernel
\be
 K_{int}(\omega, \omega_i) \equiv  \frac{1}{b \sqrt{\pi}}
 e^{-[x (\omega) -x(\omega_i)]^2/b^2}\frac{d x}{ d\omega_i}\;,
\ee
with the smoothening function defined as
\be
x(\omega) =\frac{1}{2}\;\tanh\left (\frac{\omega}{T_Q}\right )\;
\ln\left( \left ( \frac{\omega}{T_Q} \right)^2  +e^\gamma \right )\;,
\label{eq:kernel_interpolation}
\ee
by choosing $\gamma\approx T_Q$ e.g.
This formula interpolates smoothly along the real axis for all frequencies. It has the property
that for frequencies smaller than the 'quantum temperature', $|\omega|\ll
T_Q$, the broadening is Gaussian, while for $|\omega|\gg
T_Q$ it becomes log-Gaussian. In this scheme, for zero temperature
calculations the quantum temperature $T_Q$ must be the smallest energy scale 
in the problem, on the other hand for finite temperature calculations,
the quantum temperature must be set to be in the range of 
the temperature itself,  $T_Q\sim 0.5T \div T$. In both interpolative methods
there is a fitting parameter $b$, which controls the widths of kernels and 
in general is $\Lambda$ dependent, which should be in the range of 
$ b \sim \sqrt{\Lambda} / 2 $ (for $\Lambda = 2$, $b=0.5\div 0.9$).

\chapter{Installation and technical support}

\section{ Memory and disk space requirements}

The resources required by the DM-NRG code depend a lot on the
type of the ongoing calculation. Increasing the number of block
states that are kept during a run, the CPU time needed for
performing the calculation,  the necessary memory, as well as the
used disk space may increase substantially. As an example in Table
\ref{table:memory} we present some estimates for a  simple model.
The following information was extracted by running the code on an
Intel(R) Core(TM)2 Duo CPU T7250  @ 2.00GHz processor with 4MB CPU
cache. We have used the  CentOS 5.1 distribution for the Linux
operating system. The code was run by  using an input file for the
Anderson model with  U(1) $\times$ SU(2) symmetries  that comes
with the distribution. We made a run for a flat band and performed
50 iterations.  The compiler used was gcc version 4.1.2.

\begin{table}[htb]
%\begin{tabular}{@{\hspace{1mm}}lc@{\hspace{5mm}}c@{\hspace{4mm}}c@{\hspace{1mm}}}
\begin{tabular}{ccccccc}
\hline \hline
%\multicolumn{1}{c}{$U$} &  & $\Delta_{\rm{c}}$ & $\Delta_{\rm{s}}$ \\
Model name      & number of    & CPU-type & CPU time & disk space     & memory\\
            & block states &          & (sec)  & (MB) &   (MB) \\
\hline \hline
Anderson model &    50 & Intel Core(TM)2 Duo & 1.27          &  12    & 4    \\
   U(1) $\times$ SU(2)                    & 100 & CPU T7250 @ 2.00GHz & 3.56          &  25     &  7  \\
   symmetries                    & 200 & with 4MB cache & 14.39         &  69    & 12    \\
                                   & 500 && 156.26     &  341    & 18    \\
\hline
\end{tabular}
\caption{CPU time, memory and disk space requirements.}
\label{table:memory}
\end{table}

Important note: when temporary files are also generated in ASCII format by setting
{\em \bf text\_swap\_files\_flag = ON } in
$<$SECTION-FLAGS$>$ of the input file  {\tt input.dat},
approximately  four times larger disk space is required than the ones
shown in the table.

\section{Compilation environment}

The code can be compiled with either {\tt g++} or with  {\tt icc} (intel c++ compiler). The Intel compiler,
{\tt icc} is available for free for non-commercial  and non-academic use after signing a license agreement.

\section{Libraries}

The Flexible-DM-NRG code is linked with state of the art libraries.
The following list of libraries (packages) are needed to have
the {\tt flexible-dmnrg} package properly installed. The easiest
way to install these packages is by using the {\tt rpm  - RPM package manager} method.

For further information and links to the websites from where these libraries can be downloaded
 please refer to the web page of the code
{\tt http://www.phy.bme.hu/$\sim$dmnrg/index.html}.

List of required libraries:

$\bullet$
{\tt libg2c} -- Fortran 77 Libraries

$\bullet$
{\tt blas} -- Basic Linear Algebra Subprograms

$\bullet$
{\tt lapack} -- The LAPACK libraries for numerical linear algebra.

$\bullet$
{\tt  gmp} -- A GNU arbitrary precision library

$\bullet$
{\tt  gsl} -- GNU Scientific Library

$\bullet$
{\tt  gsl-devel} -- GNU Scientific Library - development files

\section{Automatic installation}

Once the libraries were successfully installed,  run the
following shell script in the main directory of the code:

{\tt ./configure}

This command must run without errors. It checks your system, searches for the
necessary libraries, and sets the library paths.
An error message returned while running {\tt ./configure} means that
some packages might be missing or not properly installed.
If the {\tt ./configure} has run without errors then the {\tt make.sys} and
{\tt make.rules} and {\tt  config.h} files  have been generated. These files now contain
the system variables needed for the installation
\footnote {The automatic installation checks  for the {\tt g++} and {\tt gcc}
compilers and does not check for the {\tt icc}. To run the code with
{\tt icc}, manual intervention is necessary (see section Manual Installation),
and the {\tt CXX} environmental variable needs to be modified from  {\tt g++}
to  {\tt icc}.}.

Next run the following command:

{\tt make all}

If everything  goes smooth and no error messages were generated, then
the code  is successfully installed.
If for some reasons the automatic configuration does not work, please, follow the instructions
in the next subsection and install the code manually. If you do not succeed,
feel free to contact us.

\section{Manual installation}
Sometimes, for some mysterious reasons, the {\tt autoconf} does not work
properly and some of the  libraries are not detected. In this case, the {\tt make.sys} file, that can
be found in the main folder of the code, needs to be edited manually. Altogether
there are ten shell variables that need to be edited.

$\bullet$
{\tt CXX} - the name of the C++ compiler, is usually {\tt g++} or  {\tt icc}.

$\bullet$
{\tt C} - the name of the C compiler, usually  {\tt gcc}.

$\bullet$
{\tt CC\_FLAGS} - flags for the g++ or icc compiler (this variable can be left
empty at this stage). You can set here the optimization level to speed up the code.

$\bullet$
{\tt GSL\_FLAGS} - path to the gsl include files, usually {\tt -I/usr/include}.

$\bullet$
{\tt GMP\_FLAGS} - path to the gmp include files, usually {\tt -I/usr/include}.

$\bullet$
{\tt FLIBS} - fortran libraries, usually {\tt g2c} or  {\tt gfortran}.

$\bullet$
{\tt LAPACK\_LIBS} - lapack library, usually {\tt -llapack}.
For static libraries the full path name needs to be provided sometimes:  {\tt /lib/liblapack.a}.

$\bullet$
{\tt BLAS\_LIBS} - blas library, usually {\tt -lblas}.
For static libraries the full path name needs to be provided sometimes:  {\tt /lib/libblas.a}.

$\bullet$
{\tt GSL\_LIBS} - gsl library, usually {\tt -lgsl -lgslcblas}.
For static libraries the full path name needs to be provided sometimes: {\tt /lib/libgsl.a}.

$\bullet$
{\tt GMP\_LIBS} - gmp library, usually {\tt -lgmp}.
For static libraries the full path name needs to be provided sometimes: {\tt /lib/libgmp.a}.

Once the {\tt make.sys} file is properly edited run

{\tt make all}

If the compilation runs without errors then the code  is successfully installed and
all the binaries have been generated.

\section{Generated binaries}
After a successful installation the following binaries are generated:

\begin{itemize}
\item {\bf fnrg} - the main program that does the  NRG/DM-NRG
calculations.
\item {\bf sfb} -  utility that does the spectral function
broadening.
\item {\bf he} -  utility that computes the hopping for an
energy-dependent density of states.
\item {\bf es} - utility to generate/analyze the energy spectrum.
\item {\bf cgc} - utility to generate/compute the Clebsch-Gordan
coefficients

\end{itemize}

To run the code the {\tt input.dat} file  must be set properly.
For further details, refer to the next sections of the manual.
Example input files are provided with the code  for the single band Anderson
model with $ U(1)$  charge and spin symmetries
and  for  the Kondo model with combinations of
$ U(1)$ and $ SU(2)$ symmetries. These  'ready to run' input files are
available in the {\tt ./input\_files} folder.  
Furthermore, we also provided the
Mathematica files which were used to compute the content of these input
files. Some of the technical details of these calculations were summarized in Chapter~\ref{ch:NRG_intro}.

\section{BUG reports}

We have made this code publicly  available, so if you think 
you have found a bug in the code, please let us know so we can fix
it for future releases.

Please send any bug report to:  {\bf \tt dmnrg@neumann.phy.bme.hu}.
Use {\bf  BUG REPORT} in the subject area.

Please include the followings in any report:

$\bullet$ The  version number of the code. 

$\bullet$ input.dat file so we can test the code with it.  

$\bullet$ A description of what is wrong. If the results are incorrect, in what way. If you get a crash, etc. 

$\bullet$ Please do not send core dumps, of executables.

$\bullet$ The name of the compiler that was used and its version.

$\bullet$ The output from running {\tt ./configure} 

$\bullet$ If the bug is related to configuration, then the contents of {\tt config.log}.

If the bug report is good than we will do our best to fix it. If the bug report is poor 
we will not do much about it, probably just ask for a better report.

If you think something in this manual is unclear, or downright incorrect, or if the language needs
to be improved, please send us a note to the same address.

\section{Some hints for developers}
\subsection{Organization of the code}
The components of the program are located in three main directories.
Directory {\tt ./src/} contains the {\tt *.cc} source files,
{\tt ./include/} the corresponding {\tt .h} files and {\tt ./input\_files/} the user
defined input files. The directory {\em ./doc/} contains the documentation of the
code in {\em LATEX} and {\em HTML} forms. Table \ref{table:main_dir_struct} gives
a hint on the structure of the code.

\begin{table}[htb]
%\begin{tabular}{@{\hspace{1mm}}lc@{\hspace{5mm}}c@{\hspace{4mm}}c@{\hspace{1mm}}}
\begin{tabular}{lllr}
\hline \hline
%\multicolumn{1}{c}{$U$} &  & $\Delta_{\rm{c}}$ & $\Delta_{\rm{s}}$ \\
\hline
./&{\tt  Makefile}   &       &     \\
./&{\tt DOXYFILE}   &       & configuration for on-line documention    \\
  &{\tt src/ }      & {\tt *.cc } & source code (source files)    \\
  &{\tt include/}   & {\tt *.h  } & source code (header files)   \\
  &{\tt input\_files/} & {\tt *.dat}  & directory for input files \\
  &{\tt results/}   &       & output directory of the code \\
%  & lib/       &       & external libraries required by the code   \\
  & {\tt doc/}       &  {\tt nrg\_user\_guide.tex}      & documentation of the code      \\
  &            & {\tt  html/} & on-line documention   \\
  &            & {\tt tex/}  & on-line documention   \\

\hline \hline
\end{tabular}
\caption{Directory structure of the code.}
\label{table:main_dir_struct}
\end{table}

The other files in the main directory are configurations files needed for properly installing
the code.

\subsection{On-line documentation of the code}

The documentation of the source code is managed with the use of
the {\tt doxygen} a free software under {\tt Linux}. At present
version 1.4.6 is used. The configuration of {\tt doxygene} is set
in file {\em Doxyfile} and the program is called from shell as
{\tt doxywizard}. Description of the classes, variables and
functions is included directly in the source code (in the {\tt *.cc} and
{\tt *.h} files), therefore, all comments about new variables,
functions, classes etc. are documented right away. The on-line
documentation file {\tt ./doc/html/....html} can be opened with any
WEB-browser. The DM-NRG reference manual in pdf form ({\tt refman.pdf})
can be generated by the command: {\tt latex
./doc/latex/refman.tex}.

All the information about the functions (classes) i.e., what it does,
how to call, description of the input and output variables must be
located in the main body of the functions, otherwise, this information
will not be accessible in the output files generated by doxygene!
The hierarchy of the classes, dependencies and calling functions are
generated automatically by {\tt doxygen}.

%%%%%%%%%%%%%%%%%%%%%%%%%%%%%%%%%%%%%%%%%%%%%%%%%%%%%%%%%%%%%%%%%%%%%%%%%%%%%%%%
%       Version management                                                     %
%%%%%%%%%%%%%%%%%%%%%%%%%%%%%%%%%%%%%%%%%%%%%%%%%%%%%%%%%%%%%%%%%%%%%%%%%%%%%%%%

\subsection{Version management}

Version management of the code is carried out with the use of
{\tt Subversion system(svn)} a free software under {\tt Linux}.
This program is similar to {\tt Control Version System (cvs)}
but has several additional features. The WEB page of the svn
interface can be accessed by the members of the
{\em flexible DM-NRG} developers group only.

The management of this manual -- in {\tt LATEX} form--, however,
has to be done manually. There are no scripts yet, that would
automatically update the manual once the code has been changed.
However, since all directories and files -- including this manual
as well -- are checked in to the {\tt svn} it is straightforward
to update the manual and all changes of it are documented by the
{\tt svn} itself.

\chapter{Using the DM-NRG code}

%---------------------- Section%------------------------%

        %---------------------- Section -----------------------------------------------%
\section{Main features of the code}

If the installation was successful then, to run the code for your
problem of interest, you first need an input file,  {\tt
input.dat}.  The structure and content of this file shall be
described in detail later in this chapter, and a few examples are
also provided with the code in the library, {\tt ./input\_files}.
Furthermore, we provided some {\it Mathematica} files in the {\tt
mathematica\_files} folder, which were used to generate these
input files. For the generation of the irreducible matrix elements
it is sometimes important to know, what is the definition of the
Clebsch-Gordan coefficients used by the code. For this purpose, a
utility called  {\bf cgc} is provided in the package that computes
the Clebsch-Gordan coefficients.

When running the code, all the results are
written into the folder {\tt ./results}. The content of these files
shall also be described later in this chapter.
Finally, these results can further be analyzed with the utilities
{\bf sfb} -- real axis spectral function generator and
{\bf es} -- energy spectrum generator, also provided with the
code, and shall be described later.

\subsubsection{Main features included in version 1.0.0:}

\begin{itemize}
 \item Zero and finite temperature calculations of the spectral function of
   not necessarily  the same operators.

\item Static averages.

\item NRG type calculations or  DM-NRG computations (using the
complete basis set).

\item The code uses a single input file that can be easily
modified. Examples are available for Anderson as well as Kondo
Hamiltonians in the presence of $U(1)$ and $SU(2)$ symmetries for spin
as well as charge symmetries.

\item
Flat density of states as well as  arbitrary energy-dependent density of
states can be considered.  In the latter case
arbitrary precision routines  are used to generate the parameters of the
Wilson chain.

\item
Utility for broadening the spectral function is provided to compute  the imaginary and the real parts of the retarded Green's function
\begin{equation}
G_{A,B^{\dagger}}^{R}\left ( t \right ) = -i \left < \left [
A(t),\, B^{\dagger}\left (0 \right)\right ]_\pm \right >
\Theta(t)\;, \label{eq:green_function}
\end{equation}
with the operators $A$ and $B$ not necessarily identical.

\item
Utility for computing the on-site energies as well as the hoppings
for a non-uniform density of states is provided.

\item
Utility to generate/analyze  the  finite size spectrum is provided.
\end{itemize}

\subsubsection{Utilities  provided in the package:}

\begin{itemize}
\item {\bf he} - generates the hopping amplitudes along the chain
and the on-site energies when an energy-dependent density of
states is used. (This utility provides input files that are read
directly in the {\bf fnrg} binary code).

\item
{\bf sfb} -  does the broadening of the spectral function and computes the real
part of the Green's function by a Hilbert transform. This utility uses the outputs of the {\bf fnrg} to generate these data.

\item
{\bf notebook examples} - there are some extra {\it Mathematica 6.0} notebook examples where the matrices for Hamiltonians as well as for some hopping and spectral operators are computed. These notebooks can easily be extended to other impurity models.

\item
{\bf es} -  generates the energy spectrum. This utility uses the output of the {\bf fnrg} to generate the data.

\item
{\bf cgc} - an interactive utility, which generates the
Clebsch-Gordan coefficients used by the code.

\end{itemize}

\section{Running the code}

In this section we shall give some information
 on how exactly to run the code. At this point we suppose that
the code was successfully compiled and linked properly with the required libraries.
 If this is the case there should be five binary codes in the main folder,
{\bf fnrg}, {\bf sfb}, {\bf he},  {\bf cgc}, and {\bf es}.
Depending on the type of calculation that is performed, the following steps need to be followed:

\subsection{ Flat density of states}

Most NRG calculations are done for a flat density of
states, $\varrho(\omega)= \frac{1}{2}$, with $\omega\in[-1,1]$.
In this case there is no need for the {\bf he} routine, since the hoppings along
the chain are computed on the fly. In this case
 the on-site energies are all zero as a consequence of
electron-hole symmetry.

For this type of calculation, first
the {\em hoppings\_on\_site\_energies\_flag} in the
$<$SECTION-FLAGS$>$ needs to be set to 'OFF' in the file {\tt input.dat}.
Then the {\bf fnrg} must be  run without any arguments in the
command line. The name of the input file
must always be {\tt input.dat}, since the code looks for this particular
file to read in the data.

Outputs are  saved in a folder in {\tt results/}.
Results for the spectral function can be generated by running the
{\bf sfb} binary without any argument after exiting normally from the run.
Note that {\bf sfb} reads the same input file, {\tt input.dat}, so do not
modify it between the two runs except for the part
$<$SECTION-SPECTRAL\_FUNCTION\_BROADENING$>$.
Depending on the flags set  in $<$SECTION-SPECTRAL\_FUNCTION\_BROADENING$>$,
spectral functions,  Green's functions and static averages are computed.
Note that the real part of a Green's function can only be computed if
the corresponding spectral function has previously been computed.

\subsection{ Energy dependent density of states}

The user can also provide an arbitrary density of states (DOS) for the code.
In the present
version of the code, the DOS is read from a file, {\tt dos.dat}.
This file must be in the folder,
{\tt ./dos\_mapping}, and it must be provided by the user.
This file must contain at least two columns, one of them being the energy, and
the others being the corresponding density of states. There should be as many
density of states columns as hopping operators are defined in the input file.

There are some constraints that the DOS must satisfy
\begin{itemize}
\item The  DOS needs to be rescaled such that all energies are in the
interval,  $E\in [-1, 1]$. In other words, the energy unit of the
NRG calculation (=1) must be larger than the range of the density of states.

\item
The DOS needs to be normalized to 1.

\item
The energies must be ordered.

\item
Each density of states column corresponds to a given hopping operator
in the {\tt input.dat} file. The columns must be ordered accordingly to the order
of the hopping operators in the {\tt input.dat} file.
\end{itemize}

There is no further requirement for the mesh, both uniform and
non-uniform meshes can be provided.

When the NRG calculation is performed with an energy-dependent
 DOS, the hoppings and the on-site energies must be computed first.
This must be done interactively, using the binary {\bf he}.
When running {\bf he}, one needs to specify an iteration number.
This can be larger than the
number of iterations  specified  in the {\tt input.dat} file.
After a successful run of  {\bf he},
the hopping amplitudes  and the on-site energies are saved
in files having descriptive names
such as {\tt hopping\_couplings\_*.dat} and {\em on\_site\_energies\_*.dat}
in the folder {\tt ./dos\_mapping/}.

Next, the  {\tt input.dat} file needs to be set-up: There, make
sure that the {\tt  hoppings\_on\_site\_energies\_flag} in the
$<$SECTION-FLAGS$>$ is set to 'ON' for this type of run.

Once the hopping amplitudes
and the on-site energies were computed, the same steps
must be followed as in the case of
a flat density of states. The code
 {\bf fnrg} must be run without arguments.
\\
\\
{\tt Note:} 
The routine {\bf he} is using arbitrary precision libraries for computing the 
hoppings and the on-site energies, however due to the accumulation
of the errors during the calculations sometimes the {\bf he} utility
may give unexpected results for the hoppings. Therefore, we advise 
the user to check the results for the hoppings before pluging them 
into {\bf fnrg} for any kind of nrg calculations.

\section{Initialization and the input file} \label{ch:init}

In the initialization step -- the zero'th NRG iteration step --
one has to define the Hamiltonian. Therefore, for the iterative
process, among others, the user has to provide the following input parameters:
\begin{enumerate}
\item One needs to specify the number ({\em symmetry\_no}) and type of symmetries.

\item One needs to specify the
 number ({\em block\_state\_number}) of initial {\em block states}, i.e.,  multiplets that span the
Hilbert space for $n=0$ (we denote this number in this manual by $K$).
One also needs to define the  quantum numbers $ \underline{Q}_{u}$
of the block states, $\left | u, \underline{Q}_{u}, \underline{Q}_u^z \right >_0$.

\item
{\em Block Hamiltonian matrix } - The matrix
elements of the $K\times K$ Hamiltonian matrix $H_0$, defined on
the $K$ {\em block states}.

\item
{\em block hopping operators} - The reduced matrix elements of all
hopping operators between the block states. These are $K\times K$ matrices.

\item
{\em spectral operators} - The reduced matrix elements of the  operators,
whose spectral functions we compute. These are also  $K\times K$ matrices.

\item
{\em local state number} - The number of multiplets that span the local Hilbert
space  (we label this number $L$).

\item
{\em local states} - Quantum numbers $\underline{q}_{\mu}$
of the local states,  $\left | \mu, \underline{q}_{\mu}, \underline{q}_{\mu}^z \right >$.

\item
{\em local hopping operators } - The reduced matrix elements of
the hopping operators. These are $L\times L$ matrices.

\end{enumerate}

This list is rather incomplete, and  explains only briefly  the
meaning of the most basic variables. In addition, many other, less
relevant  parameters must be specified for the NRG process
(broadening, maximum iteration number, etc.). All these input
parameters are read from  file, {\tt input.dat}, which must be
constructed for every model separately. Some examples of such
input files can be found in the folder, {\tt ./input\_files}, and also
in the appendix of this manual.  A detailed description of these
parameters shall be provided in forthcoming subsections.

\vskip0.3cm
{\bf \large Structure of the input file}
\vskip0.3cm

The input file is organized along sections, and contains a lot of
comments. As a general rule, any line that  starts with the symbol
'\#' is treated as a comment line in these input files and is
ignored by the code. You can thus add additional comment lines for
yourself using this convention.

The input file has the following overall structure (that we shall describe in detail in
subsequent subsections):
\begin{enumerate}
\item
$<$SECTION-PARAMETERS$>$
\item
$<$SECTION-FLAGS$>$
\item
$<$SECTION-SYMMETRIES$>$
\item
$<$SECTION-BLOCK\_STATES$>$
\item
$<$SECTION-LOCAL\_STATES$>$
\item
$<$SECTION-LOCAL\_STATES\_SIGNS$>$
\item
$<$SECTION-BLOCK\_HAMILTONIAN$>$

  \begin{enumerate}
  \item
  $<$BLOCK\_HAMILTONIAN\_TERM$>$

    \begin{enumerate}
    \item
    $<$BLOCK\_HAMILTONIAN\_NAME$>$
    \item
    $<$BLOCK\_HAMILTONIAN\_COUPLING$>$
    \item
    $<$BLOCK\_HAMILTONIAN\_REPRESENTATION\_INDEX$>$
    \item
    $<$BLOCK\_HAMILTONIAN\_MATRIX$>$
    \end{enumerate}
  \end{enumerate}

\item
$<$SECTION-HOPPING\_OPERATORS$>$
  \begin{enumerate}
  \item
  $<$HOPPING\_OPERATOR$>$
    \begin{enumerate}
    \item
    $<$HOPPING\_OPERATOR\_NAME$>$
    \item
    $<$HOPPING\_OPERATOR\_REPRESENTATION\_INDEX$>$
    \item
    $<$HOPPING\_OPERATOR\_SIGN$>$
    \item
    $<$HOPPING\_OPERATOR\_MATRIX$>$
    \end{enumerate}
  \end{enumerate}

\item
$<$SECTION-SPECTRAL\_OPERATORS$>$
  \begin{enumerate}
  \item
  $<$SPECTRAL\_OPERATOR$>$
    \begin{enumerate}
    \item
    $<$SPECTRAL\_OPERATOR\_NAME$>$
    \item
    $<$SPECTRAL\_OPERATOR\_REPRESENTATION\_INDEX$>$
    \item
    $<$SPECTRAL\_OPERATOR\_SIGN$>$
    \item
    $<$SPECTRAL\_OPERATOR\_MATRIX$>$
    \end{enumerate}
  \end{enumerate}

\item
$<$SECTION-STATIC\_OPERATORS$>$
  \begin{enumerate}
  \item
  $<$STATIC\_OPERATOR$>$
    \begin{enumerate}
    \item
    $<$STATIC\_OPERATOR\_NAME$>$
    \item
    $<$STATIC\_OPERATOR\_REPRESENTATION\_INDEX$>$
    \item
    $<$STATIC\_OPERATOR\_SIGN$>$
    \item
    $<$STATIC\_OPERATOR\_MATRIX$>$
    \end{enumerate}
  \end{enumerate}

\item
$<$SECTION-SPECTRAL\_FUNCTION$>$
\item
$<$SECTION-SPECTRAL\_FUNCTION\_BROADENING$>$

\item
$<$SECTION-LOCAL\_HOPPING\_OPERATORS$>$
  \begin{enumerate}
  \item
  $<$LOCAL\_HOPPING\_OPERATOR$>$
    \begin{enumerate}
    \item
    $<$LOCAL\_HOPPING\_OPERATOR\_NAME$>$
    \item
    $<$LOCAL\_HOPPING\_OPERATOR\_REPRESENTATION\_INDEX$>$
    \item
    $<$LOCAL\_HOPPING\_OPERATOR\_SIGN$>$
    \item
    $<$LOCAL\_HOPPING\_OPERATOR\_MATRIX$>$
    \item
    $<$LOCAL\_ON\_SITE\_ENERGY\_MATRIX$>$
    \end{enumerate}
  \end{enumerate}

\end{enumerate}

%%%%%%%%%%%%%%%%%%%Detailed description of each section %%%%%%%%%%%%%%%%%%%%%%%%

\subsection{$<$SECTION-PARAMETERS$>$}

In this section, the main parameters of the given model and also those of the  DM-NRG method
are set. The order of the parameters does not matter as long as none of them is missing.

\subsubsection{Model parameters}

$\bullet$
{\bf model:}
The name of the model under investigation. It can be any string. This name
will be automatically  included in the name of the folder, where output
data  are saved.

$\bullet$
{\bf lambda:}
The value for the logarithmic discretization parameter, $\Lambda$. It is usually
fixed between 2 and 3.
%The coupling strengths of the hopping
%operators and the energy spectra  are both rescaled by $\sqrt{\Lambda}$ in
%every NRG iteration step.

$\bullet$
{\bf symmetry\_no:}
The number of the symmetries used. The present version of the code handles
$U(1)$, $SU(2)$,  $Z(2)$, $charge\_SU(2)$ symmetries.
The properties of the symmetries
are set in $<$SECTION-SYMMETRIES$>$ by
defining their names and setting
upper and lower bounds on their representation indices. Later versions of the code
will include $SU(3)$ symmetry and crystal field symmetries too.

$\bullet$
{\bf coupling\_no:} The number of independent coupling constants.
In general, the local Hamiltonian $H_0$ can contain several couplings. The
on-site part of the Anderson
model, e.g., can be written as
$$
H_0 = U (n-1)^2 + (\epsilon_d+U/2) \; n + V \sum_\sigma (d^\dagger_\sigma
f_{0,\sigma} + h.c.)\;,
$$
i.e. it contains three different couplings, $U$, $\epsilon_d$ and $V$. These
couple to the operators,$(n-1)^2$, $ n $, and $(d^\dagger_\sigma
f_{0,\sigma} + h.c.)$.
 Thus the local Hamiltonian, $H_0$, can be written as  a sum
of $coupling\_no=3$ independent terms,
\begin{equation}
H_0 = \sum_{\delta=1}^{coupling\_no} G_\delta\; {\cal H}_\delta \;,
\end{equation}
where ${\cal H}_\delta$'s denote the independent local terms and
$G_\delta$ stand for the corresponding couplings.
The matrix elements of ${\cal H}_\delta$ are specified
in $<$SECTION-BLOCK\_HAMILTONIAN$>$.

{\tt Important note}:
The number of $<$BLOCK\_HAMILTONIAN\_TERM$>$ items
must agree with the 'coupling\_no' set in this section.

$\bullet$
{\bf spectral\_operator\_no:}
The number of the operators for which one wants to compute the spectral functions.
The detailed parametrization of these operators is carried out in
$<$SECTION-SPECTRAL\_OPERATORS$>$. The number of  $<$SPECTRAL\_OPERATOR$>$ items
must agree with the 'spectral\_operator\_no' set in this section.

$\bullet$
{\bf static\_operator\_no:}
The number of operators for which static averages are computed.
The detailed parametrization of these operators is carried out in
$<$SECTION-STATIC\_OPERATORS$>$. The number of  $<$STATIC\_OPERATOR$>$ items
there must agree with the 'static\_operator\_no' set in this section.

$\bullet$
{\bf hopping\_operator\_no:}
The number of independent  hopping operators.
The detailed parametrization of these operators is carried out in
$<$SECTION-HOPPING\_OPERATORS$>$ and $<$SECTION-LOCAL\_HOPPING\_OPERATORS$>$.

$\bullet$
{\bf block\_state\_no:}
The  number of multiplets that span the Hilbert space for iteration $n=0$.
The detailed parametrization of the states is carried out in
$<$SECTION-BLOCK\_STATES$>$ by setting the representation indices
of all the states corresponding to all the symmetries used during the
calculation.

$\bullet$
{\bf local\_state\_no:}
The number of local multiplets, added in each iteration.
In the present version of the code this is the same for all sites,
 $0 < i \leq N$.
The detailed parametrization of local states is carried out in
$<$SECTION-LOCAL\_STATES$>$.

$\bullet$ {\bf local\_coupling\_no:} The number of independent
terms in the on-site Hamiltonian operators, ${\cal H}_n$.  In the
presence of electron-hole symmetry of the conduction band and
without any other correlations (superconducting like) the on-site
terms are all zero. If we have only on-site energies, $\xi_n$,
then {\bf local\_coupling\_no}$=0$.

$\bullet$
{\bf spectral\_function\_no:}
The number of the spectral functions that should be calculated.
One can compute several spectral functions within the same run, at the expense
of using more memory.
The detailed parametrization of the operators is carried out in
$<$SECTION-SPECTRAL\_OPERATORS$>$.

\subsubsection{DM-NRG parameters}

$\bullet$
{\bf max\_state\_no:}
The upper limit for the number of kept block multiplets.

$\bullet$
{\bf iteration\_no:}
The maximum number of the NRG iteration steps, $N$. The program terminates once
this value is reached.

$\bullet$
{\bf interval\_no:}
The energy grid used in each NRG iteration step
when spectral functions are calculated. Note that in DM-NRG a very large
number of delta peaks is generated in every iteration. Therefore, these delta
peaks are stored on a finite mesh.  In the present version of the code,
the difference between
the maximum and minimum of the energy spectrum at each iteration is divided
by {\em interval\_no} to obtain the size of the mesh,  $d\omega$. Typically
$interval\_no\sim 1000$.

$\bullet$
{\bf degeneracy\_threshhold:}
If two states are such that the difference between their energies 
is less than  {\tt degeneracy\_threshhold} these states are considered to be degenerate. 

When discarding states, those lowest in energy are kept while the  higher energy states
are discarded, such that at the next iteration no more than {\tt max\_state\_no} are kept. 
Degenerate states are block discarded. 

$\bullet$
{\bf temperature:}
The temperature used for the calculation. When the temperature is fixed to
zero,
$temperature=0$,
the code runs by using the number of iterations provided by the user.
In case of a finite temperature, $temperature>0$,  the variable {\em
  iteration\_no}  is determined and modified
during the run to correspond to  the  temperature specified,
using the well-known relationship, $T_n \simeq (1+\frac{1}{\Lambda})\,\Lambda^{-n/2}$.

%%%%%%%%%%%%%%%%%%%%%%%%%%%%%%%%%%%%%%%%%%%%%%%%%%%%%%%%%%%%%%%%%%%%%%%%%%%%%%%%

\subsection{$<$SECTION-FLAGS$>$}

In this section various flags controlling the DM-NRG method are
set.

$\bullet$
{\bf dmnrg\_flag:}
This flag controls whether to perform the backward sweep
and use the DM-NRG method or not. Allowed tags are 'ON' or 'OFF'.
In case of 'ON', the DM-NRG backward sweep is carried out and the complete set
of eigenstates is used to calculate the spectral functions.
In case of 'OFF' the usual NRG procedure is performed only.

$\bullet$
{\bf text\_swap\_files\_flag:}
This flag controls whether to generate various temporary scratch files as
readable text files in addition to the binary files.  Allowed tags are 'ON' or 'OFF'.

$\bullet$
{\bf hoppings\_on\_site\_energy\_flag:}
This flag is important when energy-dependent density of states are considered.
This flag controls whether the hoppings and the on-site energies are read from
a file or not.
If set to 'ON', the hoppings as well as the on-site energies must  be computed
 using the '{\bf he}' utility that comes with
the code. This utility generates some data files in the {\tt ./dos\_mapping}
folder, which are next read by the {\bf fnrg} binary.
When a flat band is considered,
this flag can be set to 'OFF' and then the hoppings are generated on the fly.

%%%%%%%%%%%%%%%%%%%%%%%%%%%%%%%%%%%%%%%%%%%%%%%%%%%%%%%%%%%%%%%%%%%%%%%%%%%%%%%%

\subsection{$<$SECTION-SYMMETRIES$>$}

In this section the types of  symmetries used can be set.
As many symmetries must be specified as were set by the  'symmetry\_no'
variable.
Each line must contain the name of a symmetry,
and a minimum and maximum value for the corresponding representation index. Those states
for which these bounds are exceeded are simply discarded,
and  to avoid such uncontrolled truncation, a large value ($\simeq 20$) should be used for
most symmetries.
The possible symmetry types and the corresponding tags that the present version of
the code can handle are as follows:
'$U(1)$', '$SU(2)$', '$Z(2)$', '$charge\_SU(2)$'. Other symmetries will also be
included in later versions of the code.

%$\bullet$
%symmetry name  min value   max value

%%%%%%%%%%%%%%%%%%%%%%%%%%%%%%%%%%%%%%%%%%%%%%%%%%%%%%%%%%%%%%%%%%%%%%%%%%%%%%%%

\subsection{$<$SECTION-BLOCK\_STATES$>$}

In this section the quantum numbers  of the block states are  set.
A matrix of dimensions $block\_state\_no \times symmetry\_no$
with integer entries enumerating
 the representation indices of the
block states  must be provided.
The number of rows has to be equal
to the value 'block\_state\_no' set in $<$SECTION-PARAMETERS$>$.
Integers in a  row $u$ ($u=1$, $\dots$, $block\_state\_no$)
correspond to the representation indices (quantum
numbers) of the state $u$. Thus each row contains
 $symmetry\_no$ integer entries. Note that it is necessary to provide the double of
the usual quantum numbers in case of $SU(2)$ symmetries in order to
be able to work with integer numbers.

It is important that the states must be {\em ordered} by
symmetries: The ordering is done with respect to the first
symmetry, then with respect to the second symmetry used and so on.
While reading the input file, the code checks if the quantum
numbers are in ascending order, and if not, then an error message
is generated and the run stops.

%%%%%%%%%%%%%%%%%%%%%%%%%%%%%%%%%%%%%%%%%%%%%%%%%%%%%%%%%%%%%%%%%%%%%%%%%%%%%%%%

\subsection{$<$SECTION-LOCAL\_STATES$>$}

In this section the quantum numbers are set.
An integer type matrix of dimensions local\_state\_no$\times$symmetry\_no
must be provided.
The number of rows must be equal
to the value 'local\_state\_no' set in $<$SECTION-PARAMETERS$>$.
Otherwise, the rules are the same as for $<$SECTION-BLOCK\_STATES$>$.

%%%%%%%%%%%%%%%%%%%%%%%%%%%%%%%%%%%%%%%%%%%%%%%%%%%%%%%%%%%%%%%%%%%%%%%%%%%%%%%%

\subsection{$<$SECTION-LOCAL\_STATES\_SIGNS$>$}

In this section the signs of the local states must be set, and
a vector of length 'local\_state\_no', containing $\pm 1$-s  must be provided.
The sign of the local state is {\bf -1 / +1} if it contains  an {\bf odd / even} number
of fermions.

%%%%%%%%%%%%%%%%%%%%%%%%%%%%%%%%%%%%%%%%%%%%%%%%%%%%%%%%%%%%%%%%%%%%%%%%%%%%%%%%

\subsection{$<$SECTION-BLOCK\_HAMILTONIAN$>$}

This section contains the definition of the interaction Hamiltonians, ${\cal H}_\delta$.

\begin{itemize}
\item
{\bf $<$BLOCK\_HAMILTONIAN\_TERM$>$}
tag denotes the beginning of the definition of a new
interaction Hamiltonian term.

\item
{\bf $<$BLOCK\_HAMILTONIAN\_NAME$>$}
is a character string of maximum length char[256] to define the name of the ${\cal H}_\delta$ term. It could be any string.

\item
{\bf $<$BLOCK\_HAMILTONIAN\_COUPLING$>$}
a real*8 number, which sets the coupling strength, ($G_\delta$),
corresponding to ${\cal H}_\delta$. In this line you also specify the {\em name}
(notation) of the coupling. By writing a line "J = 0.736", e.g.,  you teach
the code  that this  coupling is called J, and its value is 0.736.

\item
{\bf $<$BLOCK\_HAMILTONIAN\_REPRESENTATION\_INDEX$>$}
\\
the quantum numbers of the Hamiltonian term,
an integer  vector of length 'symmetry\_no'. Under any circumstances
the quantum numbers must be 0 for all the symmetries, otherwise the problem is
badly posed. The representation index of the Hamiltonian may look therefore
redundant  at first, but to have a uniform treatment of all  operators in
the input file,  we decided to keep this line.

\item {\bf $<$BLOCK\_HAMILTONIAN\_MATRIX$>$} This is a matrix of
dimension block\_state\_no $\times$ block\_state\_no. This matrix
contains the matrix elements of the Hamiltonian $H_\delta$.
between the states (multiplets)  enumerated in
$<$SECTION-BLOCK\_STATES$>$. Some {\it Mathematica 6.0} 
notebooks are provided in directory {\tt mathematica\_files}
which can be used to generate  these  matrices in a routine way.
\end{itemize}

The above series of entries must be repeated for {\em every} local Hamiltonian term,
$H_\delta$. The number of these structures  must be equal to {\bf coupling\_no} defined in  $<$SECTION-PARAMETERS$>$.

%%%%%%%%%%%%%%%%%%%%%%%%%%%%%%%%%%%%%%%%%%%%%%%%%%%%%%%%%%%%%%%%%%%%%%%%%%%%%%%%

\subsection{$<$SECTION-HOPPING\_OPERATORS$>$}
This section provides information on the hopping operators.
As a rule, we provide  information only on creation
operators. No information on annihilation operators is needed.

\begin{itemize}
\item
{\bf $ <$HOPPING\_OPERATOR$>$}
tag denotes the beginning of the definition of a new
hopping operator.

\item
{\bf $<$HOPPING\_OPERATOR\_NAME$>$}
is a character string with maximum length char[256] to define the name of the corresponding
hopping operator. (Could be something like ``spin\_up\_hopping'', e.g.)

\item
{\bf $ <$HOPPING\_OPERATOR\_REPRESENTATION\_INDEX$>$}
the quantum numbers of the hopping operator, an integer vector of dimension 1 $\times$ 'symmetry\_no' must be given.
\item
{\bf $ <$HOPPING\_OPERATOR\_SIGN$>$}
The sign is -1 for a fermionic operator and 1 for a bosonic operator.

\item {\bf $ <$HOPPING\_OPERATOR\_MATRIX$>$} Contains the reduced
matrix elements of the hopping operator ($\sim f^\dagger_0$)
between the block states specified  in
$<$SECTION-BLOCK\_STATES$>$. It is a  block\_state\_no $\times$
block\_state\_no matrix.  We provided a few  $Mathematica$
notebooks which you can use to generate these matrix elements in a
simple way.
\end{itemize}

The above series of entries must be repeated as many times as many hopping
operators one has. The number of these structures  must thus  agree with the
'hopping\_operator\_no' set in $<$SECTION-PARAMETERS$>$.

%%%%%%%%%%%%%%%%%%%%%%%%%%%%%%%%%%%%%%%%%%%%%%%%%%%%%%%%%%%%%%%%%%%%%%%%%%%%%%%%

\subsection{$<$SECTION-SPECTRAL\_OPERATORS$>$}

This section provides information on the spectral operators, i.e., operators,
whose spectral functions are computed.
As a rule, we provide information only on the 'creation
operators', and  no information on annihilation operators is given
(see Section \ref{sec:DM-NRG} for a precise definition). The following
structure must be repeated for every spectral operator. Clearly, the number of
structures must be equal to the parameter {\bf  spectral\_operator\_no} set in
section $<$SECTION-PARAMETERS$>$.

\begin{itemize}
\item
{\bf $<$SPECTRAL\_OPERATOR$>$}
tag denotes the beginning of the definition of a new
spectral operator.

\item
{\bf $<$SPECTRAL\_OPERATOR\_NAME$>$}
is a character string with maximum length char[256] to define the name of the corresponding
spectral operator. (E.g., ``spin operator'')

\item
{\bf $<$SPECTRAL\_OPERATOR\_REPRESENTATION\_INDEX$>$}
the quantum numbers of the spectral operator. An
integer like vector of dimension 1 $\times$ symmetry\_no is expected.

\item
{\bf $<$SPECTRAL\_OPERATOR\_SIGN$>$}
The sign is -1 for a fermionic-like operator and 1 for a bosonic operator.

\item {\bf $<$SPECTRAL\_OPERATOR\_MATRIX$>$} A matrix of
dimensions block\_state\_no $\times$ block\_state\_no
 is expected, containing the reduced matrix elements of the spectral operator
between the block states specified  in $<$SECTION-BLOCK\_STATES$>$.  We
included in this package a few examples of   $Mathematica$ notebooks, which can be used to
generate these matrices relatively easily.

\end{itemize}

%%%%%%%%%%%%%%%%%%%%%%%%%%%%%%%%%%%%%%%%%%%%%%%%%%%%%%%%%%%%%%%%%%%%%%%%%%%%%%%%

\subsection{$<$SECTION-STATIC\_OPERATORS$>$}

This section provides information on the
operators for which static averages are evaluated. The following
structure must be repeated for every static operator.

\begin{itemize}
\item
{\bf $<$STATIC\_OPERATOR$>$}
tag denotes the beginning of the definition of a
static operator.

\item
{\bf $<$STATIC\_OPERATOR\_NAME$>$}
is a character string with maximum length char[256], defines the name
of the static operator being defined.

\item
{\bf $<$STATIC\_OPERATOR\_REPRESENTATION\_INDEX$>$}
quantum numbers of the static operator,
an integer type vector of dimension 1 $\times$ symmetry\_no is expected.

\item
{\bf $<$STATIC\_OPERATOR\_SIGN$>$}
This sign is -1 for a fermionic and 1 for a bosonic operator.

\item {\bf $<$STATIC\_OPERATOR\_MATRIX$>$} Contains the reduced
matrix elements of the static operator between the block states
specified  in $<$SECTION-BLOCK\_STATES$>$. It is a
block\_state\_no $\times$ block\_state\_no matrix. For the
calculation of these matrix elements, one can use the mathematica
codes provided with the package.
\end{itemize}

The number of structures shown above must agree with the variable
'static\_operator\_no' set in  $<$SECTION-PARAMETERS$>$.

\subsection{$<$SECTION-LOCAL\_HAMILTONIAN$>$}

\begin{itemize}
\item
{\bf $<$LOCAL\_HAMILTONIAN\_TERM$>$}
tag denotes the beginning of the definition of a
local Hamiltonian, i.e. a term  that occurs in ${\cal H}_i$,
$$
{\cal H}_i = \sum_{\alpha=1}^{local\_coupling\_no}
g_{\alpha,i}\; {\cal H}_{i,\alpha} \;,
$$
with ${\cal H}_{i,\alpha} = {\cal H}_{\alpha}(\{\underline C_{i,\lambda}\})$.
In other words, the local Hamiltonian ${\cal H}_{i,\alpha}$ depends on the
site index $i$ only through the creation operators acting at site
$i$, otherwise its structure is fixed. A trivial example of such operators
are the  energies, $\xi_\lambda \leftrightarrow g$,
$\underline C_{i,\lambda} \underline C_{i,\lambda}^\dagger \leftrightarrow
{\cal H}_{i,\alpha} $.
In the present version of the code,
we only handle {\em uniform local couplings}, $g_{\alpha,i} = g_{\alpha}$,
however, later versions shall also be able to treat a user-specified
coupling  set, $g_{\alpha,i}$.
Note that the on-site energy terms  are specified under the section
$<$SECTION-LOCAL\_HOPPING\_OPERATORS$>$, not here. Here you can define
additional local terms in the Hamiltonian, e.g.,
superconducting pairing terms.

\item
{\bf $<$LOCAL\_HAMILTONIAN\_NAME$>$}
is a character string with maximum length char[256],
defines the name of the ${ H}_i$ term.

\item
{\bf $<$LOCAL\_HAMILTONIAN\_COUPLING$>$}
is a real*8 number to set the coupling strength,
$g_\alpha$, corresponding to the local term, ${\cal H}_{i,\alpha}$.
In the present version of the code the coupling
strength does not change along the Wilson chain.

\item
{\bf $<$LOCAL\_HAMILTONIAN\_REPRESENTATION\_INDEX$>$}
the quantum numbers of the Hamiltonian term,
an integer type vector with dimension 1 $\times$ 'symmetry\_no'.
Similar to the Hamiltonian, these quantum numbers must be all 0,
otherwise the Hamiltonian is not invariant under the
symmetry assumed.

\item {\bf $<$LOCAL\_HAMILTONIAN\_MATRIX$>$} It is a
local\_state\_no $\times$ local\_state\_no matrix, defined on the
basis given in $<$SECTION-LOCAL\_STATES$>$. This contains the
(reduced) matrix elements of the coupling Hamiltonians, ${\cal
H}_{\alpha}$.
\end{itemize}

%%%%%%%%%%%%%%%%%%%%%%%%%%%%%%%%%%%%%%%%%%%%%%%%%%%%%%%%%%%%%%%%%%%%%%%%%%%%%%%%

\subsection{$<$SECTION-LOCAL\_HOPPING\_OPERATORS$>$}
Here we specify matrix elements of the local hopping operators
between the  local states enumerated in $<$SECTION-LOCAL\_STATES$>$.
As a rule, we provide the information only for the creation
operators.

\begin{itemize}
\item
{\bf $<$LOCAL\_HOPPING\_OPERATOR$>$}
 denotes the beginning of the definition of a
hopping operator.

\item
{\bf $<$LOCAL\_HOPPING\_OPERATOR\_NAME$>$}
is a character string with maximum length char[256]. Defines the name
of the  local hopping operator.

\item
{\bf $<$LOCAL\_HOPPING\_OPERATOR\_REPRESENTATION\_INDEX$>$}
the quantum numbers of the local hopping operator, an integer type vector
with dimension 1 $\times$ 'symmetry\_no' must be given.

\item
{\bf $<$LOCAL\_HOPPING\_OPERATOR\_SIGN$>$}
This sign is -1 for a fermionic and 1 for a bosonic operator.

\item {\bf $<$LOCAL\_HOPPING\_OPERATOR\_MATRIX$>$} Contains the
reduced matrix elements of the static operator between the local
states specified  in $<$SECTION-LOCAL\_STATES$>$. It is a
local\_state\_no $\times$ local\_state\_no matrix. For the
calculation of these matrix elements, one can use the mathematica
codes provided with the package.

\item {\bf $<$LOCAL\_ON\_SITE\_ENERGY\_MATRIX$>$} Contains the
reduced matrix elements of the on-site energy term between the
local states specified  in $<$SECTION-LOCAL\_STATES$>$. It is a
local\_state\_no $\times$ local\_state\_no matrix. For the
calculation of these matrix elements, one can use the mathematica
codes provided with the package.
\end{itemize}

The number of structures shown above must agree with the variable
'hopping\_operator\_no'  set in the $<$SECTION-PARAMETERS$>$.

%%%%%%%%%%%%%%%%%%%%%%%%%%%%%%%%%%%%%%%%%%%%%%%%%%%%%%%%%%%%%%%%%%%%%%%%%%%%%%%%

\subsection{$<$SECTION-SPECTRAL\_FUNCTION$>$}

In this section one specifies the pairs of spectral operators, whose spectral
function is computed. This section must contain 'spectral\_function\_no'
rows, each  of them corresponding to a given pair of operators.

To compute  the spectral function of the Green's function of the
form
\begin{equation}
G_{A,B^{\dagger}}^{R}\left ( t \right ) = -i \left < \left [
A(t),\, B^{\dagger}\left (0 \right)\right ]_{\xi} \right
>\Theta(t), \label{eq:green_function}
\end{equation}
one must add a row of the form
$$
\{\mbox{name\_of\_}A^\dagger; \mbox{name\_of\_}B^\dagger\}\;,
$$
where '$\mbox{name\_of\_}A^\dagger$' and '$\mbox{name\_of\_}B^\dagger$' are
the names of the 'creation operators' set up under
$<$SECTION-SPECTRAL\_OPERATORS$>$.

\subsection{$<$SECTION-SPECTRAL\_FUNCTION\_BROADENING$>$}

Here you specify the parameters of the broadening, used to evaluate
the spectral functions.  Note that the code {\bf fnrg} does not generate the spectral
functions, you need to run  the utility {\bf sfb} for that
 immediately after the run. This part of the input file is thus used by
{\bf sfb}. Note that one can improve/change the broadening parameters
after running the code too, and then run {\bf sfb} without running
{\bf fnrg} again.
For details on the broadening, see Section \ref{sec:DM-NRG}).

The following  parameters must be set in this section:

\begin{itemize}
\item
{\bf broadening\_method:}
the method that is used for the broadening of the spectral function along the real axis.
 Two methods are available: LOG\_GAUSS  and INTERPOLATIVE\_LOG\_GAUSS.

\item
{\bf broadening\_parameter:}
The value of the broadening parameter
 (half-width of the log Gauss function in both the log-Gauss and
 interpolative log-Gauss type of broadening). This parameter is used no matter what method is used.

\item
{\bf broadening\_energy\_minim:}
The minimum energy for which the broadening is performed
($broadening\_energy\_minim<|\omega|<broadening\_energy\_maxim$).
This parameter is used for both methods. This quantity must be positive.

\item
{\bf broadening\_energy\_maxim:} The maximum energy for which the broadening
is performed ($broadening\_energy\_minim<|\omega|<broadening\_energy\_maxim$).
This quantity must be positive.

\item
{\bf broadening\_grid\_mesh:} The number of points along the real axis,
for which the broadened spectral function
is computed. The binary {\bf sfb} generates a logarithmic mesh between
[-broadening\_energy\_maxim, -broadening\_energy\_minim]
and [broadening\_energy\_minim, broadening\_energy\_maxim]
for which the spectral function is computed.

\item
{\bf quantum\_temperature:}
This parameter is used only with the INTERPOLATIVE\_LOG\_GAUSS method. It sets
the scale below which a linear interpolation is used rather than a logarithmic
interpolation.

\item
{\bf spectral\_function\_nrg\_even:}
This is a flag that is relevant for NRG runs only. Allowed tags are 'YES' or 'NO':
In case of 'YES' the spectral function is computed from the even NRG
iterations, and a corresponding data file is generated.
 If set to 'NO', the code skips this calculation.

\item
{\bf spectral\_function\_nrg\_odd:}
This is a flag that is relevant for NRG runs only. Allowed tags are 'YES' or 'NO':
In case of 'YES' the spectral function is also computed from odd NRG
iterations.
If set to 'NO', the code skips this calculation.

\item
{\bf spectral\_function\_dmnrg:}
This is a flag parameter that controls whether the spectral function from the
 full NRG spectrum using the density matrix formalism is computed.
Allowed tags are 'YES' or 'NO'. In case of 'YES' the DM-NRG calculation is
also performed.

\item
{\bf static\_average\_dmnrg:}
Flag parameter that controls whether the static average of
the form $\langle A^{\dagger}B\rangle$ is also computed by {\bf sfb}
for the two spectral operators, in addition to the  calculation of the spectral
functions (see the notations  in the previous section).
Allowed tags are 'YES' or 'NO'.

\item {\bf green\_function\_nrg\_even:} This flag  controls
whether the real part of the Green's function is computed from the
even part of the spectral function generated from the nrg data, by
performing a Hilbert transform. Allowed tags are 'YES' or 'NO'.
The real  part of the Green's function can be computed only after
the calculation of the spectral function, i.e. if the {\bf
spectral\_function\_nrg\_even} flag is set to YES.

\item
{\bf green\_function\_nrg\_odd:}
This flag  controls whether the real part of the Green's function
is computed from the odd part of the
spectral function generated from the nrg data, by performing a Hilbert transform.
Allowed tags are 'YES' or 'NO'. The real  part of the Green's function can be
computed only after the calculation of the spectral function, i.e.
if the  {\bf spectral\_function\_nrg\_odd} flag is set to YES.

\item
{\bf green\_function\_dmnrg:}
This flag
controls whether the real part of the Green's function
is computed from the dmnrg spectral function by performing a Hilbert transform.
Allowed tags are 'YES' or 'NO'. The real  part of the Green's function can be
computed only after the calculation of the spectral function, i.e.
if the  {\bf spectral\_function\_dmnrg} flag is set to YES.

\item
{\bf green\_function\_grid\_mesh:}
The number of points on a log-mesh used for doing the Hilbert transform.
Once the spectral function for a combination of spectral
 operators is computed, the real part of corresponding Green's
function is generated by performing a Hilbert transformation. To
compute the Hilbert transform,  a logarithmic mesh is generated.
The energy limits for the real/imaginary part of the Green's
functions are the same as those for the spectral function , {\it
i.e.} [-broadening\_energy\_maxim, -broadening\_energy\_minim] and
[broadening\_energy\_minim, broadening\_energy\_maxim]. The energy
values for which the Green's function is computed correspond to
those for the spectral function. Cubic splines are used for the
interpolation of the spectral function. Therefore,  for too small
green\_function\_grid\_mesh values sometimes oscillations are
observed in the energy dependence of the real
 part of the Green's function. In this case, $green\_function\_grid\_mesh$
must be increased  until the oscillations vanish. Increasing this parameter
has only a small impact on computing time.

\end{itemize}

%%%%%%%%%%%%%%%%%%%%%%%%%%%%%%%%%%%%%%%%%%%%%%%%%%%%%%%%%%%%%%%%%%%%%%%%%%%%%%%%

%---------------------- Section ------------------------%
\vfill \eject

\section{Outputs of the DM-NRG code}

\subsection{Output directory structure}

The output of the code has the following directory structure:
%\begin{table}{ccc}
%\end{table}
%{\em ./results/model_dependent_name/data/

\begin{table}[htb]
{\small
%\begin{tabular}{@{\hspace{1mm}}lc@{\hspace{5mm}}c@{\hspace{4mm}}c@{\hspace{1mm}}}
\begin{tabular}{llll}
\hline \hline
%\multicolumn{1}{c}{$U$} &  & $\Delta_{\rm{c}}$ & $\Delta_{\rm{s}}$ \\
\hline
{\tt  ./results/}&                              &              & all results of the code    \\
{\tt ./results}/& {\tt automatic\_directory\_name/}              &              & all data for a given run   \\
          &  {\tt (code generated)}            &              &\\
          & {\tt log\_file.dat  }             &          &   \\
          &                              &{\tt input.dat} &  \\
          &                              &{\tt spectral\_function\_dmnrg\_* }    &  \\
          &                              &{\tt spectral\_function\_even\_* }    & \\
         &                              &{\tt spectral\_function\_odd\_*  }   &  \\
          &                              &{\tt green\_function\_dmnrg\_* }&  \\
          &                              &{\tt green\_function\_even\_*} &  \\
          &                              &{\tt green\_function\_odd\_*} &  \\
          &                              &{\tt static\_average\_dmnrg* }    &  \\
          &                              &{\tt energy\_levels\_even* }    & \\
          &                              &{\tt energy\_levels\_odd*}     & \\
%G          &                              &data/         &   m\_star.dat\\
          &                              &{\tt data/ }        &{\tt spectral\_operators/spectral\_operator\_*.dat} \\
          &                              &              &{\tt static\_operators/static\_operator\_*.dat}\\
          &                              &              &{\tt  sectors/sectors\_*.dat} \\
          &                              &              & {\tt transformations/transformation\_*.dat} \\
          &                              &              &{\tt  states/states\_*.dat} \\
          &                              &              & {\tt delta\_peaks/delta\_peaks\_*.dat}\\
          &                              &              &{\tt spectral\_functions/spectral\_function\_*.dat}\\
          &                              &               &{\tt mapping/dos.dat}\\
          &                              &              &{\tt mapping/hopping\_couplings*.dat}\\
          &                              &              & {\tt  mapping/on\_site\_energies*.dat}\\
          &                              &               &{\tt mapping/density\_of\_states*.dat}\\
          &                              &               &{\tt m\_star.dat}\\
\hline \hline
\end{tabular}
}
\caption{Output directory structure of the code. Directory {\tt automatic\_directory\_ name/} is
  generated automatically from the parameters in the input file. The ``*'' in the
  file names also stands for a list of the parameters (names of the couplings
  etc.) generated automatically.}
\label{table:dir_struct}
\end{table}

All outputs of the code, which are saved on disk,  are located in the directory
{\tt results}.
In this directory the program generates automatically a directory
corresponding to the parameters of the given run. The name of this directory,
{\tt automatic\_directory\_name/},
contains the values of the following input variables: {\em model, lambda,
max\_state\_no,
iteration\_no,
temperature,
symmetry\_no,
symmetry\_types,
coupling\_no},
and a file {\tt log\_file.dat} containing the log/error messages during the run.

The {\tt automatic\_directory\_name/}, folder has the following structure:

$\bullet$
{\bf input.dat} is a copy of the original input file.

$\bullet$
Various {\bf *.dat} files contain the final results for the {\bf spectral functions},
{\bf  static averages}, {\bf green's functions} and the {\bf energy levels}.
The names of these files are descriptive, and contain some additional tags
such as, e.g.,  the names of quantum operators for which the spectral functions
have been computed.

After running  {\bf fnrg},  only a directory {\bf data} containing
the raw data is generated. The spectral functions, Green's
functions, static averages files are generated only later by
running the {\bf sfb} utility. The energy level files are also
generated separately, by running the {\bf es} utility.

$\bullet$
Directory {\bf data} contains the data files  generated by {\bf fnrg}.
Since there are hundreds of files, this directory is  divided into
subdirectories with descriptive names and containing the following files:
{\tt 

\hspace{1cm} mapping/

\hspace{1cm} sectors/sectors\_*.dat

\hspace{1cm} states/states\_*.dat

\hspace{1cm} delta\_peaks/delta\_peaks\_dmnrg\_*

\hspace{1cm} delta\_peaks/delta\_peaks\_nrg\_*

\hspace{1cm} spectral\_operators/spectral\_operator\_*

\hspace{1cm} static\_operators/static\_average\_dmnrg\_*

\hspace{1cm} static\_operators/static\_average\_nrg\_even\_*

\hspace{1cm} static\_operators/static\_average\_nrg\_odd\_*

\hspace{1cm} transformations/transformation\_*.dat

\hspace{1cm} spectral\_functions/spectral\_function\_*.dat

\hspace{1cm} m\_star.dat
}
Here again,  *'s stand for operator names and code-generated tags such as the
iteration number etc.

\subsection{Detailed description of the output files in the folder \\
{\tt ./results/automatic\_directory\_name/data}}

This folder contains the original data generated/used by the code
{\bf fnrg} and the utilities {\bf he} and {\bf sfb}. It contains
the following files/folders:

\vspace{0.3cm}
$\bullet$ {\bf mapping/}:  This folder contains information on the
hopping parameters and on site energies, in case a user-defined density of states
has been used.  This directory is a copy of the directory
{\it dos\_mapping}, located in the main directory.  It contains the
following files:

%\item[]{\bf dos.dat}: This is the original DOS provided by the user, from which the hopping and
%on-site energies are generated by the utility {\bf he}. The first column
%contains the energy and there is hopping\_operator\_no additional columns, one
%for every hopping operator's density of states.

%\item[] {\bf density\_of\_states\_iteration\_(interation\_no).dat}
%contain data for the density of states on a logarithmic mesh.
%These files are generated from {\it dos.dat} by the utility
%{\bf he} (hopping energies), by using an interpolation formula.
%Data for positive and negative energies are saved separately.

\vspace{0.3cm}$\bullet$
 {\bf mapping/hopping\_couplings*.dat}:
Contains the data for the hopping couplings along the chain as computed by the
{\bf he} (hopping energies) utility.
Each line contains information for one iteration, the first column being the
iteration number and
subsequent columns representing the values of the hopping amplitudes along the Wilson
chain.  Each column corresponds to a hopping operator, so the
number of columns has to agree with the 'hopping\_operator\_no'. The tag
``self-consistency''  in the file name has no relevance here, it was included
for later purposes.

\vspace{0.3cm}$\bullet$
 {\bf mapping/on\_site\_energies*.dat}
contains the data for the on-site energies along the chain as computed by the
{\bf he} (hopping energies) utility.
Each line contains information for each iteration, the first column being the
iteration number and the subsequent columns being the values of the on-site
energies.   Each column corresponds to a hopping operator, so the
number of columns has to agree with the 'hopping\_operator\_no'.
 The tag ``self-consistency''in the file name has no relevance here, it was
 included for later purposes.

\vspace{0.3cm}$\bullet$
{\bf sectors/sectors\_(iteration\_no).dat} The code is organized using so-called sectors.
A given sector is characterized by a list of representation labels,
specifying the symmetries of the states of that sector. Similarly,
operators are also organized using the notion of sectors.  The files
{\tt sectors\_(iteration\_no).dat} contain the information on the size and
symmetry of sectors in iteration  {\it iteration\_no}.

This file has two separate parts. The first part describes the sectors
in the new basis at iteration {\it iteration\_no}.
The first row of the file contains the total number of the sectors in the
new basis ({\it new\_block\_sector\_no}). The subsequent {\it
  new\_block\_sector\_no} lines  contain the information for each sector (from 0
to {\it new\_block\_sector\_no-1}):  the
representation indices of the sectors, the dimensions of the sectors,
the first states of the sectors among the list of all the basis states (multiplets), and
the degeneracies of the sectors.
Note that the sum of the dimensions multiplied with the corresponding degeneracy
values must be equal to the
total number of basis states of the new block state basis.

The second part of the files  {\tt sectors\_(iteration\_no).dat} contains the
same information on the states kept from the previous iteration (iteration {\it iteration\_no-1}).

\vspace{0.3cm}$\bullet$
{\bf states/states\_(iteration\_no).dat}:  These files
contain the information for the new  basis states at  iteration {\it
  iteration\_no}, and the   states kept from the previous iteration (iteration {\it iteration\_no}-1).
States know about their parent states, and this information is also displayed
in these files.

The first row of the file contains the total number of the new basis states
({\it new\_block\_state\_no})
The next {\it new\_block\_state\_no}  lines contain the information on the
multiplets after the diagonalization has been done. Each line contains the number (label) of the sector
the state belongs to, the index of the state within this sector,  the list of representation indices,
the labels of the parent and local states the state was constructed from, the sign of the
state (this is 0 for block states), whether the state is kept (K) or
eliminated (E) after truncation, and finally the eigen-energy of the state.
Note that the sector number cannot be less than zero and it must be smaller than
{\it new\_block\_sector\_no}, specified in file {\em
  sectors\_(iteration\_no).dat}. Also  the variable {\it sector\_index}
must be non-negative and less than the dimension of the sector given also in file
{\tt sectors\_(iteration\_no).dat}.

The second part of the file {\tt states/states\_(iteration\_no).dat} gives
the same information on the states kept from the previous iteration.

\vspace{0.3cm}$\bullet$
{\bf transformations/transformation\_(iteration\_no).dat}: These files
contain the information on the matrix $O_n$
that is used to diagonalize the Hamiltonian, and to transform the hopping operators at
site {\it iteration\_no},  and the
spectral operators (acting at site 0) to the new basis as
\begin{equation}
A_n\to O_n A_{n} O_n^\dagger
\end{equation}

These files are only generated if the flag {\it text\_swap\_files\_flag} in the
file {\tt input.dat} is set   ON.

The first row of the file contains the total number of the sectors in the new basis
in iteration {\it iteration\_no},
and this is followed by three terms for each sector: the sector number, the size of the
sector, and finally the real*8 matrix of size  sector\_size $\times$ sector\_size,
transforming the given sector.

\vspace{0.3cm}$\bullet$
{\bf transformations/transformation\_binary\_(iteration\_no).dat} contains the information for the $O$
matrix performing the diagonalization at iteration {\it iteration\_no}
in a binary form.
These files are only generated if the flag {\it binary\_swap\_files\_flag} in the
file {\tt input.dat} is set to  ON.

\vspace{0.3cm}$\bullet$ {\bf
spectral\_operators/spectral\_operator\_({\em
operator\_name})\_(iteration\_no).dat} contains the matrix
elements of a spectral operator acting at site 0 of the chain (for
example $d_0^{\dagger}$ or $S_z$) in a sector-decomposed form,
after the diagonalization in iteration step {\it iteration\_no}
has been done.  These operators (which are 'creation operators')
are defined and initialized in the {\tt input.dat} file,
$<$SECTION-SPECTRAL\_OPERATORS$>$. These files are only generated
if the flag {\it text\_swap\_files\_flag} in the file {\tt
input.dat} is set   ON.

The matrix of a spectral operator can be decomposed to sector\_no
by sector\_no small blocks. Depending on the quantum numbers of
the operator, several sectors (actually most of them) contain only
zero elements. These are considered as non-active sectors. The
first line of the file contains the total number of the active
sectors (those with proper change of quantum numbers). Then for
each active sector the row and column index of the sector is given
and is followed by the row and column dimension of the sector, and
the matrix elements of the operator in that sector.

\vspace{0.3cm}$\bullet$
{\bf spectral\_operators/spectral\_operator\_binary\_({\em operator
    name})\_(iteration\_no).dat}: These files
contain the same information as the files {\bf spectral\_operator\_({\em operator name})\_*.dat},
excepting that the data are in a binary form.
These files are only generated if the flag {\it binary\_swap\_files\_flag} in the
file {\tt input.dat} is set to  ON.

\vspace{0.3cm}$\bullet$
{\bf static\_operators/static\_operator\_({\em operator
    name})\_(iteration\_no).dat}:
They contain the matrix elements of a static operator acting at
site 0 of the chain (for example $n_0$)  in a sector-decomposed
form, after the diagonalization in iteration step {\it
  iteration\_no} has been done.  These operators (which are 'creation operators')
are read from the {\tt input.dat} file,
$<$SECTION-STATIC\_OPERATORS$>$. These files are only generated if
the flag {\it text\_swap\_files\_flag} in the file {\tt input.dat}
is set   ON.

Static operators must have a block-diagonal structure by symmetry, therefore
only these sectors are active. These files are organized otherwise in the
same way as the files {\tt spectral\_operator\_*.dat}.

These files are only generated if the flag {\it text\_swap\_files\_flag} in the
file {\tt input.dat} is set to  ON. Furthermore corresponding binary files
are also generated
if the flag {\it binary\_swap\_files\_flag} in the
file {\tt input.dat} is set to  ON.

\vspace{0.3cm}$\bullet$
{\bf delta\_peaks/delta\_peaks\_dmnrg\_({\em operator 1})\_({\em operator  2})\_({\em
    iteration\_no }).dat}
contain the raw spectral function peaks for the spectral function of {\em
  operator 1} and {\em operator  2}, generated from iteration  {\it iteration\_no}.
These files  contain two columns: the first one is the energy and
the second one the contribution to the amplitude of the spectral
function at that energy in iteration {\em iteration\_no}. The
calculation is performed at a temperature $T$ specified in the
input file, {\tt input.dat}.   These output files are used by the
utility {\bf sfb} to compute the DM-NRG spectral functions,
combining the data from all  iterations.

\vspace{0.3cm}$\bullet$
{\bf delta\_peaks/delta\_peaks\_nrg\_({\em operator name 1})\_({\em operator name 2})\_({\em iteration\_no}) .dat}
contains peaks for the spectral function of {\em
  operator 1} and {\em operator  2}, generated from iteration  {\it iteration\_no}.
These files contain two columns, the first one being the
renormalization energy and the second one the amplitude of the
spectral function at the corresponding energy  and  iteration {\em
iteration\_no}. The calculation is performed at a temperature $T$
specified in the input file, {\tt input.dat}.  These files are
used to compute the NRG spectral function  combining data from
either even or odd iterations.

\vspace{0.3cm}$\bullet$ {\bf
spectral\_functions/spectral\_function\_dmnrg\_({\em operator
1})\_({\em operator 2})\\ \_({\em broadening\_par})\_({\em
iteration}) .dat} contains two columns, the first one being the
energy and the second one the real axis value of the spectral
function corresponding to the two operators from the filename at
the iteration {\em iteration\_no}. Data for negative as well as
positive energies are saved into this file. These particular files
are relatively important for analyzing the contribution of each
iteration to the total spectral function.

 \vspace{0.3cm}$\bullet$
{\bf static\_operators/static\_average\_nrg\_even\_(operator\_name).dat}
and \\{\bf static\_average\_nrg\_odd\_(operator\_name).dat}:
These files contain  average of a  static operator as a function of
temperature, computed by the standard NRG prescription, from the even/odd
iterations, respectively.

\vspace{0.3cm}$\bullet$
{\bf static\_operators/static\_average\_dmnrg\_(operator\_name).dat} and \\ {\bf
  static\_average\_dmnrg\_(operator1)\_(operator2).dat}: These files contain
the expectation values of the static operators and of the pairs of spectral
operators specified in the file {\tt input.dat}, computed using the DM-NRG
procedure at a temperature $T$, also specified in the file {\tt input.dat}.

$\bullet$
{\bf m\_star.dat}:
     m\_star is the value of the iteration\_index at which
     truncation of the states begins (new\_state\_no $>$ max\_state\_no), i.e.,
     for iterations larger then m\_star
     the states, transformation matrices, spectral operators have to be saved.
     m\_star remains fixed for the rest of the NRG steps. This file is set
     on input and by function {\tt get\_m\_star()} in {\tt cut\_off.cc} and read by
     various functions when spectral functions are generated later.

\subsection{Analyzing the data}

There are several utilities that we provide to analyze the data you
generated.
\begin{itemize}
\item First of all,  spectral functions and their Hilbert transforms can be generated/analyzed
by the utility {\bf sfb}. The broadening parameters must be set in the file
{\tt input.dat} sitting in the main folder. You can set there the broadening
method, as described in Chapter~\ref{ch:init}. (See also the last section of Chapter
\ref{ch:NRG_intro}.) Depending on the parameters set this utility generates
the following files.
\begin{itemize}
 \item[] {\bf spectral\_function\_dmnrg\_({\em operator1})\_({\em
       operator2})\_({\em broadening}).dat}\\
   These files contain the broadened spectral function of operators   {\em operator1}
   and {\em operator2}. Correspondingly, in case the Hilbert transform flag is ON
   in the file {\tt input.dat}, the real part of the corresponding
   Green's function is also generated in file   {\bf green\_function\_dmnrg\_*.dat}.

 \item[] {\bf spectral\_function\_even\_({\em operator1})\_({\em
       operator2})\_({\em broadening}).dat} and \\
   {\bf spectral\_function\_odd\_({\em operator1})\_({\em
       operator2})\_({\em broadening}).dat}:

   These files contain the broadened spectral functions of   {\em operator1}
   and {\em operator2}, as computed by the traditional NRG method. Data are
   collected from even and odd iterations respectively. If the Hilbert transform flag is ON
   in the file {\tt input.dat}, then the real parts of the corresponding
   Green's functions are also generated and stored in the files   {\bf
     green\_function\_even\_*.dat} and {\bf
     green\_function\_odd\_*.dat}.
\end{itemize}

\item The energy spectrum (finite size spectrum) can be analyzed by running
  the utility {\bf es}. While running this utility you can specify quantum
  number filtering as well as the number of states that you want to keep track
  of. This is a useful utility to obtain the fixed point spectrum of a model,
  identify the corresponding conformal field theory, the dimensions of the
  irrelevant operators, or to extract Fermi liquid parameters. The results are
  stored separately for even and odd iterations in the files
{\bf energy\_levels\_even*} and {\bf energy\_levels\_odd*}.
\end{itemize}

\label{ch:output}

%---------------------- Section -----------------------------------------------%

%%%%%%%%%%%%%%%%%%%%%%%%%%%%%%%%%%%%%%%%%%%%%%%%%%%%%%%%%%%%%%%%%%%%%%%%%%%%%%%%
%                      CHAPTER                                                 %
%%%%%%%%%%%%%%%%%%%%%%%%%%%%%%%%%%%%%%%%%%%%%%%%%%%%%%%%%%%%%%%%%%%%%%%%%%%%%%%%

%%%%%%%%%%%%%%%%%%%%%%%%%%%%%%%%%%%%%%%%%%%%%%%%%%%%%%%%%%%%%%%%%%%%%%%%%%%%%%%%
%       BUG Reports                                                            %
%%%%%%%%%%%%%%%%%%%%%%%%%%%%%%%%%%%%%%%%%%%%%%%%%%%%%%%%%%%%%%%%%%%%%%%%%%%%%%%%

%%%%%%%%%%%%%%%%%%%%%%%%%%%%%%%%%%%%%%%%%%%%%%%%%%%%%%%%%%%%%%%%%%%%%%%%%%%%%%%%
%       License agreements                                        %
%%%%%%%%%%%%%%%%%%%%%%%%%%%%%%%%%%%%%%%%%%%%%%%%%%%%%%%%%%%%%%%%%%%%%%%%%%%%%%%%

%%%%%%%%%%%%%%%%%%%%%%%%%%%%%%%%%%%%%%%%%%%%%%%%%%%%%%%%%%%%%%%%%%%%%%%%%%%%%%%%
%       Acknowledgement                                                        %
%%%%%%%%%%%%%%%%%%%%%%%%%%%%%%%%%%%%%%%%%%%%%%%%%%%%%%%%%%%%%%%%%%%%%%%%%%%%%%%%
\chapter * {Acknowledgments}
\addcontentsline{toc}{chapter}{Acknowledgments}

We are deeply indebted to a number of people for the useful discussions 
and comments that helped us to create this flexible code. 
First of all, we would like to thank Laszlo Borda, with whom  
we created the first  NRG code in Hungary many years ago, and whose 
experience, advices and constant support were crucial to write 
this flexible code. We benefited a lot from discussions with 
other NRG experts too, including Mikito Koga, Ralf Bulla, Frithjof Anders, 
and Walter Hofstetter. We are also indebted to  Jan von Delft 
and  Andreas Weichselbaum for many long and constructive discussions. 
Finally, we would like to thank Laszlo Udvardi for
taking care of our computer cluster at the TU Budapest (BUTE), where 
most of the work has been carried out. 

This work was supported by the Hungarian Research Fund (OTKA)
under Grant Nos. NF61726,  T046303, K68340, and K73361. 
We are also grateful to the Institute of Mathematics of the
BUTE for providing  access to their computer cluster
(supported by OTKA under grant No. 63066). 
C.P.M. was partially supported by the Romanian Grant 
No. CNCSIS 780/2007. I.W. acknowledges support from 
the Foundation for Polish Science.

\appendix
\chapter{Input file for the Kondo model with $U_{{\rm charge}}(1) \times U_{\rm spin}(1)$ symmetries}
{\small
{\tt
\# The comments must start with the symbol \# .\\
\# Empty lines are ignored.\\
\\
\\
\#\#\#\#\#\#\#\#\#\#\#\#\#\#\#\#\#\#\#\#\#\#\#\#\#\#\#\#\#\#\#\#\#\#\#\#\#\#\#\#\#\#\#\#\#\#\#\#\#\#\#\#\#\#\#\#\#\\
		$<$SECTION-PARAMETERS$>$   \\
\#\#\#\#\#\#\#\#\#\#\#\#\#\#\#\#\#\#\#\#\#\#\#\#\#\#\#\#\#\#\#\#\#\#\#\#\#\#\#\#\#\#\#\#\#\#\#\#\#\#\#\#\#\#\#\#\#\\
\\
\# Explanation of variables\\
\\
\# 'model' represents the type of the model that is \\
\#  used for the calculations. It can be any name. \\
\\
\#  'lambda' represents the value for the logarithmic \\
\#  discretization parameter. Usually it is fixed between 2 and 3. \\
\\
\#  'max\_state\_no' is the number of kept states  \\
\#  after an iteration.\\
\\
\#  'iteration\_no' is the number of iterations performed \\
\#  along the NRG run. Typically it is set to 40 - 70.\\
\\
\#  'symmetry\_no' specifies how many symmetries \\
\#  are used. The symmetries  are characterized \\
\#  in $<$SECTION-SYMMETRIES$>$ \\
\\
\#  'coupling\_no' specifies how many couplings \\
\#  are present in the interaction Hamiltonian.  \\
\\
\#  'local\_coupling\_no' is the number of \\
\#  Hamiltonians at the sites we add to the Wilson chain\\
\\
\#  'spectral\_operator\_no' is the number of spectral operators \\
\#  acting on site 0, we keep track of.\\
\#  These spectral operators are characterized\\
\# in section $<$SECTION-SPECTRAL-OPERATORS$>$\\	
\\
\#  'static\_operator\_no' is the number of static operators \\
\#  acting on site 0, we keep track of.\\
\#  These static operators are characterized\\
\#  in section $<$SECTION-STATIC-OPERATORS$>$\\	
\\
\#  'hopping\_operator\_no' is the number of hopping operators. \\
\#   These hopping operators are characterized\\
\#  in section $<$SECTION-HOPPING-OPERATORS$>$\\	
\\
\#  'block\_state\_no' specifies the number \\
\#  of initial states/multiplets. These states are characterized\\
\#  by the representation indices and are listed\\
\#  in section $<$SECTION-BLOCK\_STATES$>$\\
\\
\\
\#\#\#\#\#\#\#\#\#\#\#\#\#\#\#\#\#\#\#\#\#\#\#\#\#\#\#\#\#\#\#\#\#\#\#\#\#\#\#\#\#\#\#\#\#\#\#\#\#\#\#\#\#\#\#\#\#\\
\\
\#\#\#\#\#\#\#\#\#\#\#\#\#\#\#\#\#\#\#\#\#\#\#\#\#\#\#\#\#\#\#\#\#\#\#\#\#\#\#\#\#\#\#\#\#\#\#\#\#\#\#\#\#\#\#\#\#\\
\\
model = kondo\_model  \# the name of the present model \\
\\
lambda = 2.0   \# discretization parameter\\
\\
max\_state\_no = 100  \# maximum number of kept multiplets\\
\\
iteration\_no = 20  \# allowed number of iterations\\
\\
symmetry\_no = 2  \# the number of symmetries  \\
\\
coupling\_no = 2   \# no of Hamiltonian terms, i.e. {\cal J} and {\cal B} in this case\\
\\
spectral\_operator\_no = 3   \# no of operators for which correlation functions are computed \\
\\
static\_operator\_no = 1  \# no of operators for which the static average is computed. \\
\\
hopping\_operator\_no = 2  \# the number of hopping operators\\
\\
block\_state\_no = 8  \# the number of initial block states\\
\\
local\_state\_no = 4   \# the number of local states added in each iteration\\
\\
local\_coupling\_no = 1  \# the number of Hamiltonian terms on the added site\\
\\
spectral\_function\_no = 3 \#  the number of spectral functions that need to be generated \\
\\
interval\_no = 1000 \# mash grid between the maximum and minimum energy at each iteration \\
\\
degeneracy\_threshhold = 1e-6  \# energy thresh-hold for discarding states \\
\\
temperature = 0.0         \# temperature used for the run\\
\\
\\
\#\#\#\#\#\#\#\#\#\#\#\#\#\#\#\#\#\#\#\#\#\#\#\#\#\#\#\#\#\#\#\#\#\#\#\#\#\#\#\#\#\#\#\#\#\#\#\#\#\#\#\#\#\#\#\#\\
		$<$/SECTION-PARAMETERS$>$   \\
\#\#\#\#\#\#\#\#\#\#\#\#\#\#\#\#\#\#\#\#\#\#\#\#\#\#\#\#\#\#\#\#\#\#\#\#\#\#\#\#\#\#\#\#\#\#\#\#\#\#\#\#\#\#\#\#\\
\\
\\
\#\#\#\#\#\#\#\#\#\#\#\#\#\#\#\#\#\#\#\#\#\#\#\#\#\#\#\#\#\#\#\#\#\#\#\#\#\#\#\#\#\#\#\#\#\#\#\#\#\#\#\#\#\#\#\#\\
		$<$SECTION-FLAGS$>$   \\
\#\#\#\#\#\#\#\#\#\#\#\#\#\#\#\#\#\#\#\#\#\#\#\#\#\#\#\#\#\#\#\#\#\#\#\#\#\#\#\#\#\#\#\#\#\#\#\#\#\#\#\#\#\#\#\#\\
\\
\\
\# dmnrg\_flag  controls whether the backward procedure \\
\# has to be done and the full set of eigen-states be \\
\# used for the calculation of the spectral function. \\
\# ON - backwards iteration is done.\\
\# OFF - only the nrg calculation is performed.\\
\\
\# text\_swap\_files\_flag = flag that fixes if the text mode is \\
\# used to save the files. The files will be saved in binary mode anyway. \\
\# ON - the files are saved in the text mode and kept on the disk.\\
\# OFF= the files are not saved in the text mode. \\
\\
\#  binary\_swap\_files\_flag = flag that controls whether the \\
\#  unnecessary binary files will be removed or not not after the calculation is done.\\
\#  ON - the files will be kept on the disk.\\
\#  OFF- the files will be removed from the disk.\\
\\
\# hoppings\_on\_site\_energies\_flag = flag that controls \\
\# the reading of the hoppings and the on-site energies. \\
\# ON - the hoppings and the on site energies \\
\# are read from the files in the  results/mapping/ folder\\
\# These files are generated by the {\bf he} utility.\\
\# OFF - the hoppings are computed on the fly \\
\# assuming a flat density of states (DOS = 0.5) on [-1, 1],\\
\# and the on-site energies are set to zero. 
\\
\\
dmnrg\_flag = ON \\
text\_swap\_files\_flag = OFF \\
binary\_swap\_files\_flag = ON \\
hoppings\_on\_site\_energies\_flag = OFF \\
\\
\\
\#\#\#\#\#\#\#\#\#\#\#\#\#\#\#\#\#\#\#\#\#\#\#\#\#\#\#\#\#\#\#\#\#\#\#\#\#\#\#\#\#\#\#\#\#\#\#\#\#\#\#\#\#\#\#\#\\
		$<$/SECTION-FLAGS$>$   \\
\#\#\#\#\#\#\#\#\#\#\#\#\#\#\#\#\#\#\#\#\#\#\#\#\#\#\#\#\#\#\#\#\#\#\#\#\#\#\#\#\#\#\#\#\#\#\#\#\#\#\#\#\#\#\#\#\\
\\
\\
\#\#\#\#\#\#\#\#\#\#\#\#\#\#\#\#\#\#\#\#\#\#\#\#\#\#\#\#\#\#\#\#\#\#\#\#\#\#\#\#\#\#\#\#\#\#\#\#\#\#\#\#\#\#\#\#\#\\
		$<$SECTION-SYMMETRIES$>$\\
\#\#\#\#\#\#\#\#\#\#\#\#\#\#\#\#\#\#\#\#\#\#\#\#\#\#\#\#\#\#\#\#\#\#\#\#\#\#\#\#\#\#\#\#\#\#\#\#\#\#\#\#\#\#\#\#\#\\
\\
\#  As many symmetries must be defined below  \\
\#  as set by the  variable 'symmetry\_no' \\
\#  in the previous section. \\
\\
\#  The possible symmetry types are the following:\\
\#  U(1), SU(2), Z(2), charge\_SU(2).\\
\#  \\
\#  Next the limits for the representation index \\
\#  are set. All states with smaller/larger\\
\#  representation indices will be cut off\\
\\
\#  For U(1) symmetries the representations are labeled\\
\# by integers (2 S\_z) and for SU(2) by non-negative\\
\# integers (2S).\\
\\
\#  For the Z(2) symmetry the representations are labeled\\
\#  by integers \\
\#  0 - even parity states\\
\#  1 - odd parity states\\
\\
\\
\#\#\#\#\#\#\#\#\#\#\#\#\#\#\#\#\#\#\#\#\#\#\#\#\#\#\#\#\#\#\#\#\#\#\#\#\#\#\#\#\#\#\#\#\#\#\#\#\#\#\#\#\#\#\#\#\# \\
\\
\\
\# charge symmetry corresponding to Q \\
U(1)     -10     10      \\
\\
\# spin symmetry corresponding to 2 S\_z\\
U(1)     -20     20     \\
\\
\\
\#\#\#\#\#\#\#\#\#\#\#\#\#\#\#\#\#\#\#\#\#\#\#\#\#\#\#\#\#\#\#\#\#\#\#\#\#\#\#\#\#\#\#\#\#\#\#\#\#\#\#\#\#\#\#\#\# \\
		$<$/SECTION-SYMMETRIES$>$\\
\#\#\#\#\#\#\#\#\#\#\#\#\#\#\#\#\#\#\#\#\#\#\#\#\#\#\#\#\#\#\#\#\#\#\#\#\#\#\#\#\#\#\#\#\#\#\#\#\#\#\#\#\#\#\#\#\# \\
\\
\\
\#\#\#\#\#\#\#\#\#\#\#\#\#\#\#\#\#\#\#\#\#\#\#\#\#\#\#\#\#\#\#\#\#\#\#\#\#\#\#\#\#\#\#\#\#\#\#\#\#\#\#\#\#\#\#\#\#\\
		$<$SECTION-BLOCK\_STATES$>$\\
\#\#\#\#\#\#\#\#\#\#\#\#\#\#\#\#\#\#\#\#\#\#\#\#\#\#\#\#\#\#\#\#\#\#\#\#\#\#\#\#\#\#\#\#\#\#\#\#\#\#\#\#\#\#\#\#\#\\
\# This section gives the representation indices of the \\
\# block states of the first iteration. \\
\# The number of rows has to be equal \\
\# with the 'block\_state\_no' in the first section, \\
\# and the number of columns to the 'symmetry\_no'.\\
\# Each column contains indices corresponding to a symmetry.\\
\#\#\#\#\#\#\#\#\#\#\#\#\#\#\#\#\#\#\#\#\#\#\#\#\#\#\#\#\#\#\#\#\#\#\#\#\#\#\#\#\#\#\#\#\#\#\#\#\#\#\#\#\#\#\#\#\#\\
$
\begin{array}{ll}
-1&-1\\
-1&1\\
0&-2\\
0&0\\
0&0\\
0&2\\
1&-1\\
1&1\\
\end{array}
$\\
\#\#\#\#\#\#\#\#\#\#\#\#\#\#\#\#\#\#\#\#\#\#\#\#\#\#\#\#\#\#\#\#\#\#\#\#\#\#\#\#\#\#\#\#\#\#\#\#\#\#\#\#\#\#\#\#\#\\
		$<$/SECTION-BLOCK\_STATES$>$\\
\#\#\#\#\#\#\#\#\#\#\#\#\#\#\#\#\#\#\#\#\#\#\#\#\#\#\#\#\#\#\#\#\#\#\#\#\#\#\#\#\#\#\#\#\#\#\#\#\#\#\#\#\#\#\#\#\#\\
\\
\\
\#\#\#\#\#\#\#\#\#\#\#\#\#\#\#\#\#\#\#\#\#\#\#\#\#\#\#\#\#\#\#\#\#\#\#\#\#\#\#\#\#\#\#\#\#\#\#\#\#\#\#\#\#\#\#\#\#\\
		$<$SECTION-LOCAL\_STATES$>$\\
\#\#\#\#\#\#\#\#\#\#\#\#\#\#\#\#\#\#\#\#\#\#\#\#\#\#\#\#\#\#\#\#\#\#\#\#\#\#\#\#\#\#\#\#\#\#\#\#\#\#\#\#\#\#\#\#\#\\
\# This section gives  the representation indices \\
\# of the local states. The number of rows has to be equal\\ 
\# with the 'local\_state\_no' in the first section\\
\# and the number of columns to the 'symmetry\_no'.\\
\# Each column contains indices corresponding to a symmetry.\\
\#\#\#\#\#\#\#\#\#\#\#\#\#\#\#\#\#\#\#\#\#\#\#\#\#\#\#\#\#\#\#\#\#\#\#\#\#\#\#\#\#\#\#\#\#\#\#\#\#\#\#\#\#\#\#\#\\
$
\begin{array}{ll}
-1    & 0\\
0     & -1\\
0     &  1\\
1     &  0\\
\end{array}
$\\
\#\#\#\#\#\#\#\#\#\#\#\#\#\#\#\#\#\#\#\#\#\#\#\#\#\#\#\#\#\#\#\#\#\#\#\#\#\#\#\#\#\#\#\#\#\#\#\#\#\#\#\#\#\#\#\#\\
		$<$/SECTION-LOCAL\_STATES$>$\\
\#\#\#\#\#\#\#\#\#\#\#\#\#\#\#\#\#\#\#\#\#\#\#\#\#\#\#\#\#\#\#\#\#\#\#\#\#\#\#\#\#\#\#\#\#\#\#\#\#\#\#\#\#\#\#\#\\
\\
\\
\#\#\#\#\#\#\#\#\#\#\#\#\#\#\#\#\#\#\#\#\#\#\#\#\#\#\#\#\#\#\#\#\#\#\#\#\#\#\#\#\#\#\#\#\#\#\#\#\#\#\#\#\#\#\#\#\\
		$<$SECTION-LOCAL\_STATES\_SIGNS$>$\\
\#\#\#\#\#\#\#\#\#\#\#\#\#\#\#\#\#\#\#\#\#\#\#\#\#\#\#\#\#\#\#\#\#\#\#\#\#\#\#\#\#\#\#\#\#\#\#\#\#\#\#\#\#\#\#\#\\
\# These are the signs of the added states.\\ 
\# The number of signs has to be equal with \\
\# the number of 'local\_state\_no'. \\
\# All the signs must be enumerated in one line.\\ 
\# For even electron number, the sign is 1, \\
\# while for odd electron number the sign is -1.\\
\#\#\#\#\#\#\#\#\#\#\#\#\#\#\#\#\#\#\#\#\#\#\#\#\#\#\#\#\#\#\#\#\#\#\#\#\#\#\#\#\#\#\#\#\#\#\#\#\#\#\#\#\#\#\#\#\#\\
\\
1 -1 -1  1\\
\\
\#\#\#\#\#\#\#\#\#\#\#\#\#\#\#\#\#\#\#\#\#\#\#\#\#\#\#\#\#\#\#\#\#\#\#\#\#\#\#\#\#\#\#\#\#\#\#\#\#\#\#\#\#\#\#\#\#\\
		$<$/SECTION-LOCAL\_STATES\_SIGNS$>$\\
\#\#\#\#\#\#\#\#\#\#\#\#\#\#\#\#\#\#\#\#\#\#\#\#\#\#\#\#\#\#\#\#\#\#\#\#\#\#\#\#\#\#\#\#\#\#\#\#\#\#\#\#\#\#\#\#\#\\
\\
\\
\#\#\#\#\#\#\#\#\#\#\#\#\#\#\#\#\#\#\#\#\#\#\#\#\#\#\#\#\#\#\#\#\#\#\#\#\#\#\#\#\#\#\#\#\#\#\#\#\#\#\#\#\#\#\#\#\#\\
\\
		$<$SECTION-BLOCK\_HAMILTONIAN$>$\\
\\
\# The total Hamiltonian will be written as a sum\\
\# of the Hamiltonian terms.\\
$
H_0 = \sum_{\delta}^{coupling\_no} G_\delta H_\delta
$\\
\# There has to be as many $<$HAMILTONIAN\_TERM$>$ items\\
\# as 'coupling\_no' were set in the first section.\\
\# The first tag is the name of the Hamiltonian term,\\ 
\# the second one gives the representation indices and\\
\# at last comes the matrix elements that will be\\
\# read into the sectors.\\
\\
\\
\#   first term in the Hamiltonian \\
\\
		$<$BLOCK\_HAMILTONIAN\_TERM$>$\\
\\
$<$BLOCK\_HAMILTONIAN\_NAME$>$	\\
H\_Kondo\\
$<$/BLOCK\_HAMILTONIAN\_NAME$>$\\
\\
$<$BLOCK\_HAMILTONIAN\_COUPLING$>$\\
\\
\# J - exchange coupling \\
J = 0.5\\
\\
$<$/BLOCK\_HAMILTONIAN\_COUPLING$>$\\
\\
$<$BLOCK\_HAMILTONIAN\_REPRESENTATION\_INDEX$>$\\
\\
0	0\\
\\
$<$/BLOCK\_HAMILTONIAN\_REPRESENTATION\_INDEX$>$\\
\\
$<$BLOCK\_HAMILTONIAN\_MATRIX$>$\\
\\
$
\begin{array}{llllllll}
0	&0	&0	&0	&0	&0	&0	&0\\	
0	&0	&0	&0	&0	&0	&0	&0\\	
0	&0	&0.25	&0	&0	&0	&0	&0\\	
0	&0	&0	&-0.75	&0	&0	&0	&0\\	
0	&0	&0	&0	&0.25	&0	&0	&0\\	
0	&0	&0	&0	&0	&0.25	&0	&0\\
0	&0	&0	&0	&0	&0	&0	&0\\	
0	&0	&0	&0	&0	&0	&0	&0
\end{array}	
$\\
\\
$<$/BLOCK\_HAMILTONIAN\_MATRIX$>$\\
\\
		$<$/BLOCK\_HAMILTONIAN\_TERM$>$\\
\\
\\
\#   second term in the Hamiltonian\\
\\
		$<$BLOCK\_HAMILTONIAN\_TERM$>$\\
\\
$<$BLOCK\_HAMILTONIAN\_NAME$>$	\\
H\_Zeeman\\
$<$/BLOCK\_HAMILTONIAN\_NAME$>$\\
\\
$<$BLOCK\_HAMILTONIAN\_COUPLING$>$\\
\\
\# B - external magnetic field\\
B = 0.01\\
\\
$<$/BLOCK\_HAMILTONIAN\_COUPLING$>$\\
\\
$<$BLOCK\_HAMILTONIAN\_REPRESENTATION\_INDEX$>$\\
\\
0	0\\
\\
$<$/BLOCK\_HAMILTONIAN\_REPRESENTATION\_INDEX$>$\\
\\
$<$BLOCK\_HAMILTONIAN\_MATRIX$>$\\
\\
$
\begin{array}{llllllll}
-0.5	&0	&0	&0	&0	&0	&0	&0\\
0	&0.5	&0	&0	&0	&0	&0	&0\\
0	&0	&-0.5	&0	&0	&0	&0	&0\\
0	&0	&0	&0	&0.5	&0	&0	&0\\
0	&0	&0	&0.5	&0	&0	&0	&0\\
0	&0	&0	&0	&0	&0.5	&0	&0\\
0	&0	&0	&0	&0	&0	&-0.5	&0\\
0	&0	&0	&0	&0	&0	&0	&0.5\\
\end{array}
$\\
\\
$<$/BLOCK\_HAMILTONIAN\_MATRIX$>$\\
\\
		$<$/BLOCK\_HAMILTONIAN\_TERM$>$\\
\\
\\
\#\#\#\#\#\#\#\#\#\#\#\#\#\#\#\#\#\#\#\#\#\#\#\#\#\#\#\#\#\#\#\#\#\#\#\#\#\#\#\#\#\#\#\#\#\#\#\#\#\#\#\#\#\#\#\#\\
		$<$/SECTION-BLOCK\_HAMILTONIAN$>$\\
\#\#\#\#\#\#\#\#\#\#\#\#\#\#\#\#\#\#\#\#\#\#\#\#\#\#\#\#\#\#\#\#\#\#\#\#\#\#\#\#\#\#\#\#\#\#\#\#\#\#\#\#\#\#\#\#\\
\\
\#\#\#\#\#\#\#\#\#\#\#\#\#\#\#\#\#\#\#\#\#\#\#\#\#\#\#\#\#\#\#\#\#\#\#\#\#\#\#\#\#\#\#\#\#\#\#\#\#\#\#\#\#\#\#\#\\
\\

		$<$SECTION-HOPPING\_OPERATORS$>$\\
\\
\# There must be as many $<$HOPPING\_OPERATOR$>$ items \\
\# as 'hopping\_operator\_no' were set in the first section\\
\# The matrices must contain the Reduced Matrix Elements.
\\
\#\#\#\#\#\#\#\#\#\#\#\#\#\#\#\#\#\#\#\#\#\#\#\#\#\#\#\#\#\#\#\#\#\#\#\#\#\#\#\#\#\#\#\#\#\#\#\#\#\#\#\#\#\#\#\#\\
\\
\# here is the first hopping operator\\
\\
		$<$HOPPING\_OPERATOR$>$\\
\\
$<$HOPPING\_OPERATOR\_NAME$>$\\
\\
f\_N\_dagger\_up   \\
\\
$<$/HOPPING\_OPERATOR\_NAME$>$\\
\\
$<$HOPPING\_OPERATOR\_REPRESENTATION\_INDEX$>$\\
\\
1 1\\
\\
$<$/HOPPING\_OPERATOR\_REPRESENTATION\_INDEX$>$\\
\\
$<$HOPPING\_OPERATOR\_SIGN$>$\\
\\
\# The sign is 1 for bosonic operators and -1 for fermionic operators.\\
-1\\
\\
$<$/HOPPING\_OPERATOR\_SIGN$>$\\
\\
$<$HOPPING\_OPERATOR\_MATRIX$>$\\
\\
$
\begin{array}{llllllll}
0		&0	&0	&0		&0		&0	&0	&0\\
0		&0	&0	&0		&0		&0	&0	&0\\
0		&0	&0	&0		&0		&0	&0	&0\\
-0.707106781186	&0	&0	&0		&0		&0	&0	&0\\
0.707106781186	&0	&0	&0		&0		&0	&0	&0\\
0		&1	&0	&0		&0		&0	&0	&0\\
0		&0	&1	&0		&0		&0	&0	&0\\	
0		&0	&0	&0.707106781186	&0.707106781186	&0	&0	&0
\end{array}
$\\
\\

$<$/HOPPING\_OPERATOR\_MATRIX$>$\\
\\
\\		$<$/HOPPING\_OPERATOR$>$\\
\\
\# here is the second hopping operator\\
\\
		$<$HOPPING\_OPERATOR$>$\\
\\
$<$HOPPING\_OPERATOR\_NAME$>$\\
\\
f\_N\_dagger\_down   \\
\\
$<$/HOPPING\_OPERATOR\_NAME$>$\\
\\
$<$HOPPING\_OPERATOR\_REPRESENTATION\_INDEX$>$\\
\\
1 -1\\
\\
$<$/HOPPING\_OPERATOR\_REPRESENTATION\_INDEX$>$\\
\\
$<$HOPPING\_OPERATOR\_SIGN$>$\\
\\
\# The sign is 1 for bosonic operators and -1 for fermionic operators.\\
-1\\
\\
$<$/HOPPING\_OPERATOR\_SIGN$>$\\
\\
$<$HOPPING\_OPERATOR\_MATRIX$>$\\
\\
$
\begin{array}{llllllll}
0	&0		&0	&0		&0		&0	&0	&0\\
0	&0		&0	&0		&0		&0	&0	&0\\
1	&0		&0	&0		&0		&0	&0	&0\\
0	&0.707106781186	&0	&0		&0		&0	&0	&0\\
0	&0.707106781186	&0	&0		&0		&0	&0	&0\\
0	&0		&0	&0		&0		&0	&0	&0\\
0	&0		&0	&0.707106781186	&-0.707106781186	&0	&0	&0\\
0	&0		&0	&0		&0		&-1	&0	&0
\end{array}
$\\
\\

$<$/HOPPING\_OPERATOR\_MATRIX$>$\\
\\
\\		$<$/HOPPING\_OPERATOR$>$\\
\\
\\
\#\#\#\#\#\#\#\#\#\#\#\#\#\#\#\#\#\#\#\#\#\#\#\#\#\#\#\#\#\#\#\#\#\#\#\#\#\#\#\#\#\#\#\#\#\#\#\#\#\#\#\#\#\#\#\#\\
		$<$/SECTION-HOPPING\_OPERATORS$>$\\
\#\#\#\#\#\#\#\#\#\#\#\#\#\#\#\#\#\#\#\#\#\#\#\#\#\#\#\#\#\#\#\#\#\#\#\#\#\#\#\#\#\#\#\#\#\#\#\#\#\#\#\#\#\#\#\#\\
\\
\\
\\
\#\#\#\#\#\#\#\#\#\#\#\#\#\#\#\#\#\#\#\#\#\#\#\#\#\#\#\#\#\#\#\#\#\#\#\#\#\#\#\#\#\#\#\#\#\#\#\#\#\#\#\#\#\#\#\#\\
		$<$SECTION-SPECTRAL\_OPERATORS$>$\\
\#\#\#\#\#\#\#\#\#\#\#\#\#\#\#\#\#\#\#\#\#\#\#\#\#\#\#\#\#\#\#\#\#\#\#\#\#\#\#\#\#\#\#\#\#\#\#\#\#\#\#\#\#\#\#\#\\
\# In this section we set up the matrices for the \\
\# operators for which the spectral function is computed\\
\# Matrices must contain the reduced matrix elements.\\
\#\#\#\#\#\#\#\#\#\#\#\#\#\#\#\#\#\#\#\#\#\#\#\#\#\#\#\#\#\#\#\#\#\#\#\#\#\#\#\#\#\#\#\#\#\#\#\#\#\#\#\#\#\#\#\\
\\
\# here comes the first spectral operator.
\\

		$<$SPECTRAL\_OPERATOR$>$\\
\\
$<$SPECTRAL\_OPERATOR\_NAME$>$\\
\\
S\_z\\
\\
$<$/SPECTRAL\_OPERATOR\_NAME$>$\\
\\
$<$SPECTRAL\_OPERATOR\_REPRESENTATION\_INDEX$>$\\
0 0 \\
$<$/SPECTRAL\_OPERATOR\_REPRESENTATION\_INDEX$>$\\
\\
$<$SPECTRAL\_OPERATOR\_SIGN$>$\\
\\
\# The sign is 1 for bosonic operators and -1 for fermionic operators.\\
1\\
\\
$<$/SPECTRAL\_OPERATOR\_SIGN$>$\\
\\
$<$SPECTRAL\_OPERATOR\_MATRIX$>$\\
\\
$
\begin{array}{llllllll}
-0.5	&0	&0	&0	&0	&0	&0	&0\\
0	&0.5	&0	&0	&0	&0	&0	&0\\
0	&0	&-0.5	&0	&0	&0	&0	&0\\
0	&0	&0	&0	&0.5	&0	&0	&0\\
0	&0	&0	&0.5	&0	&0	&0	&0\\
0	&0	&0	&0	&0	&0.5	&0	&0\\
0	&0	&0	&0	&0	&0	&-0.5	&0\\
0	&0	&0	&0	&0	&0	&0	&0.5\\
\end{array}
$\\
\\
$<$/SPECTRAL\_OPERATOR\_MATRIX$>$\\
\\
		$<$/SPECTRAL\_OPERATOR$>$\\
\\
\# here comes the second spectral operator.
\\
		$<$SPECTRAL\_OPERATOR$>$\\
\\
$<$SPECTRAL\_OPERATOR\_NAME$>$\\
\\
f\_0\_dagger\_up\\
\\
$<$/SPECTRAL\_OPERATOR\_NAME$>$\\
\\
$<$SPECTRAL\_OPERATOR\_REPRESENTATION\_INDEX$>$\\
1 1 \\
$<$/SPECTRAL\_OPERATOR\_REPRESENTATION\_INDEX$>$\\
\\
$<$SPECTRAL\_OPERATOR\_SIGN$>$\\
\\
\# The sign is 1 for bosonic operators and -1 for fermionic operators.\\
-1\\
\\
$<$/SPECTRAL\_OPERATOR\_SIGN$>$\\
\\
$<$SPECTRAL\_OPERATOR\_MATRIX$>$\\
\\
$
\begin{array}{llllllll}
0		&0	&0	&0		&0		&0	&0	&0\\
0		&0	&0	&0		&0		&0	&0	&0\\
0		&0	&0	&0		&0		&0	&0	&0\\
-0.707106781186	&0	&0	&0		&0		&0	&0	&0\\
0.707106781186	&0	&0	&0		&0		&0	&0	&0\\
0		&1	&0	&0		&0		&0	&0	&0\\
0		&0	&1	&0		&0		&0	&0	&0\\
0		&0	&0	&0.707106781186	&0.707106781186	&0	&0	&0\\
\end{array}
$\\
\\
$<$/SPECTRAL\_OPERATOR\_MATRIX$>$\\
\\
		$<$/SPECTRAL\_OPERATOR$>$\\
\\
\# here comes the third spectral operator.
\\

		$<$SPECTRAL\_OPERATOR$>$\\
\\
$<$SPECTRAL\_OPERATOR\_NAME$>$\\
\\
f\_0\_dagger\_down\\
\\
$<$/SPECTRAL\_OPERATOR\_NAME$>$\\
\\
$<$SPECTRAL\_OPERATOR\_REPRESENTATION\_INDEX$>$\\
\\
1 -1\\
\\
$<$/SPECTRAL\_OPERATOR\_REPRESENTATION\_INDEX$>$\\
\\
$<$SPECTRAL\_OPERATOR\_SIGN$>$\\
\\
\# The sign is 1 for bosonic operators and -1 for fermionic operators.\\
-1\\
\\
$<$/SPECTRAL\_OPERATOR\_SIGN$>$\\
\\
$<$SPECTRAL\_OPERATOR\_MATRIX$>$\\
\\
$
\begin{array}{llllllll}
0	&0		&0	&0		&0		&0	&0	&0\\
0	&0		&0	&0		&0		&0	&0	&0\\
1	&0		&0	&0		&0		&0	&0	&0\\
0	&0.707106781186	&0	&0		&0		&0	&0	&0\\
0	&0.707106781186	&0	&0		&0		&0	&0	&0\\
0	&0		&0	&0		&0		&0	&0	&0\\
0	&0		&0	&0.707106781186	&-0.707106781186	&0	&0	&0\\
0	&0		&0	&0		&0		&-1	&0	&0\\
\end{array}
$\\
\\
$<$/SPECTRAL\_OPERATOR\_MATRIX$>$\\
\\
		$<$/SPECTRAL\_OPERATOR$>$\\
\\
\#\#\#\#\#\#\#\#\#\#\#\#\#\#\#\#\#\#\#\#\#\#\#\#\#\#\#\#\#\#\#\#\#\#\#\#\#\#\#\#\#\#\#\#\#\#\#\#\#\#\#\#\#\#\#\#\\
		$<$/SECTION-SPECTRAL\_OPERATORS$>$\\
\#\#\#\#\#\#\#\#\#\#\#\#\#\#\#\#\#\#\#\#\#\#\#\#\#\#\#\#\#\#\#\#\#\#\#\#\#\#\#\#\#\#\#\#\#\#\#\#\#\#\#\#\#\#\#\#\\
\\
\\
\#\#\#\#\#\#\#\#\#\#\#\#\#\#\#\#\#\#\#\#\#\#\#\#\#\#\#\#\#\#\#\#\#\#\#\#\#\#\#\#\#\#\#\#\#\#\#\#\#\#\#\#\#\#\#\#\\
		$<$SECTION-STATIC\_OPERATORS$>$\\
\#\#\#\#\#\#\#\#\#\#\#\#\#\#\#\#\#\#\#\#\#\#\#\#\#\#\#\#\#\#\#\#\#\#\#\#\#\#\#\#\#\#\#\#\#\#\#\#\#\#\#\#\#\#\#\#\\
\# In this section we set up the matrices for the \\
\# operators for which the static values are computed\\
\# These operators need to have representation \\
\# indices always zero!\\
\#\#\#\#\#\#\#\#\#\#\#\#\#\#\#\#\#\#\#\#\#\#\#\#\#\#\#\#\#\#\#\#\#\#\#\#\#\#\#\#\#\#\#\#\#\#\#\#\#\#\#\#\#\#\#\\
\\
		$<$STATIC\_OPERATOR$>$\\
\\
$<$STATIC\_OPERATOR\_NAME$>$\\
\\
S\_z\\
\\
$<$/STATIC\_OPERATOR\_NAME$>$\\
\\
$<$STATIC\_OPERATOR\_REPRESENTATION\_INDEX$>$\\
0  0\\
$<$/STATIC\_OPERATOR\_REPRESENTATION\_INDEX$>$\\
\\
$<$STATIC\_OPERATOR\_SIGN$>$\\
\\
\# The sign is 1 for bosonic operators and -1 for fermionic operators. \\
1\\
\\
$<$/STATIC\_OPERATOR\_SIGN$>$\\
\\
$<$STATIC\_OPERATOR\_MATRIX$>$\\
\\
$
\begin{array}{llllllll}
-0.5	&0	&0	&0	&0	&0	&0	&0\\
0	&0.5	&0	&0	&0	&0	&0	&0\\
0	&0	&-0.5	&0	&0	&0	&0	&0\\
0	&0	&0	&0	&0.5	&0	&0	&0\\
0	&0	&0	&0.5	&0	&0	&0	&0\\
0	&0	&0	&0	&0	&0.5	&0	&0\\
0	&0	&0	&0	&0	&0	&-0.5	&0\\
0	&0	&0	&0	&0	&0	&0	&0.5\\
\end{array}
$\\
\\
$<$/STATIC\_OPERATOR\_MATRIX$>$\\
\\
		$<$/STATIC\_OPERATOR$>$\\
\\
\#\#\#\#\#\#\#\#\#\#\#\#\#\#\#\#\#\#\#\#\#\#\#\#\#\#\#\#\#\#\#\#\#\#\#\#\#\#\#\#\#\#\#\#\#\#\#\#\#\#\#\#\#\#\#\#\\
		$<$/SECTION-STATIC\_OPERATORS$>$\\
\#\#\#\#\#\#\#\#\#\#\#\#\#\#\#\#\#\#\#\#\#\#\#\#\#\#\#\#\#\#\#\#\#\#\#\#\#\#\#\#\#\#\#\#\#\#\#\#\#\#\#\#\#\#\#\#\\
\\
\\
\#\#\#\#\#\#\#\#\#\#\#\#\#\#\#\#\#\#\#\#\#\#\#\#\#\#\#\#\#\#\#\#\#\#\#\#\#\#\#\#\#\#\#\#\#\#\#\#\#\#\#\#\#\#\#\#\\
		$<$SECTION-LOCAL\_HOPPING\_OPERATORS$>$\\
\#\#\#\#\#\#\#\#\#\#\#\#\#\#\#\#\#\#\#\#\#\#\#\#\#\#\#\#\#\#\#\#\#\#\#\#\#\#\#\#\#\#\#\#\#\#\#\#\#\#\#\#\#\#\#\#\\
\\
		$<$LOCAL\_HOPPING\_OPERATOR$>$\\
\\
\# here comes the first local hopping operator
\\
$<$LOCAL\_HOPPING\_OPERATOR\_NAME$>$\\
\\
f\_N+1\_dagger\_up\\
\\
$<$/LOCAL\_HOPPING\_OPERATOR\_NAME$>$\\
\\
$<$LOCAL\_HOPPING\_OPERATOR\_REPRESENTATION\_INDEX$>$\\
\\
1 1\\
\\
$<$/LOCAL\_HOPPING\_OPERATOR\_REPRESENTATION\_INDEX$>$\\
\\
$<$LOCAL\_HOPPING\_OPERATOR\_SIGN$>$\\
\\
\# The sign is 1 for bosonic operators and -1 for fermionic operators.\\
-1\\
\\
$<$/LOCAL\_HOPPING\_OPERATOR\_SIGN$>$\\
\\
$<$LOCAL\_HOPPING\_OPERATOR\_MATRIX$>$\\
\\
$
\begin{array}{llll}
0	&0	&0	&0\\
0	&0	&0	&0\\
1	&0	&0	&0\\
0	&1	&0	&0\\
\end{array}
$\\
\\
$<$/LOCAL\_HOPPING\_OPERATOR\_MATRIX$>$\\
\\
$<$LOCAL\_ON\_SITE\_ENERGY\_MATRIX$>$\\
\\
$
\begin{array}{llll}
0       &0       &0       &0\\
0       &0       &0       &0\\
0       &0       &1       &0\\
0       &0       &0       &1\\
\end{array}
$\\
\\
$<$/LOCAL\_ON\_SITE\_ENERGY\_MATRIX$>$\\
\\
		$<$/LOCAL\_HOPPING\_OPERATOR$>$\\
\\
\# here comes the second local hopping operator\\
\\
		$<$LOCAL\_HOPPING\_OPERATOR$>$\\
\\
$<$LOCAL\_HOPPING\_OPERATOR\_NAME$>$\\
\\
f\_N+1\_dagger\_down\\
\\
$<$/LOCAL\_HOPPING\_OPERATOR\_NAME$>$\\
\\
$<$LOCAL\_HOPPING\_OPERATOR\_REPRESENTATION\_INDEX$>$\\
\\
1 -1\\
\\
$<$/LOCAL\_HOPPING\_OPERATOR\_REPRESENTATION\_INDEX$>$\\
\\
$<$LOCAL\_HOPPING\_OPERATOR\_SIGN$>$\\
\\
\# The sign is 1 for bosonic operators and -1 for fermionic operators.\\
-1\\
\\
$<$/LOCAL\_HOPPING\_OPERATOR\_SIGN$>$\\
\\
$<$LOCAL\_HOPPING\_OPERATOR\_MATRIX$>$\\
\\
$
\begin{array}{llll}
0	&0	&0	&0\\
1	&0	&0	&0\\
0	&0	&0	&0\\
0	&0	&-1	&0\\
\end{array}
$\\
\\
$<$/LOCAL\_HOPPING\_OPERATOR\_MATRIX$>$\\
\\
$<$LOCAL\_ON\_SITE\_ENERGY\_MATRIX$>$\\
\\
$
\begin{array}{llll}
0       &0       &0       &0\\
0       &1       &0       &0\\
0       &0       &0       &0\\
0       &0       &0       &1\\
\end{array}
$\\
\\
$<$/LOCAL\_ON\_SITE\_ENERGY\_MATRIX$>$\\
\\
		$<$/LOCAL\_HOPPING\_OPERATOR$>$\\
\\
\\
\#\#\#\#\#\#\#\#\#\#\#\#\#\#\#\#\#\#\#\#\#\#\#\#\#\#\#\#\#\#\#\#\#\#\#\#\#\#\#\#\#\#\#\#\#\#\#\#\#\#\#\#\#\#\#\#\\
		$<$/SECTION-LOCAL\_HOPPING\_OPERATORS$>$\\
\#\#\#\#\#\#\#\#\#\#\#\#\#\#\#\#\#\#\#\#\#\#\#\#\#\#\#\#\#\#\#\#\#\#\#\#\#\#\#\#\#\#\#\#\#\#\#\#\#\#\#\#\#\#\#\#\\
\\
\#\#\#\#\#\#\#\#\#\#\#\#\#\#\#\#\#\#\#\#\#\#\#\#\#\#\#\#\#\#\#\#\#\#\#\#\#\#\#\#\#\#\#\#\#\#\#\#\#\#\#\#\#\#\#\#\\
	
		$<$SECTION-LOCAL\_HAMILTONIAN$>$

\# The total Local Hamiltonian of the added site \\
\# will be written as a sum of Local Hamiltonian terms\\
\# acting on site n:\\
$
H_n = \sum_{i}g_{i,n} O_{i,n}
$
\\
\# The number of  $<$LOCAL\_HAMILTONIAN\_TERM$>$ items\\
\# must be equal with the variable 'local\_coupling\_no'\\
\# set in the first section\\
\\
\# The first tag is the name of the local Hamiltonian term\\
\# the second one gives the representation indices\\
\#  last come the matrix elements.\\
\\
\#\#\#\#\#\#\#\#\#\#\#\#\#\#\#\#\#\#\#\#\#\#\#\#\#\#\#\#\#\#\#\#\#\#\#\#\#\#\#\#\#\#\#\#\#\#\#\#\#\#\#\#\#\#\#\#\\
\#   first term in the Local\_Hamiltonian\\
\#\#\#\#\#\#\#\#\#\#\#\#\#\#\#\#\#\#\#\#\#\#\#\#\#\#\#\#\#\#\#\#\#\#\#\#\#\#\#\#\#\#\#\#\#\#\#\#\#\#\#\#\#\#\#\\
\\
		$<$LOCAL\_HAMILTONIAN\_TERM$>$\\
\\
$<$LOCAL\_HAMILTONIAN\_NAME$>$		\\
H\_1\\
$<$/LOCAL\_HAMILTONIAN\_NAME$>$\\
\\
$<$LOCAL\_HAMILTONIAN\_COUPLING$>$\\
\\
\#irrelevant \\
0.0\\
\\
$<$/LOCAL\_HAMILTONIAN\_COUPLING$>$\\
\\
$<$LOCAL\_HAMILTONIAN\_REPRESENTATION\_INDEX$>$\\
\\
0	0\\
\\
$<$/LOCAL\_HAMILTONIAN\_REPRESENTATION\_INDEX$>$\\
\\
$<$LOCAL\_HAMILTONIAN\_MATRIX$>$\\
\\
$
\begin{array}{llll}
1       &0       &0       &0\\
0       &0       &0       &0\\
0       &0       &1       &0\\
0       &0       &0       &0\\
\end{array}
$\\
\\
$<$/LOCAL\_HAMILTONIAN\_MATRIX$>$\\
\\
		$<$/LOCAL\_HAMILTONIAN\_TERM$>$\\
\\
$<$/SECTION-LOCAL\_HAMILTONIAN$>$\\
\\
\#\#\#\#\#\#\#\#\#\#\#\#\#\#\#\#\#\#\#\#\#\#\#\#\#\#\#\#\#\#\#\#\#\#\#\#\#\#\#\#\#\#\#\#\#\#\#\#\#\#\#\#\#\#\#\#\\
		$<$SECTION-SPECTRAL\_FUNCTION$>$\\
\#\#\#\#\#\#\#\#\#\#\#\#\#\#\#\#\#\#\#\#\#\#\#\#\#\#\#\#\#\#\#\#\#\#\#\#\#\#\#\#\#\#\#\#\#\#\#\#\#\#\#\#\#\#\#\#\\
\\
\# In this section we set up the spectral function\\
\# calculation for different types of operators.\\
\# All the time we provide in the input file the \\
\# matrices for the $A^\dagger$   and $B^\dagger$  operators \\
\# but the spectral function is computed for the\\
\# combination $<$$A,B^\dagger$$>$\\
\# spectral function for $<$f\_0\_up, f\_0\_dagger\_up$>$ \\
\\
\{f\_0\_dagger\_up; f\_0\_dagger\_up\}\\
\\
\# spectral function for  $<$f\_0\_down, f\_0\_dagger\_down$>$\\
\\
\{f\_0\_dagger\_down; f\_0\_dagger\_down\}\\
\\
\#spectral function for $<$S\_z, S\_z$>$ \\
\\
\{S\_z; S\_z\}\\
\\
\#\#\#\#\#\#\#\#\#\#\#\#\#\#\#\#\#\#\#\#\#\#\#\#\#\#\#\#\#\#\#\#\#\#\#\#\#\#\#\#\#\#\#\#\#\#\#\#\#\#\#\#\#\#\#\#\\
		$<$/SECTION-SPECTRAL\_FUNCTION$>$\\
\#\#\#\#\#\#\#\#\#\#\#\#\#\#\#\#\#\#\#\#\#\#\#\#\#\#\#\#\#\#\#\#\#\#\#\#\#\#\#\#\#\#\#\#\#\#\#\#\#\#\#\#\#\#\#\#\\
\\
\\
\# The section below is not needed by the {\bf fnrg} binary.\\
\# It is needed when doing the broadening of the spectral function\\
\# with the utility {\bf sfb}.\\
\\
\\
\#\#\#\#\#\#\#\#\#\#\#\#\#\#\#\#\#\#\#\#\#\#\#\#\#\#\#\#\#\#\#\#\#\#\#\#\#\#\#\#\#\#\#\#\#\#\#\#\#\#\#\#\#\#\#\#\\
                $<$SECTION-SPECTRAL\_FUNCTION\_BROADENING$>$\\
\#\#\#\#\#\#\#\#\#\#\#\#\#\#\#\#\#\#\#\#\#\#\#\#\#\#\#\#\#\#\#\#\#\#\#\#\#\#\#\#\#\#\#\#\#\#\#\#\#\#\#\#\#\#\#\#\\
\# In this section we include broadening parameters that\\
\# are used for computing the broaden spectral function\\
\# The methods that can be used are either LOG\_GAUSS  or\\ 
\# INTERPOLATIVE\_LOG\_GAUSS. \\
\# These procedures are described in the manual.\\
\\
\# The quantum temperature is relevant only \\
\# for the INTERPOLATIVE\_LOG\_GAUSS method,  \\
\# otherwise only the 'broadening\_parameter' parameter \\
\# setting the width of the log-normal \\
\# distribution function is relevant.\\
\\
\# The quantum temperature must be always lower than T \\
\# for finite temperature calculations. \\
\# A good choice is to set the quatum temperature in the range\\
\# 0.5T - T.\\  
\\
\# The  three parameters spectral\_function\_*\\
\# specify which spectral functions are being calculated. \\
\# The accepted values are YES or NO only. \\
\\
\# The static\_average\_dmnrg flags specifies whether the expectation values \\
\# of the  static operators are computed.\\
\#The accepted values are YES or NO only. \\
\\
\# The green\_function\_dmnrg specifies whether the real part of the\\
\# Green's function is computed from the spectral function data.\\
\\
broadening\_method  = INTERPOLATIVE\_LOG\_GAUSS	\# method used either LOG\_GAUSS or INTERPOLATIVE\_LOG\_GAUSS\\
broadening\_parameter = 0.70 			\# parameter used for the log gaussian distribution\\
broadening\_energy\_minim = 1e-6		\# minimum energy for which the broadening is performed\\
broadening\_energy\_maxim = 4.0			\# maximum energy for which the broadening is performed\\
broadening\_grid\_mesh = 100                     \# grid mesh on a log scale between the minimum and maximum energies\\
quantum\_temperature =  1e-7			\# used for the INTERPOLATIVE\_LOG\_GAUSS method only.	\\
\\
spectral\_function\_nrg\_even = YES		\# spectral function for the even iterations is computed 	\\
spectral\_function\_nrg\_odd  = YES		\# spectral function for the odd iterations part is computed\\
spectral\_function\_dmnrg    = YES		\# spectral function for the dmnrg calculation is computed.\\

static\_average\_dmnrg       = YES              \# average functions similar with the case of spectral function. \\

green\_function\_nrg\_even = YES                \# calculation of the real/imaginary part of the Green's function from the even
part of the nrg  spectral function\\
green\_function\_nrg\_odd = YES                \# calculation of the real/imaginary part of the Green's function from the odd
part of the nrg  spectral function\\
green\_function\_dmnrg	   = NO		        \# calculation of the real/imaginary part of the Green's function from the spectral function\\
green\_function\_grid\_mesh = 400               \# points for the mesh when doing Hilbert transform\\
\\
\#\#\#\#\#\#\#\#\#\#\#\#\#\#\#\#\#\#\#\#\#\#\#\#\#\#\#\#\#\#\#\#\#\#\#\#\#\#\#\#\#\#\#\#\#\#\#\#\#\#\#\#\#\#\#\#\\
                $<$/SECTION-SPECTRAL\_FUNCTION\_BROADENING$>$\\
\#\#\#\#\#\#\#\#\#\#\#\#\#\#\#\#\#\#\#\#\#\#\#\#\#\#\#\#\#\#\#\#\#\#\#\#\#\#\#\#\#\#\#\#\#\#\#\#\#\#\#\#\#\#\#\#\\
\\

}
}

\chapter{Input file for the Kondo model with $U_{{\rm charge}}(1) \times SU_{\rm spin}(2)$ symmetries }
{\small
{\tt

\# The comments must start with the symbol \# .\\
\# Empty lines are ignored.\\
\\
\\
\#\#\#\#\#\#\#\#\#\#\#\#\#\#\#\#\#\#\#\#\#\#\#\#\#\#\#\#\#\#\#\#\#\#\#\#\#\#\#\#\#\#\#\#\#\#\#\#\#\#\#\#\#\#\#\#\#\\
		$<$SECTION-PARAMETERS$>$   \\
\#\#\#\#\#\#\#\#\#\#\#\#\#\#\#\#\#\#\#\#\#\#\#\#\#\#\#\#\#\#\#\#\#\#\#\#\#\#\#\#\#\#\#\#\#\#\#\#\#\#\#\#\#\#\#\#\#\\
\\
\# Explanation of variables\\
\\
\# 'model' represents the type of the model that is \\
\#  used for the calculations. It can be any name. \\
\\
\#  'lambda' represents the value for the logarithmic \\
\#  discretization parameter. Usually it is fixed between 2 and 3. \\
\\
\#  'max\_state\_no' is the number of kept states  \\
\#  after an iteration.\\
\\
\#  'iteration\_no' is the number of iterations performed \\
\#  along the NRG run. Typically it is set to 40 - 70.\\
\\
\#  'symmetry\_no' specifies how many symmetries \\
\#  are used. The symmetries  are characterized \\
\#  in $<$SECTION-SYMMETRIES$>$ \\
\\
\#  'coupling\_no' specifies how many couplings \\
\#  are present in the interaction Hamiltonian.  \\
\\
\#  'local\_coupling\_no' is the number of \\
\#  Hamiltonians at the sites we add to the Wilson chain\\
\\
\#  'spectral\_operator\_no' is the number of spectral operators \\
\#  acting on site 0, we keep track of.\\
\#  These spectral operators are characterized\\
\# in section $<$SECTION-SPECTRAL-OPERATORS$>$\\	
\\
\#  'static\_operator\_no' is the number of static operators \\
\#  acting on site 0, we keep track of.\\
\#  These static operators are characterized\\
\#  in section $<$SECTION-STATIC-OPERATORS$>$\\	
\\
\#  'hopping\_operator\_no' is the number of hopping operators. \\
\#   These hopping operators are characterized\\
\#  in section $<$SECTION-HOPPING-OPERATORS$>$\\	
\\
\#  'block\_state\_no' specifies the number \\
\#  of initial states/multiplets. These states are characterized\\
\#  by the representation indices and are listed\\
\#  in section $<$SECTION-BLOCK\_STATES$>$\\
\\
\\
\#\#\#\#\#\#\#\#\#\#\#\#\#\#\#\#\#\#\#\#\#\#\#\#\#\#\#\#\#\#\#\#\#\#\#\#\#\#\#\#\#\#\#\#\#\#\#\#\#\#\#\#\#\#\#\#\#\\
\\
\#\#\#\#\#\#\#\#\#\#\#\#\#\#\#\#\#\#\#\#\#\#\#\#\#\#\#\#\#\#\#\#\#\#\#\#\#\#\#\#\#\#\#\#\#\#\#\#\#\#\#\#\#\#\#\#\#\\
\\
model = kondo\_model  \# the name of the present model \\
\\
lambda = 2.0   \# discretization parameter\\
\\
max\_state\_no = 100  \# maximum number of kept multiplets\\
\\
iteration\_no = 20  \# allowed number of iterations\\
\\
symmetry\_no = 2  \# the number of symmetries  \\
\\
coupling\_no = 1   \# no of Hamiltonian terms, i.e. {\cal J} in this case\\
\\
spectral\_operator\_no = 2   \# no of operators for which correlation functions are computed \\
\\
static\_operator\_no = 0  \# no of operators for which the static average is computed. \\
\\
hopping\_operator\_no = 1  \# the number of hopping operators\\
\\
block\_state\_no = 4  \# the number of initial block states\\
\\
local\_state\_no = 3   \# the number of local states added in each iteration\\
\\
local\_coupling\_no = 1  \# the number of Hamiltonian terms on the added site\\
\\
spectral\_function\_no = 2 \#  the number of spectral functions that need to be generated \\
\\
interval\_no = 1000 \# mash grid between the maximum and minimum energy at each iteration \\
\\
\\
degeneracy\_threshhold = 1e-6  \# energy thresh-hold for discarding states \\
\\
temperature = 0.0         \# temperature used for the run\\
\\
\\
\#\#\#\#\#\#\#\#\#\#\#\#\#\#\#\#\#\#\#\#\#\#\#\#\#\#\#\#\#\#\#\#\#\#\#\#\#\#\#\#\#\#\#\#\#\#\#\#\#\#\#\#\#\#\#\#\\
		$<$/SECTION-PARAMETERS$>$   \\
\#\#\#\#\#\#\#\#\#\#\#\#\#\#\#\#\#\#\#\#\#\#\#\#\#\#\#\#\#\#\#\#\#\#\#\#\#\#\#\#\#\#\#\#\#\#\#\#\#\#\#\#\#\#\#\#\\
\\
\\
\#\#\#\#\#\#\#\#\#\#\#\#\#\#\#\#\#\#\#\#\#\#\#\#\#\#\#\#\#\#\#\#\#\#\#\#\#\#\#\#\#\#\#\#\#\#\#\#\#\#\#\#\#\#\#\#\\
		$<$SECTION-FLAGS$>$   \\
\#\#\#\#\#\#\#\#\#\#\#\#\#\#\#\#\#\#\#\#\#\#\#\#\#\#\#\#\#\#\#\#\#\#\#\#\#\#\#\#\#\#\#\#\#\#\#\#\#\#\#\#\#\#\#\#\\
\\
\\
\# dmnrg\_flag  controls whether the backward procedure \\
\# has to be done and the full set of eigen-states be \\
\# used for the calculation of the spectral function. \\
\# ON - backwards iteration is done.\\
\# OFF - only the nrg calculation is performed.\\
\\
\# text\_swap\_files\_flag = flag that fixes if the text mode is \\
\# used to save the files. The files will be saved in binary mode anyway. \\
\# ON - the files are saved in the text mode and kept on the disk.\\
\# OFF= the files are not saved in the text mode. \\
\\
\#  binary\_swap\_files\_flag = flag that controls whether the \\
\#  unnecessary binary files will be removed or not not after the calculation is done.\\
\#  ON - the files will be kept on the disk.\\
\#  OFF- the files will be removed from the disk.\\
\\
\# hoppings\_on\_site\_energies\_flag = flag that controls \\
\# the reading of the hoppings and the on-site energies. \\
\# ON - the hoppings and the on site energies \\
\# are read from the files in the  results/mapping/ folder\\
\# These files are generated by the {\bf he} utility.\\
\# OFF - the hoppings are computed on the fly \\
\# assuming a flat density of states (DOS = 0.5) on [-1, 1],\\
\# and the on-site energies are set to zero. 
\\
\\
dmnrg\_flag = ON \\
text\_swap\_files\_flag = OFF \\
binary\_swap\_files\_flag = ON \\
hoppings\_on\_site\_energies\_flag =OFF \\
\\
\\
\#\#\#\#\#\#\#\#\#\#\#\#\#\#\#\#\#\#\#\#\#\#\#\#\#\#\#\#\#\#\#\#\#\#\#\#\#\#\#\#\#\#\#\#\#\#\#\#\#\#\#\#\#\#\#\#\\
		$<$/SECTION-FLAGS$>$   \\
\#\#\#\#\#\#\#\#\#\#\#\#\#\#\#\#\#\#\#\#\#\#\#\#\#\#\#\#\#\#\#\#\#\#\#\#\#\#\#\#\#\#\#\#\#\#\#\#\#\#\#\#\#\#\#\#\\
\\
\\
\#\#\#\#\#\#\#\#\#\#\#\#\#\#\#\#\#\#\#\#\#\#\#\#\#\#\#\#\#\#\#\#\#\#\#\#\#\#\#\#\#\#\#\#\#\#\#\#\#\#\#\#\#\#\#\#\#\\
		$<$SECTION-SYMMETRIES$>$\\
\#\#\#\#\#\#\#\#\#\#\#\#\#\#\#\#\#\#\#\#\#\#\#\#\#\#\#\#\#\#\#\#\#\#\#\#\#\#\#\#\#\#\#\#\#\#\#\#\#\#\#\#\#\#\#\#\#\\
\\
\#  As many symmetries must be defined below  \\
\#  as set by the  variable 'symmetry\_no' \\
\#  in the previous section. \\
\\
\#  The possible symmetry types are the following:\\
\#  U(1), SU(2), Z(2), charge\_SU(2).\\
\#  \\
\#  Next the limits for the representation index \\
\#  are set. All states with smaller/larger\\
\#  representation indices will be cut off\\
\\
\#  For U(1) symmetries the representations are labeled\\
\# by integers (2 S\_z) and for SU(2) by non-negative\\
\# integers (2S).\\
\\
\#  For the Z(2) symmetry the representations are labeled\\
\#  by integers \\
\#  0 - even parity states\\
\#  1 - odd parity states\\
\\
\\
\#\#\#\#\#\#\#\#\#\#\#\#\#\#\#\#\#\#\#\#\#\#\#\#\#\#\#\#\#\#\#\#\#\#\#\#\#\#\#\#\#\#\#\#\#\#\#\#\#\#\#\#\#\#\#\#\# \\
\\
\\
\# charge symmetry corresponding to Q \\
U(1)     -10     10      \\
\\
\# spin symmetry corresponding to S\\
SU(2)     0     20     \\
\\
\\
\#\#\#\#\#\#\#\#\#\#\#\#\#\#\#\#\#\#\#\#\#\#\#\#\#\#\#\#\#\#\#\#\#\#\#\#\#\#\#\#\#\#\#\#\#\#\#\#\#\#\#\#\#\#\#\#\# \\
		$<$/SECTION-SYMMETRIES$>$\\
\#\#\#\#\#\#\#\#\#\#\#\#\#\#\#\#\#\#\#\#\#\#\#\#\#\#\#\#\#\#\#\#\#\#\#\#\#\#\#\#\#\#\#\#\#\#\#\#\#\#\#\#\#\#\#\#\# \\
\\
\\
\#\#\#\#\#\#\#\#\#\#\#\#\#\#\#\#\#\#\#\#\#\#\#\#\#\#\#\#\#\#\#\#\#\#\#\#\#\#\#\#\#\#\#\#\#\#\#\#\#\#\#\#\#\#\#\#\#\\
		$<$SECTION-BLOCK\_STATES$>$\\
\#\#\#\#\#\#\#\#\#\#\#\#\#\#\#\#\#\#\#\#\#\#\#\#\#\#\#\#\#\#\#\#\#\#\#\#\#\#\#\#\#\#\#\#\#\#\#\#\#\#\#\#\#\#\#\#\#\\
\# This section gives the representation indices of the \\
\# block states of the first iteration. \\
\# The number of rows has to be equal \\
\# with the 'block\_state\_no' in the first section, \\
\# and the number of columns to the 'symmetry\_no'.\\
\# Each column contains indices corresponding to a symmetry.\\
\#\#\#\#\#\#\#\#\#\#\#\#\#\#\#\#\#\#\#\#\#\#\#\#\#\#\#\#\#\#\#\#\#\#\#\#\#\#\#\#\#\#\#\#\#\#\#\#\#\#\#\#\#\#\#\#\#\\
$
\begin{array}{ll}
-1&1\\
0&0\\
0&2\\
1&1\\
\end{array}
$\\
\#\#\#\#\#\#\#\#\#\#\#\#\#\#\#\#\#\#\#\#\#\#\#\#\#\#\#\#\#\#\#\#\#\#\#\#\#\#\#\#\#\#\#\#\#\#\#\#\#\#\#\#\#\#\#\#\#\\
		$<$/SECTION-BLOCK\_STATES$>$\\
\#\#\#\#\#\#\#\#\#\#\#\#\#\#\#\#\#\#\#\#\#\#\#\#\#\#\#\#\#\#\#\#\#\#\#\#\#\#\#\#\#\#\#\#\#\#\#\#\#\#\#\#\#\#\#\#\#\\
\\
\\
\#\#\#\#\#\#\#\#\#\#\#\#\#\#\#\#\#\#\#\#\#\#\#\#\#\#\#\#\#\#\#\#\#\#\#\#\#\#\#\#\#\#\#\#\#\#\#\#\#\#\#\#\#\#\#\#\#\\
		$<$SECTION-LOCAL\_STATES$>$\\
\#\#\#\#\#\#\#\#\#\#\#\#\#\#\#\#\#\#\#\#\#\#\#\#\#\#\#\#\#\#\#\#\#\#\#\#\#\#\#\#\#\#\#\#\#\#\#\#\#\#\#\#\#\#\#\#\#\\
\# This section gives  the representation indices \\
\# of the local states. The number of rows has to be equal\\ 
\# with the 'local\_state\_no' in the first section\\
\# and the number of columns to the 'symmetry\_no'.\\
\# Each column contains indices corresponding to a symmetry.\\
\#\#\#\#\#\#\#\#\#\#\#\#\#\#\#\#\#\#\#\#\#\#\#\#\#\#\#\#\#\#\#\#\#\#\#\#\#\#\#\#\#\#\#\#\#\#\#\#\#\#\#\#\#\#\#\#\\
$
\begin{array}{ll}
-1    & 0\\
0     &  1\\
1     &  0\\
\end{array}
$\\
\#\#\#\#\#\#\#\#\#\#\#\#\#\#\#\#\#\#\#\#\#\#\#\#\#\#\#\#\#\#\#\#\#\#\#\#\#\#\#\#\#\#\#\#\#\#\#\#\#\#\#\#\#\#\#\#\\
		$<$/SECTION-LOCAL\_STATES$>$\\
\#\#\#\#\#\#\#\#\#\#\#\#\#\#\#\#\#\#\#\#\#\#\#\#\#\#\#\#\#\#\#\#\#\#\#\#\#\#\#\#\#\#\#\#\#\#\#\#\#\#\#\#\#\#\#\#\\
\\
\\
\#\#\#\#\#\#\#\#\#\#\#\#\#\#\#\#\#\#\#\#\#\#\#\#\#\#\#\#\#\#\#\#\#\#\#\#\#\#\#\#\#\#\#\#\#\#\#\#\#\#\#\#\#\#\#\#\\
		$<$SECTION-LOCAL\_STATES\_SIGNS$>$\\
\#\#\#\#\#\#\#\#\#\#\#\#\#\#\#\#\#\#\#\#\#\#\#\#\#\#\#\#\#\#\#\#\#\#\#\#\#\#\#\#\#\#\#\#\#\#\#\#\#\#\#\#\#\#\#\#\\
\# These are the signs of the added states.\\ 
\# The number of signs has to be equal with \\
\# the number of 'local\_state\_no'. \\
\# All the signs must be enumerated in one line.\\ 
\# For even electron number, the sign is 1, \\
\# while for odd electron number the sign is -1.\\
\#\#\#\#\#\#\#\#\#\#\#\#\#\#\#\#\#\#\#\#\#\#\#\#\#\#\#\#\#\#\#\#\#\#\#\#\#\#\#\#\#\#\#\#\#\#\#\#\#\#\#\#\#\#\#\#\#\\
\\
1 -1  1\\
\\
\#\#\#\#\#\#\#\#\#\#\#\#\#\#\#\#\#\#\#\#\#\#\#\#\#\#\#\#\#\#\#\#\#\#\#\#\#\#\#\#\#\#\#\#\#\#\#\#\#\#\#\#\#\#\#\#\#\\
		$<$/SECTION-LOCAL\_STATES\_SIGNS$>$\\
\#\#\#\#\#\#\#\#\#\#\#\#\#\#\#\#\#\#\#\#\#\#\#\#\#\#\#\#\#\#\#\#\#\#\#\#\#\#\#\#\#\#\#\#\#\#\#\#\#\#\#\#\#\#\#\#\#\\
\\
\\
\#\#\#\#\#\#\#\#\#\#\#\#\#\#\#\#\#\#\#\#\#\#\#\#\#\#\#\#\#\#\#\#\#\#\#\#\#\#\#\#\#\#\#\#\#\#\#\#\#\#\#\#\#\#\#\#\#\\
\\
		$<$SECTION-BLOCK\_HAMILTONIAN$>$\\
\\
\# The total Hamiltonian will be written as a sum\\
\# of the Hamiltonian terms.\\
$
H_0 = \sum_{\delta}^{coupling\_no} G_\delta H_\delta
$\\
\# There has to be as many $<$HAMILTONIAN\_TERM$>$ items\\
\# as 'coupling\_no' were set in the first section.\\
\# The first tag is the name of the Hamiltonian term,\\ 
\# the second one gives the representation indices and\\
\# at last comes the matrix elements that will be\\
\# read into the sectors.\\
\\
\\

\#   first term in the Hamiltonian \\
\\
		$<$BLOCK\_HAMILTONIAN\_TERM$>$\\
\\
$<$BLOCK\_HAMILTONIAN\_NAME$>$	\\
H\_Kondo\\
$<$/BLOCK\_HAMILTONIAN\_NAME$>$\\
\\
$<$BLOCK\_HAMILTONIAN\_COUPLING$>$\\
\\
\# J - exchange coupling \\
J = 0.5\\
\\
$<$/BLOCK\_HAMILTONIAN\_COUPLING$>$\\
\\
$<$BLOCK\_HAMILTONIAN\_REPRESENTATION\_INDEX$>$\\
\\
0	0\\
\\
$<$/BLOCK\_HAMILTONIAN\_REPRESENTATION\_INDEX$>$\\
\\
$<$BLOCK\_HAMILTONIAN\_MATRIX$>$\\
\\
$
\begin{array}{llll}
0	&0	&0	&0\\
0	&-0.75  &0      &0\\
0	&0      &0.25   &0\\
0	&0	&0      &0\\
\end{array}	
$\\
\\
$<$/BLOCK\_HAMILTONIAN\_MATRIX$>$\\
\\
		$<$/BLOCK\_HAMILTONIAN\_TERM$>$\\
\\
\\
\#\#\#\#\#\#\#\#\#\#\#\#\#\#\#\#\#\#\#\#\#\#\#\#\#\#\#\#\#\#\#\#\#\#\#\#\#\#\#\#\#\#\#\#\#\#\#\#\#\#\#\#\#\#\#\#\\
		$<$/SECTION-BLOCK\_HAMILTONIAN$>$\\
\#\#\#\#\#\#\#\#\#\#\#\#\#\#\#\#\#\#\#\#\#\#\#\#\#\#\#\#\#\#\#\#\#\#\#\#\#\#\#\#\#\#\#\#\#\#\#\#\#\#\#\#\#\#\#\#\\
\\
\#\#\#\#\#\#\#\#\#\#\#\#\#\#\#\#\#\#\#\#\#\#\#\#\#\#\#\#\#\#\#\#\#\#\#\#\#\#\#\#\#\#\#\#\#\#\#\#\#\#\#\#\#\#\#\#\\
\\

		$<$SECTION-HOPPING\_OPERATORS$>$\\
\\
\# There must be as many $<$HOPPING\_OPERATOR$>$ items \\
\# as 'hopping\_operator\_no' were set in the first section\\
\# The matrices must contain the Reduced Matrix Elements.
\\
\#\#\#\#\#\#\#\#\#\#\#\#\#\#\#\#\#\#\#\#\#\#\#\#\#\#\#\#\#\#\#\#\#\#\#\#\#\#\#\#\#\#\#\#\#\#\#\#\#\#\#\#\#\#\#\#\\
\\
\# here is the first hopping operator\\
\\
		$<$HOPPING\_OPERATOR$>$\\
\\
$<$HOPPING\_OPERATOR\_NAME$>$\\
\\
f\_N\_dagger   \\
\\
$<$/HOPPING\_OPERATOR\_NAME$>$\\
\\
$<$HOPPING\_OPERATOR\_REPRESENTATION\_INDEX$>$\\
\\
1 1\\
\\
$<$/HOPPING\_OPERATOR\_REPRESENTATION\_INDEX$>$\\
\\
$<$HOPPING\_OPERATOR\_SIGN$>$\\
\\
\# The sign is 1 for bosonic operators and -1 for fermionic operators.\\
-1\\
\\
$<$/HOPPING\_OPERATOR\_SIGN$>$\\
\\
$<$HOPPING\_OPERATOR\_MATRIX$>$\\
\\
$
\begin{array}{llll}
0 &0              &0                &0\\
1 &0              &0                &0\\
1 &0              &0                &0\\
0 &0.707106781186 &-1.224744871392  &0\\
\end{array}
$\\
\\
$<$/HOPPING\_OPERATOR\_MATRIX$>$\\
\\
\\		$<$/HOPPING\_OPERATOR$>$\\
\\
\#\#\#\#\#\#\#\#\#\#\#\#\#\#\#\#\#\#\#\#\#\#\#\#\#\#\#\#\#\#\#\#\#\#\#\#\#\#\#\#\#\#\#\#\#\#\#\#\#\#\#\#\#\#\#\#\\
		$<$/SECTION-HOPPING\_OPERATORS$>$\\
\#\#\#\#\#\#\#\#\#\#\#\#\#\#\#\#\#\#\#\#\#\#\#\#\#\#\#\#\#\#\#\#\#\#\#\#\#\#\#\#\#\#\#\#\#\#\#\#\#\#\#\#\#\#\#\#\\
\\
\\
\\
\#\#\#\#\#\#\#\#\#\#\#\#\#\#\#\#\#\#\#\#\#\#\#\#\#\#\#\#\#\#\#\#\#\#\#\#\#\#\#\#\#\#\#\#\#\#\#\#\#\#\#\#\#\#\#\#\\
		$<$SECTION-SPECTRAL\_OPERATORS$>$\\
\#\#\#\#\#\#\#\#\#\#\#\#\#\#\#\#\#\#\#\#\#\#\#\#\#\#\#\#\#\#\#\#\#\#\#\#\#\#\#\#\#\#\#\#\#\#\#\#\#\#\#\#\#\#\#\#\\
\# In this section we set up the matrices for the \\
\# operators for which the spectral function is computed\\
\# Matrices must contain the reduced matrix elements.\\
\#\#\#\#\#\#\#\#\#\#\#\#\#\#\#\#\#\#\#\#\#\#\#\#\#\#\#\#\#\#\#\#\#\#\#\#\#\#\#\#\#\#\#\#\#\#\#\#\#\#\#\#\#\#\#\\
\\
\# here comes the first spectral operator.
\\

		$<$SPECTRAL\_OPERATOR$>$\\
\\
$<$SPECTRAL\_OPERATOR\_NAME$>$\\
\\
S\\
\\
$<$/SPECTRAL\_OPERATOR\_NAME$>$\\
\\
$<$SPECTRAL\_OPERATOR\_REPRESENTATION\_INDEX$>$\\
0 2 \\
$<$/SPECTRAL\_OPERATOR\_REPRESENTATION\_INDEX$>$\\
\\
$<$SPECTRAL\_OPERATOR\_SIGN$>$\\
\\
\# The sign is 1 for bosonic operators and -1 for fermionic operators.\\
1\\
\\
$<$/SPECTRAL\_OPERATOR\_SIGN$>$\\
\\
$<$SPECTRAL\_OPERATOR\_MATRIX$>$\\
\\
$
\begin{array}{llll}
0.866025403784   	&0       	&0 	                &0\\
0               	&0   	        &-0.866025403784  	&0\\
0               	&0.5      	&0.707106781186   	&0\\
0   	     	        &0  	        &0	                &0.866025403784\\
\end{array}
$\\
\\
$<$/SPECTRAL\_OPERATOR\_MATRIX$>$\\
\\
		$<$/SPECTRAL\_OPERATOR$>$\\
\\
\# here comes the second spectral operator.
\\
		$<$SPECTRAL\_OPERATOR$>$\\
\\
$<$SPECTRAL\_OPERATOR\_NAME$>$\\
\\
f\_0\_dagger\\
\\
$<$/SPECTRAL\_OPERATOR\_NAME$>$\\
\\
$<$SPECTRAL\_OPERATOR\_REPRESENTATION\_INDEX$>$\\
1 1 \\
$<$/SPECTRAL\_OPERATOR\_REPRESENTATION\_INDEX$>$\\
\\
$<$SPECTRAL\_OPERATOR\_SIGN$>$\\
\\
\# The sign is 1 for bosonic operators and -1 for fermionic operators.\\
-1\\
\\
$<$/SPECTRAL\_OPERATOR\_SIGN$>$\\
\\
$<$SPECTRAL\_OPERATOR\_MATRIX$>$\\
\\
$
\begin{array}{llll}
0 &0              &0                &0\\
1 &0              &0                &0\\
1 &0              &0                &0\\
0 &0.707106781186 &-1.224744871392  &0\\
\end{array}
$\\
\\
$<$/SPECTRAL\_OPERATOR\_MATRIX$>$\\
\\
		$<$/SPECTRAL\_OPERATOR$>$\\
\\
\\
\#\#\#\#\#\#\#\#\#\#\#\#\#\#\#\#\#\#\#\#\#\#\#\#\#\#\#\#\#\#\#\#\#\#\#\#\#\#\#\#\#\#\#\#\#\#\#\#\#\#\#\#\#\#\#\#\\
		$<$/SECTION-SPECTRAL\_OPERATORS$>$\\
\#\#\#\#\#\#\#\#\#\#\#\#\#\#\#\#\#\#\#\#\#\#\#\#\#\#\#\#\#\#\#\#\#\#\#\#\#\#\#\#\#\#\#\#\#\#\#\#\#\#\#\#\#\#\#\#\\
\\
\\
\#\#\#\#\#\#\#\#\#\#\#\#\#\#\#\#\#\#\#\#\#\#\#\#\#\#\#\#\#\#\#\#\#\#\#\#\#\#\#\#\#\#\#\#\#\#\#\#\#\#\#\#\#\#\#\#\\
		$<$SECTION-LOCAL\_HOPPING\_OPERATORS$>$\\
\#\#\#\#\#\#\#\#\#\#\#\#\#\#\#\#\#\#\#\#\#\#\#\#\#\#\#\#\#\#\#\#\#\#\#\#\#\#\#\#\#\#\#\#\#\#\#\#\#\#\#\#\#\#\#\#\\
\\
		$<$LOCAL\_HOPPING\_OPERATOR$>$\\
\\
\# here comes the first local hopping operator
\\
$<$LOCAL\_HOPPING\_OPERATOR\_NAME$>$\\
\\
f\_N+1\_dagger\\
\\
$<$/LOCAL\_HOPPING\_OPERATOR\_NAME$>$\\
\\
$<$LOCAL\_HOPPING\_OPERATOR\_REPRESENTATION\_INDEX$>$\\
\\
1 1\\
\\
$<$/LOCAL\_HOPPING\_OPERATOR\_REPRESENTATION\_INDEX$>$\\
\\
$<$LOCAL\_HOPPING\_OPERATOR\_SIGN$>$\\
\\
\# The sign is 1 for bosonic operators and -1 for fermionic operators.\\
-1\\
\\
$<$/LOCAL\_HOPPING\_OPERATOR\_SIGN$>$\\
\\
$<$LOCAL\_HOPPING\_OPERATOR\_MATRIX$>$\\
\\
$
\begin{array}{lll}
0       &0	                &0\\
1       &0                      &0\\
0       &-1.4142135623731       &0\\
\end{array}
$\\
\\
$<$/LOCAL\_HOPPING\_OPERATOR\_MATRIX$>$\\
\\
$<$LOCAL\_ON\_SITE\_ENERGY\_MATRIX$>$\\
\\
$
\begin{array}{llll}
0       &0       &0 \\
0       &0       &0 \\
0       &0       &0 \\
\end{array}
$\\
\\
$<$/LOCAL\_ON\_SITE\_ENERGY\_MATRIX$>$\\
\\
		$<$/LOCAL\_HOPPING\_OPERATOR$>$\\
\\
\\
\\
\#\#\#\#\#\#\#\#\#\#\#\#\#\#\#\#\#\#\#\#\#\#\#\#\#\#\#\#\#\#\#\#\#\#\#\#\#\#\#\#\#\#\#\#\#\#\#\#\#\#\#\#\#\#\#\#\\
		$<$/SECTION-LOCAL\_HOPPING\_OPERATORS$>$\\
\#\#\#\#\#\#\#\#\#\#\#\#\#\#\#\#\#\#\#\#\#\#\#\#\#\#\#\#\#\#\#\#\#\#\#\#\#\#\#\#\#\#\#\#\#\#\#\#\#\#\#\#\#\#\#\#\\
\\
\#\#\#\#\#\#\#\#\#\#\#\#\#\#\#\#\#\#\#\#\#\#\#\#\#\#\#\#\#\#\#\#\#\#\#\#\#\#\#\#\#\#\#\#\#\#\#\#\#\#\#\#\#\#\#\#\\
	
		$<$SECTION-LOCAL\_HAMILTONIAN$>$

\# The total Local Hamiltonian of the added site \\
\# will be written as a sum of Local Hamiltonian terms\\
\# acting on site n:\\
$
H_n = \sum_{i}g_{i,n} O_{i,n}
$
\\
\# The number of  $<$LOCAL\_HAMILTONIAN\_TERM$>$ items\\
\# must be equal with the variable 'local\_coupling\_no'\\
\# set in the first section\\
\\
\# The first tag is the name of the local Hamiltonian term\\
\# the second one gives the representation indices\\
\#  last come the matrix elements.\\
\\
\#\#\#\#\#\#\#\#\#\#\#\#\#\#\#\#\#\#\#\#\#\#\#\#\#\#\#\#\#\#\#\#\#\#\#\#\#\#\#\#\#\#\#\#\#\#\#\#\#\#\#\#\#\#\#\#\\
\#   first term in the Local\_Hamiltonian\\
\#\#\#\#\#\#\#\#\#\#\#\#\#\#\#\#\#\#\#\#\#\#\#\#\#\#\#\#\#\#\#\#\#\#\#\#\#\#\#\#\#\#\#\#\#\#\#\#\#\#\#\#\#\#\#\\
\\
		$<$LOCAL\_HAMILTONIAN\_TERM$>$\\
\\
$<$LOCAL\_HAMILTONIAN\_NAME$>$		\\
H\_1\\
$<$/LOCAL\_HAMILTONIAN\_NAME$>$\\
\\
$<$LOCAL\_HAMILTONIAN\_COUPLING$>$\\
\\
\#irrelevant \\
0.0\\
\\
$<$/LOCAL\_HAMILTONIAN\_COUPLING$>$\\
\\
$<$LOCAL\_HAMILTONIAN\_REPRESENTATION\_INDEX$>$\\
\\
0	0\\
\\
$<$/LOCAL\_HAMILTONIAN\_REPRESENTATION\_INDEX$>$\\
\\
$<$LOCAL\_HAMILTONIAN\_MATRIX$>$\\
\\
$
\begin{array}{lll}
1       &0       &0\\
0       &0       &0\\
0       &0       &1\\
\end{array}
$\\
\\
$<$/LOCAL\_HAMILTONIAN\_MATRIX$>$\\
\\
		$<$/LOCAL\_HAMILTONIAN\_TERM$>$\\
\\
$<$/SECTION-LOCAL\_HAMILTONIAN$>$\\
\\
\#\#\#\#\#\#\#\#\#\#\#\#\#\#\#\#\#\#\#\#\#\#\#\#\#\#\#\#\#\#\#\#\#\#\#\#\#\#\#\#\#\#\#\#\#\#\#\#\#\#\#\#\#\#\#\#\\
		$<$SECTION-SPECTRAL\_FUNCTION$>$\\
\#\#\#\#\#\#\#\#\#\#\#\#\#\#\#\#\#\#\#\#\#\#\#\#\#\#\#\#\#\#\#\#\#\#\#\#\#\#\#\#\#\#\#\#\#\#\#\#\#\#\#\#\#\#\#\#\\
\\
\# In this section we set up the spectral function\\
\# calculation for different types of operators.\\
\# All the time we provide in the input file the \\
\# matrices for the $A^\dagger$   and $B^\dagger$  operators \\
\# but the spectral function is computed for the\\
\# combination $<$$A,B^\dagger$$>$\\
\# spectral function for $<$f\_0, f\_0\_dagger$>$ \\
\\
\{f\_0\_dagger; f\_0\_dagger \}\\
\\
\#spectral function for spin $<$S, S$>$ \\
\\
\{S; S\}\\
\\
\#\#\#\#\#\#\#\#\#\#\#\#\#\#\#\#\#\#\#\#\#\#\#\#\#\#\#\#\#\#\#\#\#\#\#\#\#\#\#\#\#\#\#\#\#\#\#\#\#\#\#\#\#\#\#\#\\
		$<$/SECTION-SPECTRAL\_FUNCTION$>$\\
\#\#\#\#\#\#\#\#\#\#\#\#\#\#\#\#\#\#\#\#\#\#\#\#\#\#\#\#\#\#\#\#\#\#\#\#\#\#\#\#\#\#\#\#\#\#\#\#\#\#\#\#\#\#\#\#\\
\\
\\
\# The section below is not needed by the {\bf fnrg} binary.\\
\# It is needed when doing the broadening of the spectral function\\
\# with the utility {\bf sfb}.\\
\\
\\
\#\#\#\#\#\#\#\#\#\#\#\#\#\#\#\#\#\#\#\#\#\#\#\#\#\#\#\#\#\#\#\#\#\#\#\#\#\#\#\#\#\#\#\#\#\#\#\#\#\#\#\#\#\#\#\#\\
                $<$SECTION-SPECTRAL\_FUNCTION\_BROADENING$>$\\
\#\#\#\#\#\#\#\#\#\#\#\#\#\#\#\#\#\#\#\#\#\#\#\#\#\#\#\#\#\#\#\#\#\#\#\#\#\#\#\#\#\#\#\#\#\#\#\#\#\#\#\#\#\#\#\#\\
\# In this section we include broadening parameters that\\
\# are used for computing the broaden spectral function\\
\# The methods that can be used are either LOG\_GAUSS  or\\ 
\# INTERPOLATIVE\_LOG\_GAUSS. \\
\# These procedures are described in the manual.\\
\\
\# The quantum temperature is relevant only \\
\# for the INTERPOLATIVE\_LOG\_GAUSS method,  \\
\# otherwise only the 'broadening\_parameter' parameter \\
\# setting the width of the log-normal \\
\# distribution function is relevant.\\
\\
\# The quantum temperature must be always lower than T \\
\# for finite temperature calculations. \\
\# A good choice is to set the quatum temperature in the range\\
\# 0.5T - T.\\  
\\
\# The  three parameters spectral\_function\_*\\
\# specify which spectral functions are being calculated. \\
\# The accepted values are YES or NO only. \\
\\
\# The static\_average\_dmnrg flags specifies whether the expectation values \\
\# of the static operators are computed.\\ 
\# The accepted values are YES or NO only.\\
\\
\# The green\_function\_dmnrg specifies whether the real part of the\\
\# Green's function is computed from the spectral function data.\\
\\
broadening\_method  = INTERPOLATIVE\_LOG\_GAUSS	\# method used either LOG\_GAUSS or INTERPOLATIVE\_LOG\_GAUSS\\
broadening\_parameter = 0.70 			\# parameter used for the log gaussian distribution\\
broadening\_energy\_minim = 1e-6		\# minimum energy for which the broadening is performed\\
broadening\_energy\_maxim = 4.0			\# maximum energy for which the broadening is performed\\
broadening\_grid\_mesh = 100                     \# grid mesh on a log scale between the minimum and maximum energies\\
quantum\_temperature =  1e-7			\# used for the INTERPOLATIVE\_LOG\_GAUSS method only.	\\
\\
spectral\_function\_nrg\_even = YES		\# spectral function for the even iterations is computed 	\\
spectral\_function\_nrg\_odd  = YES		\# spectral function for the odd iterations part is computed\\
spectral\_function\_dmnrg    = YES		\# spectral function for the dmnrg calculation is computed.\\

static\_average\_dmnrg       = YES              \# average functions similar with the case of spectral function. \\

green\_function\_nrg\_even = YES                \# calculation of the real/imaginary part of the Green's function from the even
part of the nrg  spectral function\\
green\_function\_nrg\_odd = YES                \# calculation of the real/imaginary part of the Green's function from the odd
part of the nrg spectral function\\
green\_function\_dmnrg	   = NO		        \# calculation of the real/imaginary part of the Green's function from the spectral function\\
green\_function\_grid\_mesh = 400               \# points for the mesh when doing Hilbert transform\\
\\
\#\#\#\#\#\#\#\#\#\#\#\#\#\#\#\#\#\#\#\#\#\#\#\#\#\#\#\#\#\#\#\#\#\#\#\#\#\#\#\#\#\#\#\#\#\#\#\#\#\#\#\#\#\#\#\#\\
                $<$/SECTION-SPECTRAL\_FUNCTION\_BROADENING$>$\\
\#\#\#\#\#\#\#\#\#\#\#\#\#\#\#\#\#\#\#\#\#\#\#\#\#\#\#\#\#\#\#\#\#\#\#\#\#\#\#\#\#\#\#\#\#\#\#\#\#\#\#\#\#\#\#\#\\
\\

}
}

\chapter{Input file for the Kondo model with $SU_{{\rm charge}}(2) \times SU_{\rm spin}(2)$ symmetries }
{\small
{\tt
\# The comments must start with the symbol \# .\\
\# Empty lines are ignored.\\
\\
\\
\#\#\#\#\#\#\#\#\#\#\#\#\#\#\#\#\#\#\#\#\#\#\#\#\#\#\#\#\#\#\#\#\#\#\#\#\#\#\#\#\#\#\#\#\#\#\#\#\#\#\#\#\#\#\#\#\#\\
		$<$SECTION-PARAMETERS$>$   \\
\#\#\#\#\#\#\#\#\#\#\#\#\#\#\#\#\#\#\#\#\#\#\#\#\#\#\#\#\#\#\#\#\#\#\#\#\#\#\#\#\#\#\#\#\#\#\#\#\#\#\#\#\#\#\#\#\#\\
\\
\# Explanation of variables\\
\\
\# 'model' represents the type of the model that is \\
\#  used for the calculations. It can be any name. \\
\\
\#  'lambda' represents the value for the logarithmic \\
\#  discretization parameter. Usually it is fixed between 2 and 3. \\
\\
\#  'max\_state\_no' is the number of kept states  \\
\#  after an iteration.\\
\\
\#  'iteration\_no' is the number of iterations performed \\
\#  along the NRG run. Typically it is set to 40 - 70.\\
\\
\#  'symmetry\_no' specifies how many symmetries \\
\#  are used. The symmetries  are characterized \\
\#  in $<$SECTION-SYMMETRIES$>$ \\
\\
\#  'coupling\_no' specifies how many couplings \\
\#  are present in the interaction Hamiltonian.  \\
\\
\#  'local\_coupling\_no' is the number of \\
\#  Hamiltonians at the sites we add to the Wilson chain\\
\\
\#  'spectral\_operator\_no' is the number of spectral operators \\
\#  acting on site 0, we keep track of.\\
\#  These spectral operators are characterized\\
\# in section $<$SECTION-SPECTRAL-OPERATORS$>$\\	
\\
\#  'static\_operator\_no' is the number of static operators \\
\#  acting on site 0, we keep track of.\\
\#  These static operators are characterized\\
\#  in section $<$SECTION-STATIC-OPERATORS$>$\\	
\\
\#  'hopping\_operator\_no' is the number of hopping operators. \\
\#   These hopping operators are characterized\\
\#  in section $<$SECTION-HOPPING-OPERATORS$>$\\	
\\
\#  'block\_state\_no' specifies the number \\
\#  of initial states/multiplets. These states are characterized\\
\#  by the representation indices and are listed\\
\#  in section $<$SECTION-BLOCK\_STATES$>$\\
\\
\\
\#\#\#\#\#\#\#\#\#\#\#\#\#\#\#\#\#\#\#\#\#\#\#\#\#\#\#\#\#\#\#\#\#\#\#\#\#\#\#\#\#\#\#\#\#\#\#\#\#\#\#\#\#\#\#\#\#\\
\\
\#\#\#\#\#\#\#\#\#\#\#\#\#\#\#\#\#\#\#\#\#\#\#\#\#\#\#\#\#\#\#\#\#\#\#\#\#\#\#\#\#\#\#\#\#\#\#\#\#\#\#\#\#\#\#\#\#\\
\\
model = kondo\_model  \# the name of the present model \\
\\
lambda = 2.0   \# discretization parameter\\
\\
max\_state\_no = 100  \# maximum number of kept multiplets\\
\\
iteration\_no = 20  \# allowed number of iterations\\
\\
symmetry\_no = 2  \# the number of symmetries  \\
\\
coupling\_no = 1   \# no of Hamiltonian terms, i.e. {\cal J} in this case\\
\\
spectral\_operator\_no = 2   \# no of operators for which correlation functions are computed \\
\\
static\_operator\_no = 0  \# no of operators for which the static average is computed. \\
\\
hopping\_operator\_no = 1  \# the number of hopping operators\\
\\
block\_state\_no = 3  \# the number of initial block states\\
\\
local\_state\_no = 2   \# the number of local states added in each iteration\\
\\
local\_coupling\_no = 1  \# the number of Hamiltonian terms on the added site\\
\\
spectral\_function\_no = 2 \#  the number of spectral functions that need to be generated \\
\\
interval\_no = 1000 \# mash grid between the maximum and minimum energy at each iteration \\
\\
degeneracy\_threshhold = 1e-6  \# energy thresh-hold for discarding states \\
\\
temperature = 0.0         \# temperature used for the run\\
\\
\\
\#\#\#\#\#\#\#\#\#\#\#\#\#\#\#\#\#\#\#\#\#\#\#\#\#\#\#\#\#\#\#\#\#\#\#\#\#\#\#\#\#\#\#\#\#\#\#\#\#\#\#\#\#\#\#\#\\
		$<$/SECTION-PARAMETERS$>$   \\
\#\#\#\#\#\#\#\#\#\#\#\#\#\#\#\#\#\#\#\#\#\#\#\#\#\#\#\#\#\#\#\#\#\#\#\#\#\#\#\#\#\#\#\#\#\#\#\#\#\#\#\#\#\#\#\#\\
\\
\\
\#\#\#\#\#\#\#\#\#\#\#\#\#\#\#\#\#\#\#\#\#\#\#\#\#\#\#\#\#\#\#\#\#\#\#\#\#\#\#\#\#\#\#\#\#\#\#\#\#\#\#\#\#\#\#\#\\
		$<$SECTION-FLAGS$>$   \\
\#\#\#\#\#\#\#\#\#\#\#\#\#\#\#\#\#\#\#\#\#\#\#\#\#\#\#\#\#\#\#\#\#\#\#\#\#\#\#\#\#\#\#\#\#\#\#\#\#\#\#\#\#\#\#\#\\
\\
\\
\# dmnrg\_flag  controls whether the backward procedure \\
\# has to be done and the full set of eigen-states be \\
\# used for the calculation of the spectral function. \\
\# ON - backwards iteration is done.\\
\# OFF - only the nrg calculation is performed.\\
\\
\# text\_swap\_files\_flag = flag that fixes if the text mode is \\
\# used to save the files. The files will be saved in binary mode anyway. \\
\# ON - the files are saved in the text mode and kept on the disk.\\
\# OFF= the files are not saved in the text mode. \\
\\
\#  binary\_swap\_files\_flag = flag that controls whether the \\
\#  unnecessary binary files will be removed or not not after the calculation is done.\\
\#  ON - the files will be kept on the disk.\\
\#  OFF- the files will be removed from the disk.\\
\\
\# hoppings\_on\_site\_energies\_flag = flag that controls \\
\# the reading of the hoppings and the on-site energies. \\
\# ON - the hoppings and the on site energies \\
\# are read from the files in the  results/mapping/ folder\\
\# These files are generated by the {\bf he} utility.\\
\# OFF - the hoppings are computed on the fly \\
\# assuming a flat density of states (DOS = 0.5) on [-1, 1],\\
\# and the on-site energies are set to zero. 
\\
\\
dmnrg\_flag = ON \\
text\_swap\_files\_flag = OFF \\
binary\_swap\_files\_flag = ON \\
hoppings\_on\_site\_energies\_flag = OFF \\
\\
\\
\#\#\#\#\#\#\#\#\#\#\#\#\#\#\#\#\#\#\#\#\#\#\#\#\#\#\#\#\#\#\#\#\#\#\#\#\#\#\#\#\#\#\#\#\#\#\#\#\#\#\#\#\#\#\#\#\\
		$<$/SECTION-FLAGS$>$   \\
\#\#\#\#\#\#\#\#\#\#\#\#\#\#\#\#\#\#\#\#\#\#\#\#\#\#\#\#\#\#\#\#\#\#\#\#\#\#\#\#\#\#\#\#\#\#\#\#\#\#\#\#\#\#\#\#\\
\\
\\
\#\#\#\#\#\#\#\#\#\#\#\#\#\#\#\#\#\#\#\#\#\#\#\#\#\#\#\#\#\#\#\#\#\#\#\#\#\#\#\#\#\#\#\#\#\#\#\#\#\#\#\#\#\#\#\#\#\\
		$<$SECTION-SYMMETRIES$>$\\
\#\#\#\#\#\#\#\#\#\#\#\#\#\#\#\#\#\#\#\#\#\#\#\#\#\#\#\#\#\#\#\#\#\#\#\#\#\#\#\#\#\#\#\#\#\#\#\#\#\#\#\#\#\#\#\#\#\\
\\
\#  As many symmetries must be defined below  \\
\#  as set by the  variable 'symmetry\_no' \\
\#  in the previous section. \\
\\
\#  The possible symmetry types are the following:\\
\#  U(1), SU(2), Z(2), charge\_SU(2).\\
\#  \\
\#  Next the limits for the representation index \\
\#  are set. All states with smaller/larger\\
\#  representation indices will be cut off\\
\\
\#  For U(1) symmetries the representations are labeled\\
\# by integers (2 S\_z) and for SU(2) by non-negative\\
\# integers (2S).\\
\\
\\
\#  For the Z(2) symmetry the representations are labeled\\
\#  by integers \\
\#  0 - even parity states\\
\#  1 - odd parity states\\
\\
\#\#\#\#\#\#\#\#\#\#\#\#\#\#\#\#\#\#\#\#\#\#\#\#\#\#\#\#\#\#\#\#\#\#\#\#\#\#\#\#\#\#\#\#\#\#\#\#\#\#\#\#\#\#\#\#\# \\
\\
\\
\# charge symmetry corresponding to Q \\
charge\_SU(2)     0     20      \\
\\
\# spin symmetry corresponding to S\\
SU(2)     0     20     \\
\\
\\
\#\#\#\#\#\#\#\#\#\#\#\#\#\#\#\#\#\#\#\#\#\#\#\#\#\#\#\#\#\#\#\#\#\#\#\#\#\#\#\#\#\#\#\#\#\#\#\#\#\#\#\#\#\#\#\#\# \\
		$<$/SECTION-SYMMETRIES$>$\\
\#\#\#\#\#\#\#\#\#\#\#\#\#\#\#\#\#\#\#\#\#\#\#\#\#\#\#\#\#\#\#\#\#\#\#\#\#\#\#\#\#\#\#\#\#\#\#\#\#\#\#\#\#\#\#\#\# \\
\\
\\
\#\#\#\#\#\#\#\#\#\#\#\#\#\#\#\#\#\#\#\#\#\#\#\#\#\#\#\#\#\#\#\#\#\#\#\#\#\#\#\#\#\#\#\#\#\#\#\#\#\#\#\#\#\#\#\#\#\\
		$<$SECTION-BLOCK\_STATES$>$\\
\#\#\#\#\#\#\#\#\#\#\#\#\#\#\#\#\#\#\#\#\#\#\#\#\#\#\#\#\#\#\#\#\#\#\#\#\#\#\#\#\#\#\#\#\#\#\#\#\#\#\#\#\#\#\#\#\#\\
\# This section gives the representation indices of the \\
\# block states of the first iteration. \\
\# The number of rows has to be equal \\
\# with the 'block\_state\_no' in the first section, \\
\# and the number of columns to the 'symmetry\_no'.\\
\# Each column contains indices corresponding to a symmetry.\\
\#\#\#\#\#\#\#\#\#\#\#\#\#\#\#\#\#\#\#\#\#\#\#\#\#\#\#\#\#\#\#\#\#\#\#\#\#\#\#\#\#\#\#\#\#\#\#\#\#\#\#\#\#\#\#\#\#\\
$
\begin{array}{ll}
0&0\\
0&2\\
1&1\\
\end{array}
$\\
\#\#\#\#\#\#\#\#\#\#\#\#\#\#\#\#\#\#\#\#\#\#\#\#\#\#\#\#\#\#\#\#\#\#\#\#\#\#\#\#\#\#\#\#\#\#\#\#\#\#\#\#\#\#\#\#\#\\
		$<$/SECTION-BLOCK\_STATES$>$\\
\#\#\#\#\#\#\#\#\#\#\#\#\#\#\#\#\#\#\#\#\#\#\#\#\#\#\#\#\#\#\#\#\#\#\#\#\#\#\#\#\#\#\#\#\#\#\#\#\#\#\#\#\#\#\#\#\#\\
\\
\\
\#\#\#\#\#\#\#\#\#\#\#\#\#\#\#\#\#\#\#\#\#\#\#\#\#\#\#\#\#\#\#\#\#\#\#\#\#\#\#\#\#\#\#\#\#\#\#\#\#\#\#\#\#\#\#\#\#\\
		$<$SECTION-LOCAL\_STATES$>$\\
\#\#\#\#\#\#\#\#\#\#\#\#\#\#\#\#\#\#\#\#\#\#\#\#\#\#\#\#\#\#\#\#\#\#\#\#\#\#\#\#\#\#\#\#\#\#\#\#\#\#\#\#\#\#\#\#\#\\
\# This section gives  the representation indices \\
\# of the local states. The number of rows has to be equal\\ 
\# with the 'local\_state\_no' in the first section\\
\# and the number of columns to the 'symmetry\_no'.\\
\# Each column contains indices corresponding to a symmetry.\\
\#\#\#\#\#\#\#\#\#\#\#\#\#\#\#\#\#\#\#\#\#\#\#\#\#\#\#\#\#\#\#\#\#\#\#\#\#\#\#\#\#\#\#\#\#\#\#\#\#\#\#\#\#\#\#\#\\
$
\begin{array}{ll}
0     &  1\\
1     &  0\\
\end{array}
$\\
\#\#\#\#\#\#\#\#\#\#\#\#\#\#\#\#\#\#\#\#\#\#\#\#\#\#\#\#\#\#\#\#\#\#\#\#\#\#\#\#\#\#\#\#\#\#\#\#\#\#\#\#\#\#\#\#\\
		$<$/SECTION-LOCAL\_STATES$>$\\
\#\#\#\#\#\#\#\#\#\#\#\#\#\#\#\#\#\#\#\#\#\#\#\#\#\#\#\#\#\#\#\#\#\#\#\#\#\#\#\#\#\#\#\#\#\#\#\#\#\#\#\#\#\#\#\#\\
\\
\\
\#\#\#\#\#\#\#\#\#\#\#\#\#\#\#\#\#\#\#\#\#\#\#\#\#\#\#\#\#\#\#\#\#\#\#\#\#\#\#\#\#\#\#\#\#\#\#\#\#\#\#\#\#\#\#\#\\
		$<$SECTION-LOCAL\_STATES\_SIGNS$>$\\
\#\#\#\#\#\#\#\#\#\#\#\#\#\#\#\#\#\#\#\#\#\#\#\#\#\#\#\#\#\#\#\#\#\#\#\#\#\#\#\#\#\#\#\#\#\#\#\#\#\#\#\#\#\#\#\#\\
\# These are the signs of the added states.\\ 
\# The number of signs has to be equal with \\
\# the number of 'local\_state\_no'. \\
\# All the signs must be enumerated in one line.\\ 
\# For even electron number, the sign is 1, \\
\# while for odd electron number the sign is -1.\\
\#\#\#\#\#\#\#\#\#\#\#\#\#\#\#\#\#\#\#\#\#\#\#\#\#\#\#\#\#\#\#\#\#\#\#\#\#\#\#\#\#\#\#\#\#\#\#\#\#\#\#\#\#\#\#\#\#\\
\\
-1  1\\
\\
\#\#\#\#\#\#\#\#\#\#\#\#\#\#\#\#\#\#\#\#\#\#\#\#\#\#\#\#\#\#\#\#\#\#\#\#\#\#\#\#\#\#\#\#\#\#\#\#\#\#\#\#\#\#\#\#\#\\
		$<$/SECTION-LOCAL\_STATES\_SIGNS$>$\\
\#\#\#\#\#\#\#\#\#\#\#\#\#\#\#\#\#\#\#\#\#\#\#\#\#\#\#\#\#\#\#\#\#\#\#\#\#\#\#\#\#\#\#\#\#\#\#\#\#\#\#\#\#\#\#\#\#\\
\\
\\
\#\#\#\#\#\#\#\#\#\#\#\#\#\#\#\#\#\#\#\#\#\#\#\#\#\#\#\#\#\#\#\#\#\#\#\#\#\#\#\#\#\#\#\#\#\#\#\#\#\#\#\#\#\#\#\#\#\\
\\
		$<$SECTION-BLOCK\_HAMILTONIAN$>$\\
\\
\# The total Hamiltonian will be written as a sum\\
\# of the Hamiltonian terms.\\
$
H_0 = \sum_{\delta}^{coupling\_no} G_\delta H_\delta
$\\
\# There has to be as many $<$HAMILTONIAN\_TERM$>$ items\\
\# as 'coupling\_no' were set in the first section.\\
\# The first tag is the name of the Hamiltonian term,\\ 
\# the second one gives the representation indices and\\
\# at last comes the matrix elements that will be\\
\# read into the sectors.\\
\\
\\

\#   first term in the Hamiltonian \\
\\
		$<$BLOCK\_HAMILTONIAN\_TERM$>$\\
\\
$<$BLOCK\_HAMILTONIAN\_NAME$>$	\\
H\_Kondo\\
$<$/BLOCK\_HAMILTONIAN\_NAME$>$\\
\\
$<$BLOCK\_HAMILTONIAN\_COUPLING$>$\\
\\
\# J - exchange coupling \\
J = 0.5\\
\\
$<$/BLOCK\_HAMILTONIAN\_COUPLING$>$\\
\\
$<$BLOCK\_HAMILTONIAN\_REPRESENTATION\_INDEX$>$\\
\\
0	0\\
\\
$<$/BLOCK\_HAMILTONIAN\_REPRESENTATION\_INDEX$>$\\
\\
$<$BLOCK\_HAMILTONIAN\_MATRIX$>$\\
\\
$
\begin{array}{lll}
-0.75  &0      &0\\
0      &0.25   &0\\
0	&0      &0\\
\end{array}	
$\\
\\
$<$/BLOCK\_HAMILTONIAN\_MATRIX$>$\\
\\
		$<$/BLOCK\_HAMILTONIAN\_TERM$>$\\
\\
\\
\#\#\#\#\#\#\#\#\#\#\#\#\#\#\#\#\#\#\#\#\#\#\#\#\#\#\#\#\#\#\#\#\#\#\#\#\#\#\#\#\#\#\#\#\#\#\#\#\#\#\#\#\#\#\#\#\\
		$<$/SECTION-BLOCK\_HAMILTONIAN$>$\\
\#\#\#\#\#\#\#\#\#\#\#\#\#\#\#\#\#\#\#\#\#\#\#\#\#\#\#\#\#\#\#\#\#\#\#\#\#\#\#\#\#\#\#\#\#\#\#\#\#\#\#\#\#\#\#\#\\
\\
\#\#\#\#\#\#\#\#\#\#\#\#\#\#\#\#\#\#\#\#\#\#\#\#\#\#\#\#\#\#\#\#\#\#\#\#\#\#\#\#\#\#\#\#\#\#\#\#\#\#\#\#\#\#\#\#\\
\\

		$<$SECTION-HOPPING\_OPERATORS$>$\\
\\
\# There must be as many $<$HOPPING\_OPERATOR$>$ items \\
\# as 'hopping\_operator\_no' were set in the first section\\
\# The matrices must contain the Reduced Matrix Elements.
\\
\#\#\#\#\#\#\#\#\#\#\#\#\#\#\#\#\#\#\#\#\#\#\#\#\#\#\#\#\#\#\#\#\#\#\#\#\#\#\#\#\#\#\#\#\#\#\#\#\#\#\#\#\#\#\#\#\\
\\
\# here is the first hopping operator\\
\\
		$<$HOPPING\_OPERATOR$>$\\
\\
$<$HOPPING\_OPERATOR\_NAME$>$\\
\\
f\_N\_dagger   \\
\\
$<$/HOPPING\_OPERATOR\_NAME$>$\\
\\
$<$HOPPING\_OPERATOR\_REPRESENTATION\_INDEX$>$\\
\\
1 1\\
\\
$<$/HOPPING\_OPERATOR\_REPRESENTATION\_INDEX$>$\\
\\
$<$HOPPING\_OPERATOR\_SIGN$>$\\
\\
\# The sign is 1 for bosonic operators and -1 for fermionic operators.\\
-1\\
\\
$<$/HOPPING\_OPERATOR\_SIGN$>$\\
\\
$<$HOPPING\_OPERATOR\_MATRIX$>$\\
\\
$
\begin{array}{lll}
0               &0              & -1.4142135623731\\
0               &0               &-1.4142135623731\\
0.707106781186   &-1.224744871391 &0\\
\end{array}
$\\
\\
$<$/HOPPING\_OPERATOR\_MATRIX$>$\\
\\
\\		$<$/HOPPING\_OPERATOR$>$\\
\\
\#\#\#\#\#\#\#\#\#\#\#\#\#\#\#\#\#\#\#\#\#\#\#\#\#\#\#\#\#\#\#\#\#\#\#\#\#\#\#\#\#\#\#\#\#\#\#\#\#\#\#\#\#\#\#\#\\
		$<$/SECTION-HOPPING\_OPERATORS$>$\\
\#\#\#\#\#\#\#\#\#\#\#\#\#\#\#\#\#\#\#\#\#\#\#\#\#\#\#\#\#\#\#\#\#\#\#\#\#\#\#\#\#\#\#\#\#\#\#\#\#\#\#\#\#\#\#\#\\
\\
\\
\\
\#\#\#\#\#\#\#\#\#\#\#\#\#\#\#\#\#\#\#\#\#\#\#\#\#\#\#\#\#\#\#\#\#\#\#\#\#\#\#\#\#\#\#\#\#\#\#\#\#\#\#\#\#\#\#\#\\
		$<$SECTION-SPECTRAL\_OPERATORS$>$\\
\#\#\#\#\#\#\#\#\#\#\#\#\#\#\#\#\#\#\#\#\#\#\#\#\#\#\#\#\#\#\#\#\#\#\#\#\#\#\#\#\#\#\#\#\#\#\#\#\#\#\#\#\#\#\#\#\\
\# In this section we set up the matrices for the \\
\# operators for which the spectral function is computed\\
\# Matrices must contain the reduced matrix elements.\\
\#\#\#\#\#\#\#\#\#\#\#\#\#\#\#\#\#\#\#\#\#\#\#\#\#\#\#\#\#\#\#\#\#\#\#\#\#\#\#\#\#\#\#\#\#\#\#\#\#\#\#\#\#\#\#\\
\\
\# here comes the first spectral operator.
\\

		$<$SPECTRAL\_OPERATOR$>$\\
\\
\# here comes the spectral operator.
\\
		$<$SPECTRAL\_OPERATOR$>$\\
\\
$<$SPECTRAL\_OPERATOR\_NAME$>$\\
\\
f\_0\_dagger\\
\\
$<$/SPECTRAL\_OPERATOR\_NAME$>$\\
\\
$<$SPECTRAL\_OPERATOR\_REPRESENTATION\_INDEX$>$\\
1 1 \\
$<$/SPECTRAL\_OPERATOR\_REPRESENTATION\_INDEX$>$\\
\\
$<$SPECTRAL\_OPERATOR\_SIGN$>$\\
\\
\# The sign is 1 for bosonic operators and -1 for fermionic operators.\\
-1\\
\\
$<$/SPECTRAL\_OPERATOR\_SIGN$>$\\
\\
$<$SPECTRAL\_OPERATOR\_MATRIX$>$\\
\\
$
\begin{array}{llll}
0               &0               &-1.4142135623731\\
0               &0               &-1.4142135623731\\
.707106781186   &-1.224744871391 &0\\
\end{array}
$\\
\\
$<$/SPECTRAL\_OPERATOR\_MATRIX$>$\\
\\
		$<$/SPECTRAL\_OPERATOR$>$\\
\\
\\
\# here comes the second spectral operator.
\\

		$<$SPECTRAL\_OPERATOR$>$\\
\\
		$<$SPECTRAL\_OPERATOR$>$\\
\\
$<$SPECTRAL\_OPERATOR\_NAME$>$\\
\\
S\\
\\
$<$/SPECTRAL\_OPERATOR\_NAME$>$\\
\\
$<$SPECTRAL\_OPERATOR\_REPRESENTATION\_INDEX$>$\\
0 2 \\
$<$/SPECTRAL\_OPERATOR\_REPRESENTATION\_INDEX$>$\\
\\
$<$SPECTRAL\_OPERATOR\_SIGN$>$\\
\\
\# The sign is 1 for bosonic operators and -1 for fermionic operators.\\
1\\
\\
$<$/SPECTRAL\_OPERATOR\_SIGN$>$\\
\\
$<$SPECTRAL\_OPERATOR\_MATRIX$>$\\
\\
$
\begin{array}{llll}
0               &-0.866025403784   &0\\
0.5             &0.707106781186    &0\\
0               &0                 &0.866025403784\\
\end{array}
$\\
\\
$<$/SPECTRAL\_OPERATOR\_MATRIX$>$\\
\\
		$<$/SPECTRAL\_OPERATOR$>$\\
\\
\\
\#\#\#\#\#\#\#\#\#\#\#\#\#\#\#\#\#\#\#\#\#\#\#\#\#\#\#\#\#\#\#\#\#\#\#\#\#\#\#\#\#\#\#\#\#\#\#\#\#\#\#\#\#\#\#\#\\
		$<$/SECTION-SPECTRAL\_OPERATORS$>$\\
\#\#\#\#\#\#\#\#\#\#\#\#\#\#\#\#\#\#\#\#\#\#\#\#\#\#\#\#\#\#\#\#\#\#\#\#\#\#\#\#\#\#\#\#\#\#\#\#\#\#\#\#\#\#\#\#\\
\\
\\
\#\#\#\#\#\#\#\#\#\#\#\#\#\#\#\#\#\#\#\#\#\#\#\#\#\#\#\#\#\#\#\#\#\#\#\#\#\#\#\#\#\#\#\#\#\#\#\#\#\#\#\#\#\#\#\#\\
		$<$SECTION-LOCAL\_HOPPING\_OPERATORS$>$\\
\#\#\#\#\#\#\#\#\#\#\#\#\#\#\#\#\#\#\#\#\#\#\#\#\#\#\#\#\#\#\#\#\#\#\#\#\#\#\#\#\#\#\#\#\#\#\#\#\#\#\#\#\#\#\#\#\\
\\
		$<$LOCAL\_HOPPING\_OPERATOR$>$\\
\\
\# here comes the first local hopping operator
\\
$<$LOCAL\_HOPPING\_OPERATOR\_NAME$>$\\
\\
f\_N+1\_dagger\\
\\
$<$/LOCAL\_HOPPING\_OPERATOR\_NAME$>$\\
\\
$<$LOCAL\_HOPPING\_OPERATOR\_REPRESENTATION\_INDEX$>$\\
\\
1 1\\
\\
$<$/LOCAL\_HOPPING\_OPERATOR\_REPRESENTATION\_INDEX$>$\\
\\
$<$LOCAL\_HOPPING\_OPERATOR\_SIGN$>$\\
\\
\# The sign is 1 for bosonic operators and -1 for fermionic operators.\\
-1\\
\\
$<$/LOCAL\_HOPPING\_OPERATOR\_SIGN$>$\\
\\
$<$LOCAL\_HOPPING\_OPERATOR\_MATRIX$>$\\
\\
$
\begin{array}{ll}
0                       &1.4142135623731\\
-1.4142135623731        &0\\
\end{array}
$\\
\\
$<$/LOCAL\_HOPPING\_OPERATOR\_MATRIX$>$\\
\\
$<$LOCAL\_ON\_SITE\_ENERGY\_MATRIX$>$\\
\\
$
\begin{array}{ll}
0       &0\\
0       &0\\
\end{array}
$\\
\\
$<$/LOCAL\_ON\_SITE\_ENERGY\_MATRIX$>$\\
\\
		$<$/LOCAL\_HOPPING\_OPERATOR$>$\\
\\
\\
\\
\#\#\#\#\#\#\#\#\#\#\#\#\#\#\#\#\#\#\#\#\#\#\#\#\#\#\#\#\#\#\#\#\#\#\#\#\#\#\#\#\#\#\#\#\#\#\#\#\#\#\#\#\#\#\#\#\\
		$<$/SECTION-LOCAL\_HOPPING\_OPERATORS$>$\\
\#\#\#\#\#\#\#\#\#\#\#\#\#\#\#\#\#\#\#\#\#\#\#\#\#\#\#\#\#\#\#\#\#\#\#\#\#\#\#\#\#\#\#\#\#\#\#\#\#\#\#\#\#\#\#\#\\
\\
\#\#\#\#\#\#\#\#\#\#\#\#\#\#\#\#\#\#\#\#\#\#\#\#\#\#\#\#\#\#\#\#\#\#\#\#\#\#\#\#\#\#\#\#\#\#\#\#\#\#\#\#\#\#\#\#\\
	
		$<$SECTION-LOCAL\_HAMILTONIAN$>$

\# The total Local Hamiltonian of the added site \\
\# will be written as a sum of Local Hamiltonian terms\\
\# acting on site n:\\
$
H_n = \sum_{i}g_{i,n} O_{i,n}
$
\\
\# The number of  $<$LOCAL\_HAMILTONIAN\_TERM$>$ items\\
\# must be equal with the variable 'local\_coupling\_no'\\
\# set in the first section\\
\\
\# The first tag is the name of the local Hamiltonian term\\
\# the second one gives the representation indices\\
\#  last come the matrix elements.\\
\\
\#\#\#\#\#\#\#\#\#\#\#\#\#\#\#\#\#\#\#\#\#\#\#\#\#\#\#\#\#\#\#\#\#\#\#\#\#\#\#\#\#\#\#\#\#\#\#\#\#\#\#\#\#\#\#\#\\
\#   first term in the Local\_Hamiltonian\\
\#\#\#\#\#\#\#\#\#\#\#\#\#\#\#\#\#\#\#\#\#\#\#\#\#\#\#\#\#\#\#\#\#\#\#\#\#\#\#\#\#\#\#\#\#\#\#\#\#\#\#\#\#\#\#\\
\\
		$<$LOCAL\_HAMILTONIAN\_TERM$>$\\
\\
$<$LOCAL\_HAMILTONIAN\_NAME$>$		\\
H\_1\\
$<$/LOCAL\_HAMILTONIAN\_NAME$>$\\
\\
$<$LOCAL\_HAMILTONIAN\_COUPLING$>$\\
\\
\#irrelevant \\
0.0\\
\\
$<$/LOCAL\_HAMILTONIAN\_COUPLING$>$\\
\\
$<$LOCAL\_HAMILTONIAN\_REPRESENTATION\_INDEX$>$\\
\\
0	0\\
\\
$<$/LOCAL\_HAMILTONIAN\_REPRESENTATION\_INDEX$>$\\
\\
$<$LOCAL\_HAMILTONIAN\_MATRIX$>$\\
\\
$
\begin{array}{ll}
1       &0\\
0       &0\\
\end{array}
$\\
\\
$<$/LOCAL\_HAMILTONIAN\_MATRIX$>$\\
\\
		$<$/LOCAL\_HAMILTONIAN\_TERM$>$\\
\\
$<$/SECTION-LOCAL\_HAMILTONIAN$>$\\
\\
\#\#\#\#\#\#\#\#\#\#\#\#\#\#\#\#\#\#\#\#\#\#\#\#\#\#\#\#\#\#\#\#\#\#\#\#\#\#\#\#\#\#\#\#\#\#\#\#\#\#\#\#\#\#\#\#\\
		$<$SECTION-SPECTRAL\_FUNCTION$>$\\
\#\#\#\#\#\#\#\#\#\#\#\#\#\#\#\#\#\#\#\#\#\#\#\#\#\#\#\#\#\#\#\#\#\#\#\#\#\#\#\#\#\#\#\#\#\#\#\#\#\#\#\#\#\#\#\#\\
\\
\# In this section we set up the spectral function\\
\# calculation for different types of operators.\\
\# All the time we provide in the input file the \\
\# matrices for the $A^\dagger$   and $B^\dagger$  operators \\
\# but the spectral function is computed for the\\
\# combination $<$$A,B^\dagger$$>$\\
\# spectral function for $<$f\_0, f\_0\_dagger$>$ \\
\\
\{f\_0\_dagger; f\_0\_dagger \}\\
\\
\# spectral function for spin $<$S, S$>$ \\
\\
\{S; S \}\\
\\
\#\#\#\#\#\#\#\#\#\#\#\#\#\#\#\#\#\#\#\#\#\#\#\#\#\#\#\#\#\#\#\#\#\#\#\#\#\#\#\#\#\#\#\#\#\#\#\#\#\#\#\#\#\#\#\#\\
		$<$/SECTION-SPECTRAL\_FUNCTION$>$\\
\#\#\#\#\#\#\#\#\#\#\#\#\#\#\#\#\#\#\#\#\#\#\#\#\#\#\#\#\#\#\#\#\#\#\#\#\#\#\#\#\#\#\#\#\#\#\#\#\#\#\#\#\#\#\#\#\\
\\
\\
\# The section below is not needed by the {\bf fnrg} binary.\\
\# It is needed when doing the broadening of the spectral function\\
\# with the utility {\bf sfb}.\\
\\
\\
\#\#\#\#\#\#\#\#\#\#\#\#\#\#\#\#\#\#\#\#\#\#\#\#\#\#\#\#\#\#\#\#\#\#\#\#\#\#\#\#\#\#\#\#\#\#\#\#\#\#\#\#\#\#\#\#\\
                $<$SECTION-SPECTRAL\_FUNCTION\_BROADENING$>$\\
\#\#\#\#\#\#\#\#\#\#\#\#\#\#\#\#\#\#\#\#\#\#\#\#\#\#\#\#\#\#\#\#\#\#\#\#\#\#\#\#\#\#\#\#\#\#\#\#\#\#\#\#\#\#\#\#\\
\# In this section we include broadening parameters that\\
\# are used for computing the broaden spectral function\\
\# The methods that can be used are either LOG\_GAUSS  or\\ 
\# INTERPOLATIVE\_LOG\_GAUSS. \\
\# These procedures are described in the manual.\\
\\
\# The quantum temperature is relevant only \\
\# for the INTERPOLATIVE\_LOG\_GAUSS method,  \\
\# otherwise only the 'broadening\_parameter' parameter \\
\# setting the width of the log-normal \\
\# distribution function is relevant.\\
\\
\# The quantum temperature must be always lower than T \\
\# for finite temperature calculations. \\
\# A good choice is to set the quatum temperature in the range\\
\# 0.5T - T.\\  
\\
\# The  three parameters spectral\_function\_*\\
\# specify which spectral functions are being calculated. \\
\# The accepted values are YES or NO only. \\
\\
\# The static\_average\_dmnrg flags specifies whether the expectation values \\
\# of the static operators are computed.\\ 
\# The accepted values are YES or NO only.\\
\\
\# The green\_function\_dmnrg specifies whether the real part of the\\
\# Green's function is computed from the spectral function data.\\
\\
broadening\_method  = INTERPOLATIVE\_LOG\_GAUSS	\# method used either LOG\_GAUSS or INTERPOLATIVE\_LOG\_GAUSS\\
broadening\_parameter = 0.70 			\# parameter used for the log gaussian distribution\\
broadening\_energy\_minim = 1e-6		\# minimum energy for which the broadening is performed\\
broadening\_energy\_maxim = 4.0			\# maximum energy for which the broadening is performed\\
broadening\_grid\_mesh = 100                     \# grid mesh on a log scale between the minimum and maximum energies\\
quantum\_temperature =  1e-7			\# used for the INTERPOLATIVE\_LOG\_GAUSS method only.	\\
\\
spectral\_function\_nrg\_even = YES		\# spectral function for the even iterations is computed 	\\
spectral\_function\_nrg\_odd  = YES		\# spectral function for the odd iterations part is computed\\
spectral\_function\_dmnrg    = YES		\# spectral function for the dmnrg calculation is computed.\\

static\_average\_dmnrg       = YES              \# average functions similar with the case of spectral function. \\

green\_function\_nrg\_even = YES                \# calculation of the real/imaginary part of the Green's function from the even
part of the nrg  spectral function\\
green\_function\_nrg\_odd = YES                \# calculation of the real/imaginary part of the Green's function from the odd
part of the nrg spectral function\\
green\_function\_dmnrg	   = NO		        \# calculation of the real/imaginary part of the Green's function from the spectral function\\
green\_function\_grid\_mesh = 400               \# points for the mesh when doing Hilbert transform\\
\\
\#\#\#\#\#\#\#\#\#\#\#\#\#\#\#\#\#\#\#\#\#\#\#\#\#\#\#\#\#\#\#\#\#\#\#\#\#\#\#\#\#\#\#\#\#\#\#\#\#\#\#\#\#\#\#\#\\
                $<$/SECTION-SPECTRAL\_FUNCTION\_BROADENING$>$\\
\#\#\#\#\#\#\#\#\#\#\#\#\#\#\#\#\#\#\#\#\#\#\#\#\#\#\#\#\#\#\#\#\#\#\#\#\#\#\#\#\#\#\#\#\#\#\#\#\#\#\#\#\#\#\#\#\\
\\

}
}

\chapter{License agreements}
GNU LESSER GENERAL PUBLIC LICENSE
Version 3, 29 June 2007

 Copyright (C) 2007 Free Software Foundation, Inc. $<$http://fsf.org/$>$
 Everyone is permitted to copy and distribute verbatim copies
 of this license document, but changing it is not allowed.

  This version of the GNU Lesser General Public License incorporates
the terms and conditions of version 3 of the GNU General Public
License, supplemented by the additional permissions listed below.

  0. Additional Definitions. 

  As used herein, "this License" refers to version 3 of the GNU Lesser
General Public License, and the "GNU GPL" refers to version 3 of the GNU
General Public License.

  "The Library" refers to a covered work governed by this License,
other than an Application or a Combined Work as defined below.

  An "Application" is any work that makes use of an interface provided
by the Library, but which is not otherwise based on the Library.
Defining a subclass of a class defined by the Library is deemed a mode
of using an interface provided by the Library.

  A "Combined Work" is a work produced by combining or linking an
Application with the Library.  The particular version of the Library
with which the Combined Work was made is also called the "Linked
Version".

  The "Minimal Corresponding Source" for a Combined Work means the
Corresponding Source for the Combined Work, excluding any source code
for portions of the Combined Work that, considered in isolation, are
based on the Application, and not on the Linked Version.

  The "Corresponding Application Code" for a Combined Work means the
object code and/or source code for the Application, including any data
and utility programs needed for reproducing the Combined Work from the
Application, but excluding the System Libraries of the Combined Work.

  1. Exception to Section 3 of the GNU GPL.

  You may convey a covered work under sections 3 and 4 of this License
without being bound by section 3 of the GNU GPL.

  2. Conveying Modified Versions.

  If you modify a copy of the Library, and, in your modifications, a
facility refers to a function or data to be supplied by an Application
that uses the facility (other than as an argument passed when the
facility is invoked), then you may convey a copy of the modified
version:

   a) under this License, provided that you make a good faith effort to
   ensure that, in the event an Application does not supply the
   function or data, the facility still operates, and performs
   whatever part of its purpose remains meaningful, or

   b) under the GNU GPL, with none of the additional permissions of
   this License applicable to that copy.

  3. Object Code Incorporating Material from Library Header Files.

  The object code form of an Application may incorporate material from
a header file that is part of the Library.  You may convey such object
code under terms of your choice, provided that, if the incorporated
material is not limited to numerical parameters, data structure
layouts and accessors, or small macros, inline functions and templates
(ten or fewer lines in length), you do both of the following:

   a) Give prominent notice with each copy of the object code that the
   Library is used in it and that the Library and its use are
   covered by this License.

   b) Accompany the object code with a copy of the GNU GPL and this license
   document.

  4. Combined Works.

  You may convey a Combined Work under terms of your choice that,
taken together, effectively do not restrict modification of the
portions of the Library contained in the Combined Work and reverse
engineering for debugging such modifications, if you also do each of
the following:

   a) Give prominent notice with each copy of the Combined Work that
   the Library is used in it and that the Library and its use are
   covered by this License.

   b) Accompany the Combined Work with a copy of the GNU GPL and this license
   document.

   c) For a Combined Work that displays copyright notices during
   execution, include the copyright notice for the Library among
   these notices, as well as a reference directing the user to the
   copies of the GNU GPL and this license document.

   d) Do one of the following:

       0) Convey the Minimal Corresponding Source under the terms of this
       License, and the Corresponding Application Code in a form
       suitable for, and under terms that permit, the user to
       recombine or relink the Application with a modified version of
       the Linked Version to produce a modified Combined Work, in the
       manner specified by section 6 of the GNU GPL for conveying
       Corresponding Source.

       1) Use a suitable shared library mechanism for linking with the
       Library.  A suitable mechanism is one that (a) uses at run time
       a copy of the Library already present on the user's computer
       system, and (b) will operate properly with a modified version
       of the Library that is interface-compatible with the Linked
       Version. 

   e) Provide Installation Information, but only if you would otherwise
   be required to provide such information under section 6 of the
   GNU GPL, and only to the extent that such information is
   necessary to install and execute a modified version of the
   Combined Work produced by recombining or relinking the
   Application with a modified version of the Linked Version. (If
   you use option 4d0, the Installation Information must accompany
   the Minimal Corresponding Source and Corresponding Application
   Code. If you use option 4d1, you must provide the Installation
   Information in the manner specified by section 6 of the GNU GPL
   for conveying Corresponding Source.)

  5. Combined Libraries.

  You may place library facilities that are a work based on the
Library side by side in a single library together with other library
facilities that are not Applications and are not covered by this
License, and convey such a combined library under terms of your
choice, if you do both of the following:

   a) Accompany the combined library with a copy of the same work based
   on the Library, uncombined with any other library facilities,
   conveyed under the terms of this License.

   b) Give prominent notice with the combined library that part of it
   is a work based on the Library, and explaining where to find the
   accompanying uncombined form of the same work.

  6. Revised Versions of the GNU Lesser General Public License.

  The Free Software Foundation may publish revised and/or new versions
of the GNU Lesser General Public License from time to time. Such new
versions will be similar in spirit to the present version, but may
differ in detail to address new problems or concerns.

  Each version is given a distinguishing version number. If the
Library as you received it specifies that a certain numbered version
of the GNU Lesser General Public License "or any later version"
applies to it, you have the option of following the terms and
conditions either of that published version or of any later version
published by the Free Software Foundation. If the Library as you
received it does not specify a version number of the GNU Lesser
General Public License, you may choose any version of the GNU Lesser
General Public License ever published by the Free Software Foundation.

  If the Library as you received it specifies that a proxy can decide
whether future versions of the GNU Lesser General Public License shall
apply, that proxy's public statement of acceptance of any version is
permanent authorization for you to choose that version for the
Library.

\end{document}